\documentclass[11pt,document,nofootinbib,superscriptaddress,onecolumn,preprintnumbers,balancelastpage]{article}

\pdfoutput=1
\usepackage{jheppub}


\usepackage{amsmath, amsfonts, amsthm, amssymb}
\usepackage[normalem]{ulem}

\usepackage[toc,page]{appendix}

\usepackage{bm}
\usepackage{upgreek}
\usepackage{comment}
\usepackage{ifpdf}
\usepackage{xstring}
\usepackage{setspace}
\usepackage[utf8]{inputenc}

\usepackage{hyperref}
\usepackage{graphicx}  
\usepackage{breqn}
\usepackage{mathrsfs}

\usepackage{float}
\usepackage{subfig}
\usepackage{mathtools}

\usepackage[normalem]{ulem}

\usepackage{color}

\title
{\Large \bf Hidden Sector Monopole Dark Matter with Matter Domination}

\author{Michael L. Graesser$^{1}$,}
\author{Jacek K. Osi\'{n}ski$^{1,\,2}$}

\affiliation{$^{1}$~Theoretical Division, Los Alamos National Laboratory, Los Alamos, NM 87545, USA}
\affiliation{$^{2}$~Department of Physics and Astronomy, University of New Mexico, Albuquerque, NM 87131, USA}

\emailAdd{michaelgraesser@gmail.com}
\emailAdd{jaksaosinski@gmail.com}

\preprint{LA-UR-20-25052}
\abstract{
      The thermal freeze-out mechanism for relic dark matter heavier than $O(10-100 $ TeV$)$ requires cross-sections 
      that violate perturbative unitarity. Yet the existence of dark matter heavier than these scales is certainly plausible from a particle physics perspective, pointing to the need for a 
      non-thermal cosmological history for such theories. 
      Topological dark matter is a well-motivated scenario of this kind. Here the hidden-sector dark matter can be produced in abundance through the Kibble-Zurek mechanism describing the non-equilibrium dynamics of defects 
      produced in a second order 
      phase transition. We revisit the original topological dark matter scenario, focusing on hidden-sector magnetic monopoles, and consider more general cosmological histories. 
We find that a monopole mass of order (1--$10^5$) PeV is generic for the thermal histories considered here, if monopoles are to entirely reproduce the current abundance of dark matter. In particular, in a scenario involving an early era of matter domination,
the monopole number density is always less than or equal to that in a pure radiation dominated equivalent
provided a certain condition on critical exponents is satisfied. This results in a larger monopole mass needed to account for a fixed relic abundance in such cosmologies.}

\begin{document}
\maketitle
\setcounter{page}{0}
\thispagestyle{empty}

%%%%%%%%%%%
\section{Introduction}
The period between the end of inflation and the beginning of big bang nucleosynthesis (BBN) is a natural period for the production of dark matter (DM), though it is currently inaccessible to observations. The most popular dark matter candidate has traditionally been a weakly interacting massive particle (WIMP), produced in the right abundance by thermal freeze-out in the standard thermal history of radiation domination (RD) between inflation and BBN. This standard picture is now increasingly strained, with certain models excluded by indirect searches over much of the cosmologically interesting range for the WIMP mass \cite{Abdallah:2018qtu,Baumgart:2017nsr}. Nonthermal production mechanisms, which depart from the assumptions of local thermal and chemical equilibrium of dark matter with Standard Model particles in the early Universe, and/or radiation domination, have become more widespread \cite{Baer:2014eja}. 

Spontaneous symmetry breaking in the early Universe prior to BBN provides a natural mechanism to produce interesting objects through 
an out-of-equilibrium process. Specifically, symmetry breaking via a second order phase transition can produce a large density of topological defects via the Kibble-Zurek mechanism (KZM) \cite{Kibble:1976sj, Kibble:1980mv,Zurek:1985qw,Zurek:1993ek,Zurek:1996sj}, and their density, can ``leave an immediate imprint on the Universe and will be critically important'' \cite{Zurek:1996sj}.
While the KZM theory was developed some time ago, it is only recently 
that
the theory has received firm  
experimental support, at least for describing {\em classical} second order phase transitions, as certain key predictions of the theory have been 
confirmed in laboratory settings. In particular, 
the scaling of the density of topological defects with respect to the quenching rate has been verified in a number of two- and three-dimensional materials \cite{Lin_2014,Navon167,Chomaz-2015,Beugnon_2017,PhysRevX.7.041014}.

What is of focus here, is that the KZM is a plausible nonthermal mechanism for the production of an interesting class of dark matter candidates dubbed topological dark matter \cite{Murayama:2009nj}.
A key finding of \cite{Murayama:2009nj} is that in this scenario, the dark matter mass must be of O(PeV) scale to obtain the correct relic abundance. Our main motivation for the present work is to explore the robustness of this finding, when 
other cosmological histories in the early Universe are considered.
Topological dark matter is studied by \cite{Murayama:2009nj} in the context of a standard thermal history, in which the phase transition that produces topological defects occurs during a radiation dominated era, and where the temperature of the symmetry breaking and visible sectors are 
assumed for simplicity to be equal. We explore this scenario in several different directions. 
We allow for an intervening phase of matter domination (MD) in the early Universe, during which the symmetry breaking occurs. We also allow the symmetry breaking sector to have a temperature different than that of the visible sector (VS) of Standard Model particles. For since the two sectors interact only very weakly, if at all, there is no reason to expect them to have the same temperature. 

Phases of early matter domination (EMD) in the period between inflation and BBN are a generic prediction of early Universe string constructions and are commonly achieved via moduli which acquire a pressureless equation of state and drive the Universe toward matter domination before their eventual decay \cite{Kim:1992eu,Kawasaki:1995vt,Banks:1996ea, Kim:1996wi,Hashimoto:1998ua,Asaka:1998xa,Banks:2002sd}; for a review see \cite{Kane:2015jia}. An early matter dominated era can also easily happen when a decoupled massive particle comes to dominate the energy density for some time before decaying and subsequently reheating the Universe. We will consider an era of early matter domination to be caused by either a modulus or a decoupled particle, and allow the phase transition to occur anywhere before, during, or after this era. 
Our cosmological scenario actually consists of two hidden sectors: a sector driving an early matter domination phase; and a second sector with the symmetry breaking by a second order phase transition. Couplings between these two sectors would be interesting to explore - leading to a more complicated cosmological history - but we do not do so here, simply to avoid over complicating the narrative. 

While the original work on topological dark matter \cite{Murayama:2009nj} considered the production of domain walls, strings, monopoles, or skyrmions, here we focus for simplicity on the case where the produced defects are magnetic monopoles, charged under an unbroken 
$U(1)$ left over after the phase transition.\footnote{The $U(1)$ can be broken at a much lower scale.} The abundance of magnetic monopoles charged under the $U(1)$ of electromagnetism is constrained by observations, such as the Parker limit, to be less than that required for it to account for all of the DM \cite{Parker:1970xv,Turner:1982ag}. We will therefore avoid such constraints altogether in this work by considering the simplest scenario 
in which the monopoles are not charged under electromagnetism, but instead charged under a hidden sector $U(1)$, and further, that 
the hidden sector $U(1)$ does not kinetically mix with electromagnetism, so that monopoles 
of the hidden sector do not couple to (visible sector) electromagnetic fields.\footnote{Consequences of  
kinetic mixing leading to milli-magnetically 
charged monopoles are explored in \cite{Sanchez:2011mf, Hook:2017vyc, Terning:2019bhg}.}

Our scenario begins in a radiation dominated phase after inflation, where we allow for the dominant energy 
component to be radiation in either the visible or hidden sector. As the Universe expands, each sector cools independently of the other, and we enter an early matter dominated phase caused by a modulus or by a heavy particle which has decoupled from either sector. As this phase proceeds, the dominating field continually decays into radiation in the visible sector, until the decay completes (at reheating) and we transition back to a radiation dominated phase of Standard Model particles, leading to the standard cosmology at the onset of BBN. We suppose that a second-order phase transition occurs in the hidden sector as the temperature in the hidden sector drops below some critical temperature \(T_{\rm C}^{\rm(hid)}\), resulting in a significant production of magnetic monopoles in the hidden sector due to the Kibble-Zurek mechanism. We allow the phase transition to occur at any time in the pre-BBN thermal history of our scenario. We {\em a posteriori} neglect any subsequent annihilations of monopoles due to their high mass (PeV and above) and consequently low number density. As mentioned above, we also do not consider any non-gravitational interactions between the sectors, 
other than that which provides the decay that reheats our Universe.

Our main results are shown in Figures \ref{fig:monopmod}, \ref{fig:monopdec}, and \ref{fig:parameters}. 
We generally find that hidden sector monopoles in the mass range O(1-$10^5$) PeV can be dark matter candidates, 
with values for the monopole mass giving rise to the current dark matter relic abundance correlated with other particle and cosmological parameters. Furthermore, a long intervening era of matter domination in the early Universe significantly increases the hidden sector monopole mass needed 
to obtain the current relic abundance, compared to a purely radiation-dominated history, provided that the critical exponents, defined below, satisfy \(2\nu \leq 1 + \mu\). 
An analytic argument for this observation is presented in Section \ref{n/s}, which is also confirmed by our numerical results given in subsequent sections.

We begin with an overview of monopole production via the Kibble-Zurek mechanism in Section \ref{monopoleintro}, followed by an overview of a cosmological history involving EMD in Section \ref{phiintro}. In Section \ref{n/s}, we present analytical forms for the monopole abundance in the presence of an EMD phase, including monopole production before, during, and after EMD. We then present numerical results for the cases of EMD by a modulus or a heavy decoupled particle in Sections \ref{mod} and \ref{decp} respectively.
Section \ref{sec:monopole-mass-for-relic-abundance} shows
the monopole mass and cosmological parameters that give the correct present-day 
relic abundance for dark matter, using an analytic approximation that 
we show well-describes the relic abundance obtained using numerical methods.
We conclude with a brief discussion, including a summary of important caveats to our work, in Section \ref{disc}. 

A number of detailed results are summarized in several Appendices. We include a table of notation in Appendix \ref{app:table}. 
Appendix \ref{appendix:e_f} describes 
the relation of a key cosmological parameter in our work -- the length of 
the matter-dominated phase -- to other defined cosmological 
parameters. Appendices \ref{Appendix:Freeze-out} and \ref{Appendix:Freeze-in} gather 
usual formulae for the decoupling of a relativistic particle, and Appendix \ref{Appendix:Additional-constraint} gives the constraint on cosmological parameters 
from requiring that a matter-dominated phase caused by a decoupled particle lasts at all.

%%%%%%%%%%
\section{Brief review of Kibble-Zurek mechanism theory}
\label{monopoleintro}

We now summarize 
the theory of the Kibble-Zurek mechanism 
describing the 
non-equilibrium dynamics of topological defects produced in a second order phase transition. 
We refer the reader to the original references \cite{Zurek:1985qw,Zurek:1993ek,Zurek:1996sj}
and recent review \cite{delCampo:2013nla}, which give several reasons for why (\ref{keyeqn}) shown below gives the typical distance scale between topological defects. 

In the KZM theory, a system is assumed to be driven through a second-order phase transition at temperature $T_{\rm C}$ by a quench that importantly, is 
assumed to be of a finite timescale; it is 
neither instantaneous, nor extremely long. In a cosmological context, the quench is driven by the cosmological expansion of the Universe itself, a point we return to below. 
 
If the quench is slow enough, the system has time to quasi-equilibrate and therefore as $t \rightarrow t_{\rm C}$
the correlation length continues to grow with some critical scaling, namely 
\begin{equation}
\xi(t) = \xi_0|\epsilon(t)|^{-\nu} ~,
\end{equation}
for some critical exponent $\nu$. 

The key point is that there is a time scale $t_*$ prior to the phase transition, such that for times  $t> t_*$, the correlation length exceeds the sound horizon. Subsequent to 
that time, the quench is fast compared to the timescale over which the system can respond. According 
to the KZM theory, after this cross-over time $t_*$, fluctuations become frozen, and therefore 
$\xi(t_*)$ sets the scale of the topological defects, namely 
\cite{Zurek:1985qw,Zurek:1993ek,Zurek:1996sj},
\begin{equation}
\xi(t_*)= u(t_*) |t_*-t_{\rm C}|~,
\label{keyeqn}
\end{equation}
where $u(t) = u_0 \epsilon(t)^{\mu - \nu}$, for a critical exponent $\mu$ and typical velocity $u_0$, is the characteristic velocity of perturbations in the system.\footnote{In the condensed matter literature one often 
finds a different critical exponent $z$ related to $\mu$ and $\nu$ by $\mu=z \nu$.}
%\sout{Note the characteristic velocity diverges as the critical temperature is approached, unless $\nu \leq \mu$, which is equivalent 
%to requiring $\xi \leq c \tau$.}
The characteristic correlation time scale $\tau(t)$ is then 
\begin{equation}
    \tau(t)\equiv\xi(t)/u(t)=\tau_0 \epsilon(t)^{-\mu}~ \sim \xi(t)^z,
\end{equation}
for a typical timescale $\tau_0=\xi_0/u_0$.

We now arrive at the main prediction of the KZM theory. For this finite speed quench, the frozen correlation length is then predicted to be 
\begin{equation}
    \xi(t_*) \approx \xi_{\rm 0}\bigg(\frac{\tau_{\rm Q}}{\tau_{\rm 0}}\bigg)^{\frac{\nu}{1+\mu}}~,
\end{equation}
with approximately one topological defect (monopole) produced per correlation volume \(\xi(t_*)^{-3}\) \cite{Zurek:1985qw,Zurek:1993ek,Zurek:1996sj}. The size of the frozen length scale is set by physical properties in 
$\xi_0$, $\tau_0$, and the critical exponents, and by the timescale of the quench $\tau_{\rm Q}$ set by either the laboratory conditions or by the Hubble expansion rate, depending on the context. It follows that the number density of point-like defects in $D=3$ spatial dimensions is \footnote{A more general expression for the density of $d$-dimensional defects in $D$ spatial dimensions 
can be found in \cite{delCampo:2013nla}.}
\begin{equation} 
n_{defects} \sim \xi(t_*)^{-3} \approx \tau_{\rm Q}^{-\frac{3\nu}{1+\mu}} ~.
\end{equation}
This scaling of defect density has been experimentally confirmed in a number of two and 
three dimensional condensed matter systems, such as 3-D ferroelectric crystals \cite{Lin_2014},  
2-- and 3--D Bose-Einstein condensate gases \cite{Navon167, Chomaz-2015, Beugnon_2017}, and multiferroic hexagonal manganite crystals \cite{PhysRevX.7.041014}.

A critical dynamical assumption leading to these predictions is that fluctuations in spatial regions separated by more than this correlation length are randomly oriented and, subsequent to the above cross-over time, independent of each other. 
While this is a reasonable expectation for a classical phase transition, Zurek raises a caveat for systems such as the normal-to-superfluid transition in $^4$He in which quantum mechanical effects are all important \cite{Zurek:1985qw}. Namely,  correlations between regions separated by several correlation lengths may only appear to be random and independent, but in fact could be secretly strongly correlated due to conservation laws (for the vortices 
studied in \cite{Zurek:1985qw}, notably angular momentum), in analogy to spin correlations in EPR experiments. Should this situation
occur, the predicted topological number density would be smaller and these estimates for the cosmological relic density would need to be revisited \cite{Zurek:1985qw}. 

But recent experimental results do suggest that -- at least in the case of vortex formation --
defects are indeed random and independent, reaffirming the KZM expectations. For 
the KZM theory also makes some statements about this randomness, as it specifies 
how the net winding number of 
vortices ${\cal W}$ in a 
fixed spatial region of circumference $C$ should scale with the correlation length. Specifically, 
the typical absolute value $\langle |{\cal W}| \rangle$ and dispersion 
$\sqrt{\langle {\cal W}^2 \rangle}$ are both predicted to have the same scaling at large 
$\langle |{\cal W}| \rangle \gg 1$, namely
$\sqrt{\langle {\cal W}^2 \rangle} \sim \langle |{\cal W}| \rangle
\sim \sqrt{C/\xi}$, whereas at small 
winding number the KZM predicts different scaling laws for the absolute value and dispersion of 
${\cal W}$ \cite{Zurek:1985qw,
Zurek:2013qba}. In both limits the KZM predictions for these two quantities have been 
dramatically confirmed in 3-dimensional ferroelectric crystals \cite{Lin_2014}.

In a laboratory setting, in the non-relativistic mean field approximation (i.e., Landau-Ginzburg theory), 
the potential part of the 
free-energy of a system described by an order parameter $\phi$ is approximated by the Landau-Ginzburg potential, 
\begin{equation} 
V(\phi) = (T - T_{\rm C})m|\phi|^2 + (1/2)\lambda|\phi|^4~,
\end{equation}
with the time-evolution of $\phi$ approximately described by the Gross-Pitaevskii equation, 
which is first order in time. This 
leads to the critical exponents $\mu=1$ and $\nu=1/2$, predicting $\xi(t_*) \approx \xi_0 \left(\tau_Q/\tau_0\right)^{1/4}$. 
But in a relativistic quantum field theory context the scaling laws 
are different because the equation of motion for 
$\phi$ is second-order in time. For example, 
in a cosmological context the equation of motion for 
$\phi$ leads to the critical exponents $\mu =\nu =1/2$. Here then, $\xi(t_*) \approx 
\xi_0 \left(\tau_Q/\tau_0\right)^{1/3}$
and $n_{defects} \sim\tau_{\rm Q}^{-1}$ \cite{Zurek:1996sj} \cite{Murayama:2009nj}.

As noted above, when the phase transition occurs in an expanding Universe, 
the quench time can be re-expressed in terms of the Hubble rate at the critical time as \(H_{\rm C}^{-1}\). 
To see that, first note 
that
the quench is characterized by 
\begin{equation} 
\epsilon(t) \equiv (T(t)-T_{\rm C})/T_{\rm C} ~,
\end{equation} 
where $T(t)$ is the time-dependent temperature of the system.
Close to the time of the phase transition $t_{\rm C}$, this quantity scales linearly with time, 
\begin{equation}
\epsilon(t) = (t_{\rm C}-t)/\tau_{\rm Q}~,
\end{equation}
which also defines the quenching time-scale $\tau_{\rm Q}$.
For example, in a cosmological context where the scale factor $a$ increases as $a(t)=(t/t_{\rm C})^p$, $p=2/3 ~(1/2)$ for MD (RD), 
then with $t \equiv t_{\rm C}-\Delta t$, $|\Delta t| \ll t_{\rm C}$, $\epsilon(t)=p \Delta t/t_{\rm C}$ and $\tau_{\rm Q}=t_{\rm C}/p$, or in other words, 
\begin{equation} 
\tau_{\rm Q}=H^{-1}_{\rm}(t_{\rm C})~.
\end{equation}
That is, the 
characteristic time-scale $\tau_{\rm Q}$ of the quench is always given by the 
Hubble parameter at the time of the phase transition, generalizing from the pure RD scenario given in  
\cite{Zurek:1996sj} to more general equations of state.

We take the initial correlation sizes to be set by the mass \(m_\sigma\) of the \(\sigma\) particle, which
for a pure scalar $\phi^4$ theory at weak coupling is given 
by 
$m_\sigma \simeq \lambda T_{\rm C}/4$ \cite{Dolan:1973qd}.
%(Note that in the ground state $\langle \phi^2 \rangle = T^2_{\rm C}~.$)
That is, \(\xi_{\rm 0} \approx \tau_{\rm 0} \sim  m^{-1}_\sigma \!\sim (T_{\rm C}\sqrt{\lambda})^{-1}\) \cite{Murayama:2009nj}.  Although $\mu=\nu=1/2$ is the prediction for the critical exponents in 
the approximation that the second-order phase transition is described by a weakly coupled scalar field, for our 
analysis 
we consider more general values for the critical exponents.\footnote{A KZM description of the dynamics of defects 
in the quantum phase transition of 
the quantum Ising model in one-dimension with $\mu=\nu=1$ can be found in \cite{Zurek:2006dr,Dziarmaga_2005}.} 
%whereas for a quantum phase 
%transition in a laboratory setting $(\mu=\nu=1)$, $n_{defects} \sim \tau^{-3/2}_{\rm Q}$.

In terms of cosmological quantities, the frozen correlation length is then  
\begin{equation} \label{corrlength}
    \xi(t_*) \approx m^{-1}_\sigma \left(\frac{m_\sigma}{H_{\rm C}}\right)^{\frac{\nu}{1+\mu}} = \frac{1}{T_{\rm C}\sqrt{\lambda}} \bigg(\frac{T_{\rm C}\sqrt{\lambda}}{H_{\rm C}}\bigg)^{\frac{\nu}{1+\mu}}
\end{equation}
regardless of the type of dominant energy density (matter or radiation), with the understanding that the temperature dependence of the Hubble parameter when the system is at the critical temperature, \(H_{\rm C}\), does depend on the form of the dominant energy density component.

After the phase transition is complete, the monopole number density is \(n_{\rm M} \approx \xi(t_*)^{-3}\) and the comoving number density is fixed as their abundance simply redshifts through the remaining history of the Universe. We will neglect any subsequent annihilations of monopoles because the masses needed to account for the entire current DM abundance will turn out to be quite high, with correspondingly low number densities.\footnote{The interactions between magnetic monopoles or more generally, dyons, and electric charges is a 
strongly coupled system and poorly understood. For a discussion of the annihilation rate for monopole-anti-monopole pairs in an ambient plasma, see \cite{Preskill:1979zi}.}

For a general second order phase transition, quantum or classical, in the KZM theory the frozen correlation length  
setting the density of topological defects depends only on the critical temperature of the phase transition, the typical 
timescale of the quench, and the critical exponents. For a classical Landau-Ginzburg second order phase transition, 
however, the mass 
of the defect -- here the monopole mass $m_{\rm M}$ --
is not independent of the critical temperature. 
For a 't Hooft-Polyakov monopole, 
$m_{\rm M}=h T_{\rm C}$, with 
$h$ the magnetic coupling $2 \pi/e_h$, and recall that $\phi \sim T_{\rm C}$. Thus for a classical phase transition, the monopole 
mass and critical temperature are parametrically at the same scale.
Throughout this work we will assume the monopoles are produced in the early Universe by a 
{\em classical} second order phase transition, so the implied relation 
between the critical temperature and monopole mass is an important caveat to many of our results.

But such a mass--temperature ($m-T$) relation 
is not expected to be true in general. On the contrary, one expects   
the monopole mass and critical temperature to be unrelated. The $N=2$ Seiberg-Witten theory 
\cite{Seiberg:1994rs,Seiberg:1994aj} is a prominent example 
of this kind, where near certain points on the moduli space
the low-energy theory contains nearly massless composite particles charged under a magnetic $U(1)$. Here one would like 
to know whether the theory ends up near these points as the theory is cooled through 
the phase transition, and what the order of the transition is. For the former question, 
the answer is affirmative, at least in the pure $N=2$ $SU(2)$ theory \cite{Paik:2009iz}.
The latter remains an open question.

Because of this expectation, we will indicate 
which of our results are independent of any assumption about a $m-T$ relation. The most 
important of these is the ratio 
of the monopole number density to photon entropy density, such as \eqref{nMsVScbefore}, \eqref{nMsVSreh}, and 
\eqref{nMsVSc} given below. In the low-density limit where monopole annihilations 
are negligible, these depend only on the critical temperature but not the monopole mass.

As previously mentioned, 
we will also vary the critical exponents $\mu$ and $\nu$ away from the Landau-Ginzburg value of 1/2, 
as a guide to future work.

%%%%%%%%%%
\section{Summary of the cosmological history with an early matter-dominated era} \label{phiintro}

In order to proceed, we must address the relationship between the Hubble expansion rate and the temperatures of the different radiation components of the Universe. In this section we therefore introduce the general expansion history we will be considering, define terminology, and obtain relations between the Hubble parameter and key parameters during the different eras prior to reheating. 

First, we begin with radiation domination (RD) by either the hidden or visible sector (or any combination) some time after inflation, with other energy densities comparatively negligible. In this era, the Hubble expansion rate is given by

\begin{equation} \label{HTRD}
    H^2 = \frac{\big(\rho_{\rm r}^{\rm(vis)} + \rho_{\rm r}^{\rm(hid)}\big)}{3M_{\rm P}^2} = \frac{\pi^2}{90}g_*^{\rm(hid)}(1 + f) \frac{{T^{\rm(hid)}}^4}{M^2_{\rm P}}~,
\end{equation}
where the second equation implicitly defines the factor \(f \equiv \rho_{\rm r}^{\rm(vis)}/\rho_{\rm r}^{\rm(hid)}\) as the ratio of the radiation energy densities of the visible and hidden sectors. Also, \(T^{\rm(hid)}\) is the temperature of the HS, \(g_*^{\rm(hid)}\) is the number of relativistic degrees of freedom in the HS at temperature \(T^{\rm(hid)}\), and \(M_{\rm P} \approx 2.4\times 10^{18}\,{\rm GeV}\) is the reduced Planck mass. 
In this period, the factor \((1 + f)\) is well approximated by its initial value \((1 + f_{\rm i})\) regardless of the distribution of initial radiation among the two sectors, and we will make this substitution when using \eqref{HTRD} below. 

We consider the visible and hidden sectors to have independent temperatures, each with their own \(g_*\) factors depending on the specific particle content (Standard Model for the visible sector), and we could have equivalently expressed \eqref{HTRD} in terms of visible sector quantities. The \(g_*\) factors of course depend on the temperature of their respective sector, but we will treat \(g_*^{\rm(hid)}\) as roughly constant at high temperatures in order to avoid overly specifying the details of the HS. 

We achieve early matter domination (EMD) through the presence of a scalar modulus, or by the decoupling of a heavy particle from either the hidden or visible sectors during this initial RD phase. 
In both cases we refer to the modulus and the heavy particle as \(\Phi\), and based on the context, there should not be any confusion. We 
assume that $\Phi$ couples to lighter particles through higher dimension operators suppressed by the Planck scale, with a decay rate 
\begin{equation}\label{gammaphi}
    \Gamma_\Phi \sim \frac{\alpha^2}{2 \pi}\frac{m_\Phi^3}{M_{\rm P}^2}~,
\end{equation}
where \(m_\Phi\) is the \(\Phi\) mass. We have also included a possible loop factor \(\alpha\) in the case that \(\Phi\) decay occurs predominantly through a loop, but we will set \(\alpha = 1\)  throughout unless otherwise noted. The decay is complete when \(H \approx H_{\rm RH} \equiv \Gamma_\Phi\), which marks the approximate time of reheating, and we avoid having significant amounts of left over hidden radiation by requiring \(\Phi\) to decay predominantly to the Standard Model particles,  
\begin{equation} \label{HRH}
    H^2_{\rm RH} = \frac{\pi^2}{90}g_{*\rm RH}^{\rm(vis)}\bigg(1 + \frac{1}{f_{\rm RH}}\bigg)\frac{{T_{\rm RH}^{\rm(vis)}}^4}{M^2_{\rm P}} ~,
\end{equation}
where \(T_{\rm RH}^{\rm(vis)}\) is the visible sector temperature at reheating, and \(g_{*\rm RH}^{\rm(vis)}\) is the number of relativistic degrees of freedom in the visible sector at this temperature. In order to preserve standard Big Bang Nucleosynthesis (BBN), the visible sector reheat temperature must be larger than O(10 MeV). 

The ratio of the visible sector radiation energy density to that of the HS at reheating, denoted by \(f_{\rm RH}\), depends on the duration of the EMD phase as well as the initial factor \(f_{\rm i}\), but is typically large due to our visible sector reheating requirement, and thus always satisfies \(f_{\rm RH} > 1\) and \(f_{\rm RH} >\! f_{\rm i}\) (this statement is demonstrated in Appendix \ref{appendix:e_f}). This conclusion, together with our assumption that $\Phi$ predominantly decays to SM particles, ensures that the temperature of the HS at reheating, \(T_{\rm RH}^{\rm(hid)}\), is correspondingly always smaller than that of the visible sector. We also point out that this ratio remains fixed after reheating due to the absence of any further decays. From \eqref{gammaphi} and \eqref{HRH}, we additionally see that a given choice for the visible sector reheat temperature and $\alpha$ determines a corresponding \(\Phi\) mass.

In order to have a well defined EMD phase, we assume the energy density of \(\Phi\) is large enough to dominate well-before reheating. During EMD, the scaling of the Hubble rate with the visible sector temperature is altered from a typical MD redshift relation because the visible sector is fed by the decay of \(\Phi\); however, from entropy conservation, the scaling of $H$ with the HS temperature remains unaffected: \(H^2 \propto {h_{*}}^{\rm(hid)} {T^{\rm(hid)}}^3\). Based on the initial energy density of VS radiation, there can be a phase of ordinary redshift for the VS temperature even during EMD, but once the effect of the decay wins over this dilution, the relation becomes (see (20) of \cite{Giudice:2000ex} for a derivation):
\begin{equation} \label{HTEMD}
    H = \frac{\pi\sqrt{10}}{12} \frac{g_*^{\rm(vis)}}{\sqrt{g_{*\rm RH}^{\rm(vis)}}} \frac{{T^{\rm(vis)}}^4}{{T_{\rm RH}^{\rm(vis)}}^2M_{\rm P}}.
\end{equation}
This relation is always true just before reheating, but may not start until deep within the EMD phase if the initial VS radiation energy density is large.\footnote{If the VS radiation energy density is larger than the instantaneous contribution from the decay of \(\Phi\) at a given time, the VS radiation will evolve via ordinary redshift. Once the energy density is sufficiently diluted for the decay contribution to become dominant, \eqref{HTEMD} is valid. One can see this by analyzing the system in \eqref{boltzmodVR}. For more on the effects of a large abundance of radiation during EMD, see \cite{Drees:2017iod,Allahverdi:2019jsc}.}

At the end of the EMD phase, once reheating completes, we enter the RD era with the Hubble rate given by  
\begin{equation} \label{HTVSRD}
    H^2 = \frac{\pi^2}{90}g_*^{\rm(vis)}\bigg(1 + \frac{1}{f_{\rm RH}}\bigg) \frac{{T^{\rm(vis)}}^4}{M^2_{\rm P}}~,
\end{equation}
where the factor \(f_{\rm RH}\) is large such that the visible sector is dominant, thus recovering the standard thermal history leading up to BBN. 

\section{Monopole production with an era of early matter domination} \label{n/s}

Recall that we are interested in producing monoples during a second order phase transition occurring in a hidden sector, so the critical temperature appearing in \eqref{corrlength} refers to the temperature of the hidden sector at the critical time. In this section we address monopole production in the context of the thermal history presented in the previous section. The effects of EMD on the monopole abundance can be understood regardless of the mechanism for establishing MD in this early period, and we obtain analytical expressions below that do not depend on the identity of the field \(\Phi\). In addition to the start time of EMD, what matters is that the dominant energy density component decays to visible sector radiation at a rate \(\Gamma_\Phi\), thus setting the end time of EMD. The overall effect is to slow the redshift of visible sector radiation relative to the HS such that only the visible sector is dominant after EMD even if it was not initially. Because we only consider HS magnetic monopoles, this offset in the visible sector and HS temperatures generally results in a lower number density of monopoles of a given mass, where the magnitude of the offset is determined by the duration of EMD and the initial abundances of visible and hidden radiation. 

We label the start of EMD by \(H = H_{\rm MD}\), with visible and HS temperatures \(T_{\rm MD}^{\rm(vis)}\) and \(T_{\rm MD}^{\rm(hid)}\) respectively, and the end of the EMD phase occurs when \(H \approx \Gamma_\Phi\). Recall that the visible sector reheat temperature, which we restrict to be larger than O(10 MeV) such that reheating occurs before BBN, is the primary parameter that determines the end of EMD.

\subsection{Case I: phase transition occurs before EMD}

We will start with the case where the HS phase transition occurs in the RD period before EMD, resulting in a frozen monopole number density that is redshifted through the remainder of the RD phase as well as the full EMD period. This results in considerable dilution and a need for higher monopole masses in order to maintain a fixed contribution to the energy density of the Universe. Using \eqref{corrlength} and recalling that the number density of monopoles produced in the phase transition is approximately one per correlation volume, we have 
(see Appendix \ref{app:table} for a table of notation)
\begin{eqnarray}
    (n_{\rm M})_{\rm RH}^{\rm(before)} &=&
\xi(t_*)^{-3} \left(\frac{T^{\rm (hid)}_{\rm MD}}{T_{\rm C}^{\rm(hid)}}\right)^3 \left(\frac{a_{\rm MD}}{a_{\rm RH}}\right)^3 
%\nonumber \\
%&=& 
=\xi(t_*)^{-3}  \left(\frac{H_{\rm MD}}{H_{\rm C}}\right)^{3/2} \left(\frac{\Gamma_{\rm \Phi}}{H_{\rm MD}}\right)^2 ~,
\end{eqnarray}
where the first factor in parentheses on the right-side accounts for the redshift of the 
monopole number density from the critical time to the start of EMD, and the second factor gives the redshift from the start of EMD to reheating. 
We have also defined $a_{\rm MD}$ and $a_{\rm RH}$ to be the scale factors at the onset of 
matter domination and at reheating, respectively.
At this point we do not need to redshift any further, and can obtain a fixed comoving abundance by normalizing by the visible sector entropy density at reheating, as both number density and entropy density dilute as the cube of the scale factor once the significant entropy production from reheating stops. This leads to 
\begin{eqnarray} \label{nMsVScbefore}
    \bigg(\frac{n_{\rm M}}{s^{\rm(vis)}}\bigg)_{\rm RH}^{\rm(before)} &=&
\xi(t_*)^{-3}
\left(\frac{H_{\rm MD}}{H_{\rm C}}\right)^{3/2} \left(\frac{\Gamma_{\rm \Phi}}{H_{\rm MD}}\right)^2
 /(2 \pi^2 h_{*\rm RH}^{\rm(vis)}{T_{\rm RH}^{\rm(vis)}}^3/45) \nonumber \\     
   &=& \frac{45(T_{\rm C}^{\rm(hid)}\sqrt{\lambda})^{3 - \frac{3\nu}{1+\mu}}H_{\rm C}^{\frac{3\nu}{1+\mu}}}{2\pi^2 h_{*\rm RH}^{\rm(vis)}{T_{\rm RH}^{\rm(vis)}}^3}\bigg(\frac{\Gamma_\Phi^2}{H_{\rm C}^{3/2}H_{\rm MD}^{1/2}}\bigg)~.
\end{eqnarray}

The factor \(h_*^{\rm(vis)}\) tracks the visible sector relativistic degrees of freedom for entropy and is nearly equal to \(g_*^{\rm(vis)}\) for the high temperatures in our scenario as well as the low temperature today \cite{Drees:2017iod, Tanabashi:2018oca} (it is evaluated at reheating in the expression above, as indicated by the subscript). Note that the Hubble rate at the critical time is given by \eqref{HTRD}. 

\subsection{Case II: phase transition occurs during EMD}

If the phase transition occurs during the EMD phase, the frozen monopole number density only redshifts through the remaining duration of EMD, and we have

\begin{eqnarray}
    (n_{\rm M})_{\rm RH}^{\rm(during)} &= & \xi(t_*)^{-3}  \left(\frac{a_{\rm C}}{a_{\rm RH}}\right)^3
   % \nonumber \\
    %&=& 
   = \xi(t_*)^{-3} \bigg(\frac{\Gamma_\Phi}{H_{\rm C}}\bigg)^2~.
\end{eqnarray}
Again normalizing to the visible sector entropy density at reheating, one has  
\begin{equation} \label{nMsVSreh}
    \bigg(\frac{n_{\rm M}}{s^{\rm(vis)}}\bigg)_{\rm RH}^{\rm(during)} = \frac{45(T_{\rm C}^{\rm(hid)}\sqrt{\lambda})^{3 - \frac{3\nu}{1+\mu}}H_{\rm C}^{\frac{3\nu}{1+\mu}}}{2\pi^2 h_{*\rm RH}^{\rm(vis)}{T_{\rm RH}^{\rm(vis)}}^3}\bigg(\frac{\Gamma_\Phi}{H_{\rm C}}\bigg)^2
\end{equation}
 The dependence of \(H_{\rm C}\) on the HS temperature is that of ordinary MD redshift, while the relation to the visible sector temperature is more complicated, 
 for it depends on how much visible sector radiation was present at the onset of EMD. If the visible sector energy density
 at \(H = H_{\rm MD}\) is greater than the subsequent contribution from the decay of \(\Phi\) at \(H = H_{\rm MD}\), then to evaluate $H_{\rm C}$ one 
 will need to include the effect of a period of ordinary MD redshift for the visible sector temperature as well. Once the decay contribution takes over well within the EMD phase, we have the relation \eqref{HTEMD}. We note that this modified scaling can begin much earlier, even before EMD, if the initial visible sector radiation energy density is small. 

\subsection{Case III: phase transition occurs after EMD}

Finally, if the phase transition occurs in the RD period after reheating but still before BBN, so as to leave the later evolution of the Universe unchanged, the abundance can be evaluated directly at the critical time, without need of redshifting: 
\begin{eqnarray} \label{nMsVSc}
    \bigg(\frac{n_{\rm M}}{s^{\rm(vis)}}\bigg)_{\rm C}^{\rm(after)} &=& \xi(t_*)^{-3}/
    (2 \pi^2 h_{*\rm C}^{\rm(vis)}{T_{\rm C}^{\rm(vis)}}^3/45)
   % \nonumber \\ 
    %&=&
    =\frac{45(T_{\rm C}^{\rm(hid)}\sqrt{\lambda})^{3 - \frac{3\nu}{1+\mu}}H_{\rm C}^{\frac{3\nu}{1+\mu}}}{2\pi^2 h_{*\rm C}^{\rm(vis)}{T_{\rm C}^{\rm(vis)}}^3}~.
\end{eqnarray}
This expression is also valid for a thermal history that does not involve EMD at all, where the HS radiation energy density is lower than or equal to that of the visible sector by a constant factor, as both energy densities simply redshift with time. The Hubble rate at the critical time is given by \eqref{HTVSRD} in terms of visible sector quantities, but is easily related to the corresponding HS quantities by multiplying by the square root of the constant factor. 

Finally, we note that all of the results in these three subsections are independent of any 
possible relation between the monopole mass and the critical temperature.

\subsection{Monopole production: analytic approximation at boundaries}

In this subsection we obtain analytical expressions to better understand the effect of EMD in more detail. 
The three cases of monopole production described above are separated by production at the start and end of EMD, and we can easily obtain expressions below for the monopole abundance corresponding to these boundaries. 

%\subsubsection{Monopole production at the start of EMD} 

For production at the start of EMD, the HS temperature at the critical point is \(T_{\rm C}^{\rm(hid)} \!= T_{\rm MD}^{\rm(hid)}\) with corresponding \(H_{\rm C} =\! H_{\rm MD}\). From \eqref{HTRD} and \eqref{nMsVSreh}, we obtain the frozen abundance of monopoles at reheating: 
\begin{equation} \label{n/sMDstart}
    \bigg(\frac{n_{\rm M}}{s^{\rm(vis)}}\bigg)_{\rm RH}^{\rm(start)} = \frac{45\lambda^{\frac{3}{2} - \frac{3\nu}{2(1+\mu)}}\bigg(\frac{\pi^2}{90}g_{*\rm MD}^{\rm(hid)}(1 + f_{\rm i})\bigg)^{\frac{3\nu}{2(1+\mu)} - 1}{T_{\rm C}^{\rm(hid)}}^{\frac{3\nu}{1+\mu} - 1}\Gamma_\Phi^2}{2\pi^2 h_{*\rm RH}^{\rm(vis)}{T_{\rm RH}^{\rm(vis)}}^3M_{\rm P}^{\frac{3\nu}{1+\mu} - 2}}~.
\end{equation}
Aside from the parameters of the phase transition, the final abundance is determined by the visible sector reheat temperature, the initial ratio of visible sector to HS radiation, and the monopole mass. 

%\subsubsection{Monopole production at the end of EMD} 

Monopole production at the end of EMD corresponds to a HS critical temperature of \(T_{\rm C}^{\rm(hid)} = T_{\rm RH}^{\rm(hid)}\), with \(H_{\rm C} = H_{\rm RH} = \Gamma_\Phi\). This results in a frozen monopole abundance of
\begin{equation} \label{n/sMDend}
    \bigg(\frac{n_{\rm M}}{s^{\rm(vis)}}\bigg)_{\rm RH}^{\rm(end)} = \frac{45\big(T_{\rm C}^{\rm(hid)}\sqrt{\lambda}\big)^{3 - \frac{3\nu}{1+\mu}}\Gamma_\Phi^{\frac{3\nu}{1+\mu}}}{2\pi^2 h_{*\rm RH}^{\rm(vis)}{T_{\rm RH}^{\rm(vis)}}^3}~,
\end{equation}
with the implicit relation between $\Gamma_\Phi$ and $T_{\rm RH}^{\rm(vis)}$ given by (\ref{HRH}). 
Note that this expression does not depend on the initial ratio of radiation energy densities as it only involves the time of reheating. 

Requiring EMD to start before reheating, these two expressions for production at the boundaries of EMD significantly constrain the allowed parameter space. For a realistic scenario, even the shortest EMD period will have a finite duration such that EMD is well defined, ensuring that we never quite access the limiting case where the start and end of EMD are coincident. This case, rather, corresponds to the absence of EMD altogether. 

\subsection{Present-day hidden sector monopole abundance}

We will now obtain the present day relic abundance of monopoles. In the three main cases of monopole production -- before, during, or after EMD -- as well as the two boundary cases of production at the start and end of EMD, the parameters \(\mu\), \(\nu\), and \(\lambda\), are determined by the details of the phase transition, as is the ratio \(x_{\rm M} \equiv m_{\rm M}/T_{\rm C}^{\rm(hid)}\).  
The ratio $x_{\rm M}$ is the magnetic coupling, and typically has a value of \(\mathcal{O}(10)\) \cite{Murayama:2009nj} -- we will assume \(x_{\rm M} = 50\) in our numerical results below. The current abundance of monopoles, expressed as a fractional energy density \(\Omega_{\rm M}h^2\), is related to the frozen abundance provided in the previous sections by

\begin{equation}\label{omegah2}
    \Omega_{\rm M}h^2 = \Omega_\gamma h^2 \frac{2h_{*\rm 0}^{\rm(vis)}m_{\rm M}}{3T_{\rm 0}^{\rm(vis)}}\bigg(\frac{n_{\rm M}}{s^{\rm(vis)}}\bigg)_{\rm 0} = \Omega_\gamma h^2 \frac{2h_{*\rm 0}^{\rm(vis)}m_{\rm M}}{3T_{\rm 0}^{\rm(vis)}}\bigg(\frac{n_{\rm M}}{s^{\rm(vis)}}\bigg)_{\rm RH/C}^{\rm(EMD)}~,
\end{equation}
where \(\Omega_\gamma h^2=2.47\times10^{-5}\) 
corresponds to the current photon energy density, 
$\rho^{\rm(vis)}_{\gamma,0} = 2 \pi^2 T_{\rm 0}^{\rm(vis)4}/30$. 
Also, 
\(h_{*\rm 0}^{\rm(vis)}=43/11=3.91\) is the present-day era total entropy density pre-factor, assuming three massless species of neutrinos. 
The subscript `\(0\)' labels the current era, and the 
final term labeled by `(EMD)' refers to any one of the five above cases. 
The subscript `{\rm RH}' on the final term means this quantity is evaluated 
at reheating if the phase transition 
occurs before reheating, whereas in the circumstance 
that the phase transition occurs after reheating, `{\rm C}' means the quantity is simply evaluated at 
the time of the phase transition.
In order for monopoles to constitute all of dark matter, the value of \(\Omega_{\rm M}h^2\) must reach the observed value of \(0.12\) \cite{Aghanim:2018eyx}. 

For comparison with our numerical results in subsequent sections, analytical expressions for \(\Omega_{\rm M}h^2\) can be obtained in the three main periods of our scenario by noting that 
\begin{equation}\label{HCregions}
    H^2_{\rm C} \approx 
    \begin{cases} 
        \frac{\pi^2}{90}g_{*}^{\rm(hid)}(1 + f_{\rm i})\frac{{T_{\rm C}^{\rm(hid)}}^4}{M^2_{\rm P}} & \quad {\rm (case~I:~before)}\\[5pt]
        \left(\frac{\pi^2}{90}g_{*}^{\rm(hid)}(1+f_{\rm i})\right)^{3/4}\frac{H_{\rm MD}^{1/2}{T_{\rm C}^{\rm(hid)}}^{3}}{M_{\rm P}^{3/2}} & \quad {\rm (case ~II:~during)}\\[5pt]
        \frac{\pi^2}{90}g_{*}^{\rm(hid)}(1 + f_{\rm RH})\frac{{T_{\rm C}^{\rm(hid)}}^4}{M^2_{\rm P}} & \quad {\rm (case~III:~after)}
    \end{cases}
\end{equation}
where the cases refer to monopole production before, during, or after the EMD phase. In the period before EMD, we have the RD relation \eqref{HTRD}, while in the period after EMD we have this same functional form, but with a different constant factor offsetting the visible sector and HS radiation energy densities. The expression for $H_{\rm C}$ during EMD is obtained by using entropy conservation in the hidden-sector radiation, together with redshifting during the EMD era between the start of EMD to when the temperature of the hidden sector reaches $T^{(\rm hid)}_{\rm C}$. Because the HS is not being fed by the decay of \(\Phi\), the relation is that of standard MD: \({h_*^{\rm(hid)}}^{1/3}T^{\rm(hid)} \propto H^{2/3}\). 

Next, using \eqref{nMsVScbefore}, \eqref{nMsVSreh}, \eqref{nMsVSc}, \eqref{omegah2}, and \eqref{HCregions}, one obtains analytical estimates for the monopole abundance produced in the three periods by direct substitution \footnote{Expressions at the boundaries of EMD can similarly be obtained by using \eqref{n/sMDstart} and \eqref{n/sMDend} along with the corresponding values of \(H_{\rm C}\).}
\begin{equation}\label{omegabefore}
    \frac{(\Omega_{\rm M}h^2)^{\rm(before)}}{\Omega_\gamma h^2} \approx 
\left(\frac{15\lambda^{3/4}h_{*\rm    0}^{\rm(vis)}\Gamma_\Phi^2m_{\rm M}^{5/2}}
 {\pi^2x_{\rm    M}^{3/2}h_{\rm*RH}^{\rm(vis)}{T_{\rm RH}^{\rm(vis)}}^3T_{\rm 0}^{\rm(vis)}
H_{\rm MD}^{1/2}}\right)\left(\frac{\pi^2g_{*}^{\rm(hid)}(1+f_{\rm i})m_{\rm M}^2}{90\lambda x_{\rm M}^2
M_{\rm P}^2}\right)^{\frac{3\nu}{2(1+\mu)}-\frac{3}{4}}~,
\end{equation}

\begin{equation}\label{omegaduring}
    \frac{(\Omega_{\rm M}h^2)^{\rm(during)}}{\Omega_\gamma h^2} \approx \left(\frac{15\lambda^{1/2}h_{*\rm 0}^{\rm(vis)}\Gamma_\Phi^2m_{\rm M}^2}{\pi^2x_{\rm M}h_{\rm*RH}^{\rm(vis)}{T_{\rm RH}^{\rm(vis)}}^3T_{\rm 0}^{\rm(vis)}}\right)\left(\frac{\pi^2g_{*}^{\rm(hid)}(1+f_{\rm i})H_{\rm MD}^{2/3}m_{\rm M}^{4/3}}{90\lambda^{4/3}x_{\rm M}^{4/3}M_{\rm P}^2}\right)^{\frac{9\nu}{8(1+\mu)}-\frac{3}{4}}~,
\end{equation}

\begin{equation}\label{omegaafter}
    \frac{(\Omega_{\rm M}h^2)^{\rm(after)}}{\Omega_\gamma h^2} \approx \left(\frac{15\lambda^{3/4}h_{*\rm 0}^{\rm(vis)}\Gamma_\Phi^{3/2}m_{\rm M}^{5/2}}{\pi^2x_{\rm M}^{3/2}h_{\rm*RH}^{\rm(vis)}{T_{\rm RH}^{\rm(vis)}}^3T_{\rm 0}^{\rm(vis)}}\right)\left(\frac{\pi^2g_{*}^{\rm(hid)}(1+f_{\rm RH})m_{\rm M}^2}{90\lambda x_{\rm M}^2M_{\rm P}^2}\right)^{\frac{3\nu}{2(1+\mu)}-\frac{3}{4}}~.
\end{equation}

\newpage
In the model-independent discussion of this section, the Hubble rate at the onset of EMD has been an independent parameter. In Sections \ref{mod} and \ref{decp} below, where we address two examples for establishing a period of EMD, we provide expressions for \(H_{\rm MD}\) in terms of the underlying model parameters.

It is useful to extract the functional dependence of the energy density of monopoles on the monopole mass, produced during any of the three periods of before, during, or after EMD. From \eqref{omegabefore}--\eqref{omegaafter} above, we have 
\begin{equation} \label{Omega-mass}
    \Omega_{\rm M}h^2 \propto 
\begin{cases}  m_{\rm M} \cdot
    m_{\rm M}^{\frac{3\nu}{1+\mu}} & {\rm (RD)} \\[5pt] m_{\rm M} \cdot m_{\rm M}^{\frac{3\nu}{2(1+\mu)}} & {\rm (EMD)}~.
\end{cases}
\end{equation}
Here we have factored the dependence of the energy density on the mass into an explicit factor arising from the mass itself, and an implicit factor due to the number density. The RD case applies to monopole production both before and after EMD, and we have again assumed a constant factor, \(x_{\rm M}\), between the monopole mass and $T^{\rm (hid)}_{\rm C}$. 
Note that in general, the type of cosmology in which the phase transition occurs -- here either an EMD or RD era -- affects 
the monopole energy and number densities through a different power-law dependence on the critical exponents. 

Before moving on to consider specific scenarios for establishing EMD, we can see that, 
depending on the relative sizes of the critical exponents,
the presence of an intervening EMD phase in the period before BBN can push the preferred monopole mass for DM higher than in a purely RD equivalent.
For the two prefactors in \eqref{Omega-mass} are not 
the same in each case. Fixing the phase transition parameters (\(\mu\), \(\nu\), \(\lambda\), and \(x_{\rm M}\)) as well as the monopole mass, \(m_{\rm M}\), we must first identify the equivalent RD scenario, which comes down to specifying the
constant factor \(f^{(\rm RD)}\) between the VS and HS radiation energy densities in the RD scenario. We obtain this by decreasing the duration of EMD until we arrive at the limiting RD scenario to use for comparison. If EMD is preceded by a period of RD by the VS, the limiting scenario is one which preserves the initial ratio of VS-to-HS radiation: \(f^{(\rm RD)} = f_{\rm i}\). However, if HS radiation is dominant before EMD, the limiting case is one of \(f^{(\rm RD)} = 1\) because we wish to avoid RD by the HS at the onset of BBN. 
In short, 
\begin{equation}
f^{(\rm RD)} \equiv {\rm max}(1,f_{\rm i})~,\label{fR-def}
\end{equation}
and consequently, \(f^{(\rm RD)} \geq f_{\rm i}\). 

To proceed, for all three cases
we define the ratio of the scale factors at reheating and the onset of the EMD phase 
to be
\begin{equation}\label{e_f_def}
    e_{\rm f} \equiv \frac{a_{\rm RH}}{a_{\rm MD}} \geq 1
    \,, 
\end{equation}
which we show in Appendix \ref{appendix:e_f} (see \eqref{eq:fRH-ef}) to be equivalent to  
$f^{\rm (EMD)}_{\rm RH} \simeq (1+ f_i) e_{\rm f}$, 
and because of \eqref{f_EMD/RD-2}, 
\begin{eqnarray}\label{e_f_rho}
\frac{f_{\rm RH}^{\rm(EMD)}}{f^{\rm(RD)}} & \simeq & e_{\rm f}~.
\end{eqnarray}

For all three cases, \(f^{\rm (EMD)}_{\rm RH}\) is always larger than \(f^{\rm(RD)}\) by the factor \(e_{\rm f} > 1\), 
so long as \(\Phi\) preferentially decays to the VS. The factor \(e_{\rm f}\) is fixed for a given EMD phase, regardless of the value of \(f_{\rm i}\) or the timing of the phase transition.

Using \eqref{nMsVScbefore}--\eqref{nMsVSc}, and recalling that the HS temperature redshifts as \(T^{\rm(hid)} \propto a^{-1}\) in all periods of our scenarios, be they EMD or RD, we arrive at the ratio of the current monopole abundance between an EMD and a pure RD scenario: 
\begin{equation} \label{EMD/RD}
    \frac{\Omega_{\rm M}^{(\rm EMD)}}{\Omega_{\rm M}^{(\rm RD)}} = \frac{1}{e_{\rm f}^{3/4}}
\begin{cases} 
    \Big(\frac{1+f_{\rm C}^{\rm (EMD)}}{1+f^{\rm (RD)}}\Big)^{\frac{3\nu}{2(1+\mu)}} & {\rm (case ~I:~before)} \\[5pt]
    \Big(\frac{1+f_{\rm RH}^{\rm (EMD)}}{1+f^{\rm (RD)}} \Big)^{\frac{3\nu}{2(1+\mu)}}  \left( \frac{\left(T^{\rm (hid)}_{\rm RH}\right)^{\rm (EMD)}}{T^{\rm (hid)}_{\rm C}} \right)^{\frac{3\nu}{2(1+\mu)}}& {\rm (case~II:~during)} \\
    \Big(\frac{1+f_{\rm C}^{\rm (EMD)}}{1+f^{\rm (RD)}}\Big)^{\frac{3\nu}{2(1+\mu)}} \frac{\big(h_{\rm*C}^{\rm(vis)}\big)^{\rm(RD)}}{\big(h_{\rm*C}^{\rm(vis)}\big)^{\rm(EMD)}} \Bigg(\frac{\big(g_{\rm*C}^{\rm(vis)}\big)^{\rm(EMD)}}{\big(g_{\rm*C}^{\rm(vis)}\big)^{\rm(RD)}}\Bigg)^{3/4} & {\rm (case~III:~after)}
\end{cases}
\end{equation}
As with the previous expressions \eqref{nMsVScbefore}, \eqref{nMsVSreh}, and 
\eqref{nMsVSc} given above for the ratio of monopole number density to visible sector entropy density, 
in deriving these equations we have not made use of any relationship between the monopole mass 
and the temperature of the phase transition.

In all three cases, the products involving \(f\)'s and the critical exponents are the ratios of the monopole number densities produced at the critical time between the EMD and RD scenarios. We note that 
since 
$T^{\rm (hid)}_{\rm C}$ 
and $\lambda$ appearing in the correlation length \eqref{corrlength} are fixed between the two scenarios, this 
ratio is simply 
given by the ratio of Hubble parameters $H_{\rm C}$. In the first two cases, we normalize the monopole number densities by the VS entropy density at the time of reheating (when the VS temperature is equal to the reheat temperature), accounting for the redshift factors, while in the third case, because monopole production occurs in RD after EMD, there is no need for redshifting, and we normalize by the VS entropy densities at the critical time. The factor of \(1/e_{\rm f}^{3/4}\), in the first two cases, is the ratio of the redshift factors from the time of monopole production to the time when \(T^{\rm(vis)} = T_{\rm RH}^{\rm(vis)}\) between the EMD and RD scenarios respectively, while in the third case, it, along with the terms involving the relativistic degrees of freedom, comes from the ratio of entropy densities at the critical time between the two scenarios. Note that the relativistic degrees of freedom in the VS can be different at the critical time between the EMD and RD scenarios because it is the HS critical temperature, not the visible, that is the same across the scenarios.

We note in the limit of no EMD phase, the above 
expressions for the three cases smoothly go over to $\Omega_{\rm M}^{(\rm EMD)}=\Omega_{\rm M}^{(\rm RD)}$. 
For cases I and III this statement is readily apparent, since in this limit $e_{\rm f} \rightarrow 1$, $f^{\rm (EMD)} \rightarrow f^{\rm (RD)}$, and $\big((g,h)^{\rm (vis)}_{\rm*C}\big)^{\rm (EMD)} \rightarrow \big((g,h)^{\rm (vis)}_{\rm*C}\big)^{\rm (RD)}$. To see that for case II requires one additional remark. By definition of this scenario, $(T^{\rm (hid)}_{\rm RH})^{\rm (EMD)} \leq T^{\rm (hid)}_{\rm C} \leq (T^{\rm (hid)}_{\rm MD})^{\rm (EMD)}$, so as the EMD phase disappears,
$(T^{\rm (hid)}_{\rm RH})^{\rm (EMD)} \rightarrow T^{\rm (hid)}_{\rm C}$, 
and also $(T^{\rm (hid)}_{\rm MD})^{\rm (EMD)} \rightarrow T^{\rm (hid)}_{\rm C}$. Thus in this limit, for case II we 
also have $\Omega_{\rm M}^{(\rm EMD)} \rightarrow \Omega_{\rm M}^{(\rm RD)}$.

We next discuss the conditions under which $\Omega_{\rm M}^{(\rm EMD)} \leq \Omega_{\rm M}^{(\rm RD)}$ and {\em vice versa}.
\begin{itemize}
\item For case I, of monopole production before EMD, the right-side of \eqref{EMD/RD} is always less than one. To see that, first focus on the ratio of 
$f$ factors appearing in \eqref{EMD/RD}. Recall that $f_{\rm C}^{\rm (EMD)}=f_{\rm i}  \leq f^{\rm (RD)}={\rm max}(1,f_{\rm i})$, and therefore 
\begin{equation}
\frac{1}{2} \leq   \frac{1+f_{\rm C}^{\rm (EMD)}}{1+f^{\rm (RD)}} \leq 1~. 
\end{equation}
Thus the number density of monopoles just after their production is smaller than, or at most equal to, the number density in a RD equivalent scenario. Furthermore, the factor \(e_{\rm f} >1\), and therefore the number density experiences more redshift due to the EMD phase than the RD equivalent number density, resulting in a smaller frozen abundance.
\end{itemize}
For the other two cases, whether the monopole relic abundance is larger or smaller in the EMD scenario 
compared to the RD-equivalent 
scenario depends on the relative sizes of the critical exponents, and for case II, additionally 
on the ratio of the temperature of the hidden sector at reheating to the critical temperature.
A sufficient condition for the right-side of \eqref{EMD/RD} to be less than or equal to one is 
\begin{equation}
    2 \nu \leq 1 + \mu~. \label{nu-mu-condition}
\end{equation} 
This condition can be verified by considering the relative sizes of the numerical factors involved:
\begin{itemize}
\item For case II, note that 
\begin{equation}
    \frac{1}{2}e_{\rm f} < \frac{1+f^{(\rm EMD)}_{\rm RH}}{1+f^{(\rm RD)}_{\rm RH}} < 1+e_{\rm f}
\end{equation}
so that this fraction of $f$'s is bracketed by $e_{\rm f}$. Thus for critical exponents satisfying 
\eqref{nu-mu-condition}, the factor of $e^{3/4}_{\rm f}$ in the denominator of \eqref{EMD/RD} due to the redshift is always larger than the ratio of number densities at the production time, irrespective of the relative size of $(T^{\rm (hid)}_{\rm RH})^{\rm (EMD)}$
to $T^{\rm (hid)}_{\rm C}$.
But for critical exponents violating \eqref{nu-mu-condition}, then the right-side of \eqref{EMD/RD} can in principal be larger than 1, but whether that occurs 
depends on the relative of size of 
$e_{\rm f}$ and the ratio of temperatures $(T^{\rm (hid)}_{\rm RH})^{\rm (EMD)}/T^{\rm (hid)}_{\rm C}$.
\item In the last case, of monopole production after EMD, the ratio of $f$'s is the same as for case II, 
because $f^{\rm (EMD)}_{\rm C}=f^{\rm (EMD)}_{\rm RH}$. To further simplify the analysis, assume 
that the
visible sector degrees of freedom are the same in the two scenarios when the phase 
transition occurs in the hidden sector (which may occur at different visible sector 
temperatures). Then if the critical exponents satisfy \eqref{nu-mu-condition} the ratio on the 
right-side of \eqref{EMD/RD} is always less than one.  
\end{itemize}

We therefore conclude that {\em provided the critical exponents satisfy 
 $2 \nu \leq 1 + \mu$, the current frozen monopole abundance in a scenario involving EMD is always less than or equal to that in a pure RD equivalent, for a fixed monopole mass. This, along with the mass-dependence of \eqref{Omega-mass}, results in a larger monopole mass needed to account for a fixed \(\Omega_{\rm M}h^2\) when EMD is involved. }

\newpage
\section{EMD by a modulus: numerical results}
\label{mod}

We now move to consider specific mechanisms for establishing a period of EMD, beginning with the case where the matter-dominating field \(\Phi\) is a scalar modulus with mass \(m_\Phi\) and initial amplitude \(\Phi_{\rm i} \lesssim M_{\rm P}\) \cite{Kane:2015jia}. The modulus begins to oscillate, acquiring a matter equation of state, when \(H \approx m_\Phi\), at which time its energy density is given by \(\rho_\Phi(t_{\rm i}) = (1/2)m_\Phi^2\Phi_{\rm i}^2\). This initial energy density, along with the matter-like redshift relation \(\rho_\Phi \sim a^{-3}\), determines how quickly \(\Phi\) can dominate over the background radiation energy density, be it of the hidden or visible sectors. The initial ratio of the VS radiation energy density to that of the hidden sector is given by the factor \(f_{\rm i}\). The Hubble factor during the period before EMD by \(\Phi\) is given by \eqref{HTRD}. 

The modulus amplitude, initially fixed at $\Phi_{\rm i}$, starts to oscillate once $H \simeq m_{\Phi}$, and an EMD phase  begins shortly after the energy densities of \(\Phi\) and radiation become comparable. Solving for $H \simeq m_{\Phi}$ and redshifting to this first era of matter--radiation equality, one finds 
the expansion at this time approximately corresponds to 
\begin{equation}\label{HMDmodulus}
    H_{\rm MD} \approx \frac{m_\Phi \Phi_{\rm i}^4}{36M_{\rm P}^4}~.
\end{equation}
In calculating this, we have assumed the energy density of \(\Phi\) is dominant over, as opposed to equal to, that of radiation, which results in a better agreement between our analytical calculations and numerical results shown below. For a modulus with maximal amplitude, we note that the modulus essentially dominates the energy density of the Universe as it begins to oscillate, while a smaller amplitude results in a delay. In order to successfully establish EMD, \(\Phi\) must also be sufficiently long lived such that its decay completes well after the start of EMD. The minimum value of the initial amplitude, corresponding to decay at the onset of EMD, can be estimated from \eqref{gammaphi} and \eqref{HMDmodulus} to be 
\begin{equation}
    \Phi_{\rm i} \gtrsim \bigg(\frac{36\Gamma_\Phi M_{\rm P}^4}{m_\Phi}\bigg)^{1/4} = \sqrt{6\alpha m_\Phi M_{\rm P}}/(2 \pi)^{1/4}~.
    \label{Phi:bc}
\end{equation}
For tree-level decays, a given visible sector reheat temperature determines not only the end of EMD, but also the mass of \(\Phi\) and thus the minimum amplitude to have an EMD era at all. A choice of \(\Phi_{\rm i}\), within the allowed limits, then determines how early the EMD phase starts. 

We parenthetically note that 
for a given visible sector reheat temperature, 
the inclusion of a loop factor in \(\Gamma_\Phi\) shifts the values of \(m_\Phi\) and \(\Phi_{\rm i}\) which correspond to a particular EMD duration. There is however, some degeneracy in the corresponding cosmologies. For instance, a change in initial amplitude of \(10^{-1}\) can be compensated by a change in mass of \(10^4\) and a loop factor $\alpha$ of \(10^{-6}\), such that the resulting EMD phase is unchanged, having the same $H_{\rm MD}$, $\Gamma_{\Phi}$, and boundary condition (\ref{Phi:bc}). As mentioned previously, 
we will set \(\alpha = 1\) throughout unless otherwise specified.

The evolution of the three background energy density components (that of \(\Phi\) and the radiation from the hidden and visible sectors) is governed by the following usual set of Boltzmann equations: 
\begin{align}
  &  \frac{d\rho_\Phi}{dt} + 3H\rho_\Phi = - \Gamma_\Phi \rho_\Phi~,
\label{boltzmodVR} \\
&    \frac{d\rho_{\rm r}^{\rm(vis)}}{dt} + 4H\rho_{\rm r}^{\rm(vis)} = \Gamma_\Phi \rho_\Phi ~,\\
 &   \frac{d\rho_{\rm r}^{\rm(hid)}}{dt} + 4H\rho_{\rm r}^{\rm(hid)} = 0~,
\end{align}
where \(3H^2M_{\rm P}^2 = \rho_\Phi + \rho_{\rm r}^{\rm(vis)} + \rho_{\rm r}^{\rm(hid)}\). We emphasize that, for simplicity, in the Boltzmann equations above we have taken \(\Phi\) to decay only to the visible sector,  
though it is straightforward to include branching fractions for decay to both sectors. We numerically solve this set of equations beginning in a period of RD by any combination of visible sector and HS radiation, and track the evolution sufficiently beyond reheating such that RD in the visible sector is well-established.

\begin{figure}[hb!]
    \centering
    \includegraphics[width=0.5\textwidth]{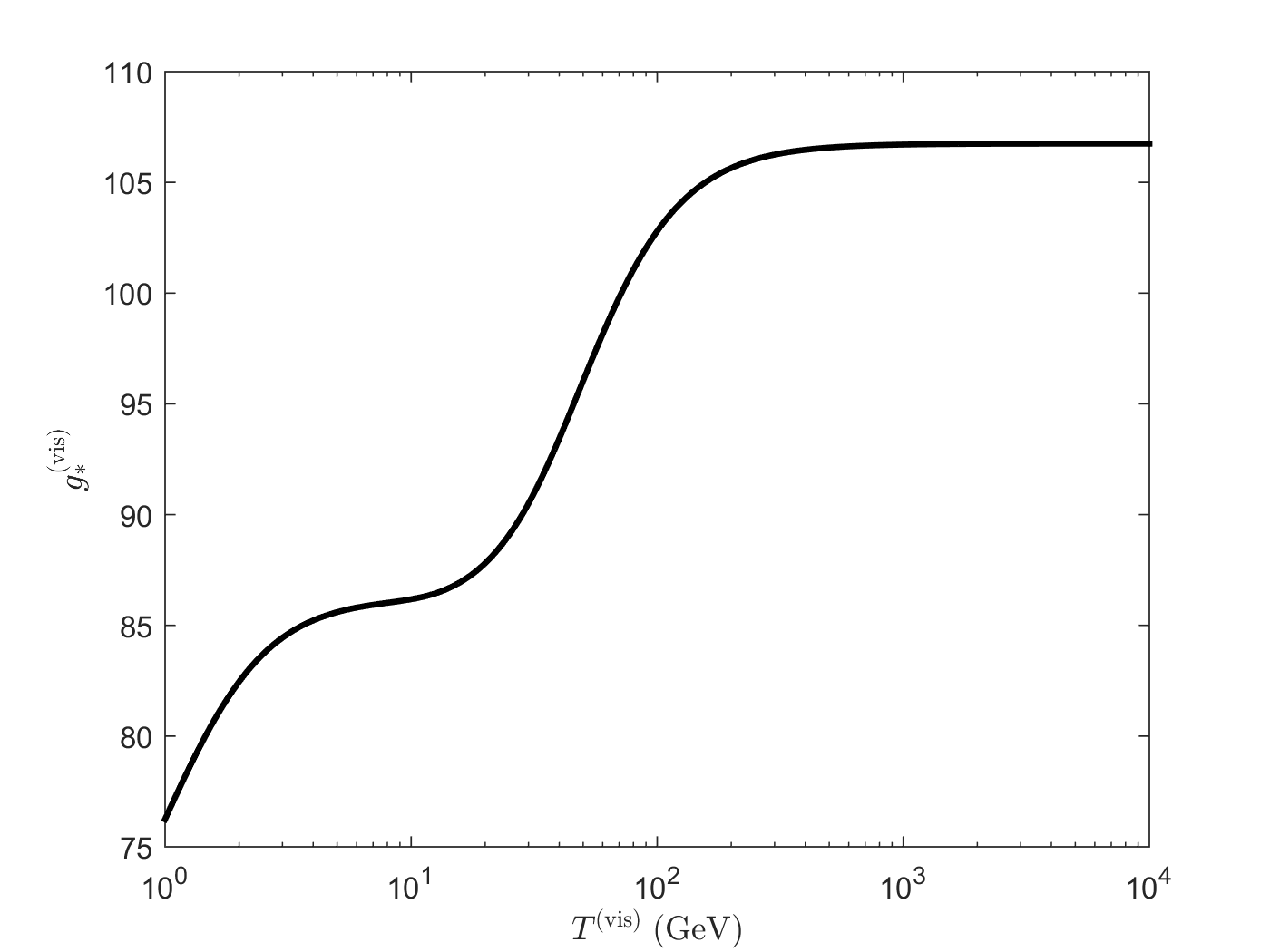}
    \caption{Temperature dependence of the relativistic degrees of freedom in the visible sector assumed in our numerical calculations for temperatures greater than \(1\;{\rm GeV}\).}
    \label{fig:gstar}
\end{figure}

In our numerical calculations, we use a smooth function to estimate the temperature dependence of the relativistic degrees of freedom for energy density in the VS, \(g_*^{\rm(vis)}\), shown in Figure \ref{fig:gstar}. At temperatures greater than $\sim$100 GeV, when all SM species are relativistic, \(g_*^{\rm(vis)}\) takes its maximum value of 106.75. As the temperature decreases, the value smoothly drops as the various particle species become nonrelativistic. We only show temperatures greater than \(1\;{\rm GeV}\) because the VS reheat temperature in our scenarios is typically larger. 
The minimum value of \(g_*^{\rm(vis)}\), corresponding to the present era, is 3.36 assuming 3 massless neutrino species. 
For the HS we assume a constant \(g_*^{\rm(hid)} = 100\).

\begin{figure}[hb!] 
    \centering
    \subfloat{\includegraphics[width=0.8\textwidth]{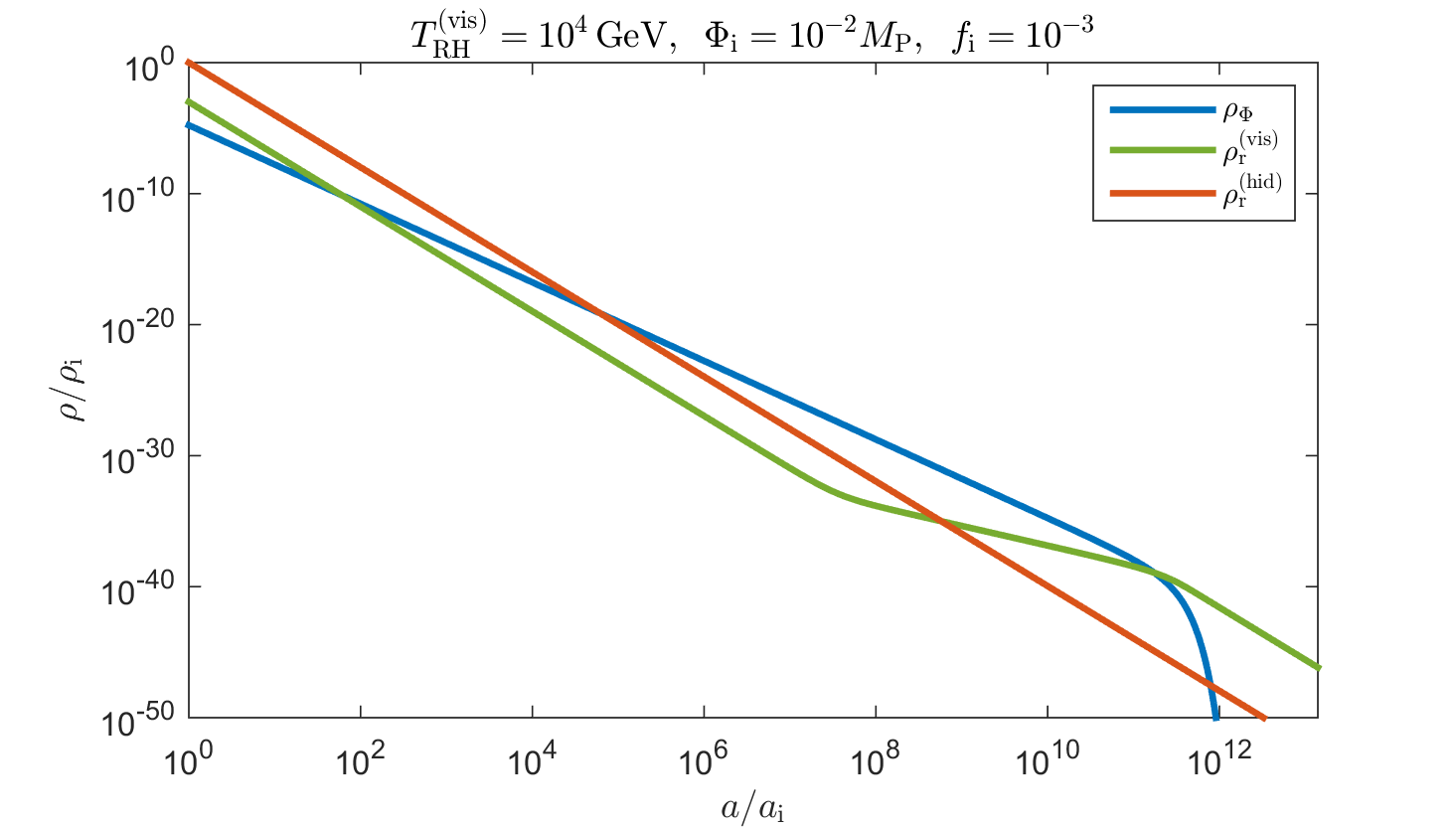}}\\
    \subfloat{\includegraphics[width=0.8\textwidth]{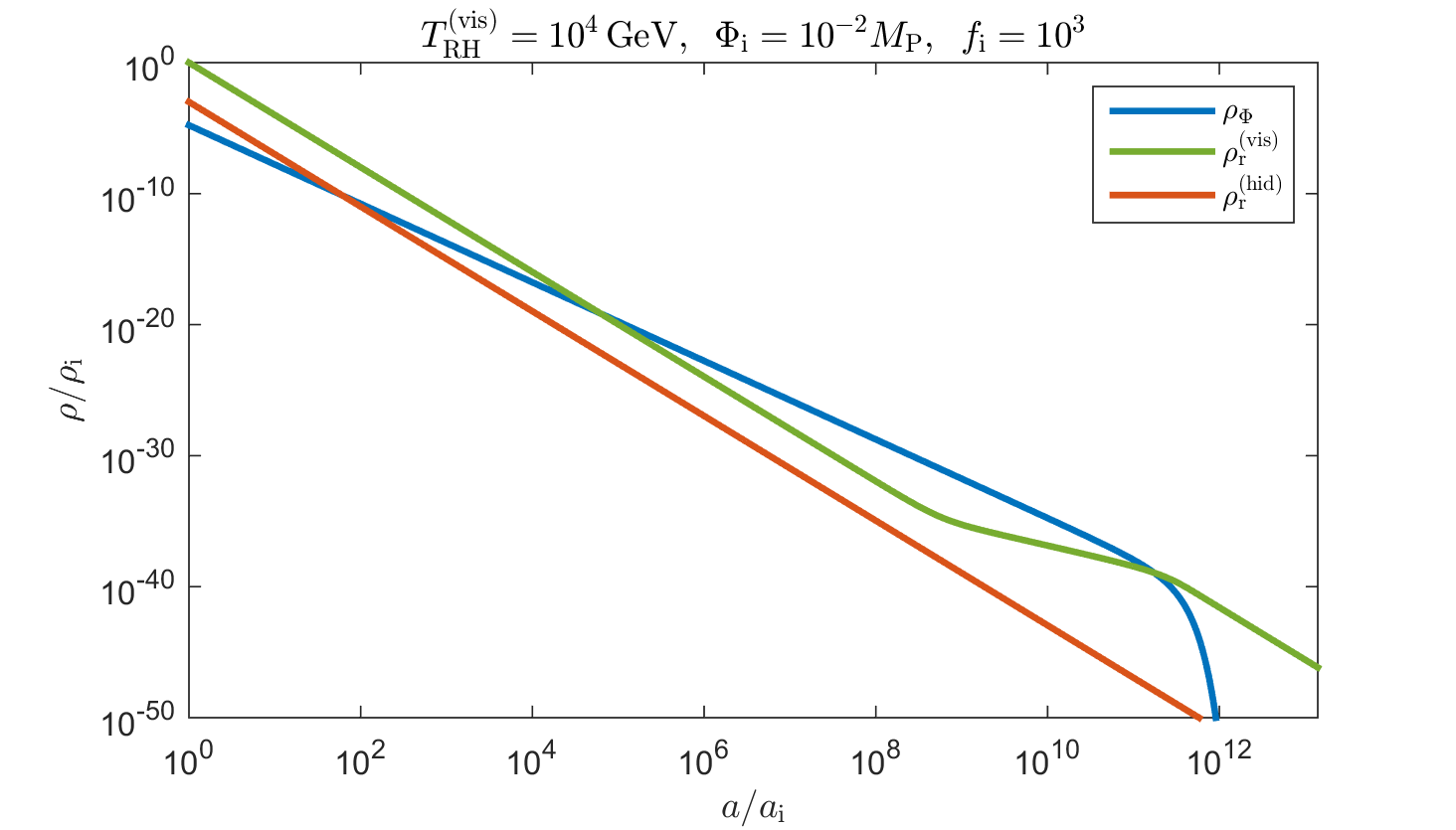}}
    \caption{Numerical evolution of the background energy density components with scale factor in the case of modulus-driven EMD. EMD begins once \(\rho_\Phi\) dominates over both radiation components, and lasts until \(\Phi\) decays. Top panel: initial RD by the hidden sector. Bottom panel: initial RD by the visible sector. A VS reheat temperature of \(T_{\rm RH}^{\rm(vis)} = 10^4\;{\rm GeV}\), with \(\alpha = 1\), results in a modulus mass of \(m_\Phi \approx 10^9\;{\rm GeV}\) essentially independent of the other parameters as long as the EMD period has a noticeable duration.}
    \label{fig:rhovstmodulus-4}
\end{figure}

Figure \ref{fig:rhovstmodulus-4} shows the energy density evolution in the two cases of initial RD by the HS (\(f_{\rm i} << 1\)) and VS (\(f_{\rm i} >> 1\)) respectively, for an example set of parameters.

We allow the phase transition of the HS to occur at any time in the background evolution, and obtain the resultant current monopole abundance from the numerical solution. This is done by evaluating (\ref{corrlength}), the equation for the correlation length at the phase transition, when the temperature of the hidden sector reaches $T^{\rm(hid)}_{\rm C}$, and then approximating the number density of monopoles at that time as $n_{\rm M}\sim \xi(t_*)^{-3}$. Subsequently, the number density is simply redshifted numerically through the EMD era and then normalized to the VS entropy density at reheating. 

\begin{figure}[ht!] 
  \centering
  \subfloat{\includegraphics[width=0.5\textwidth]{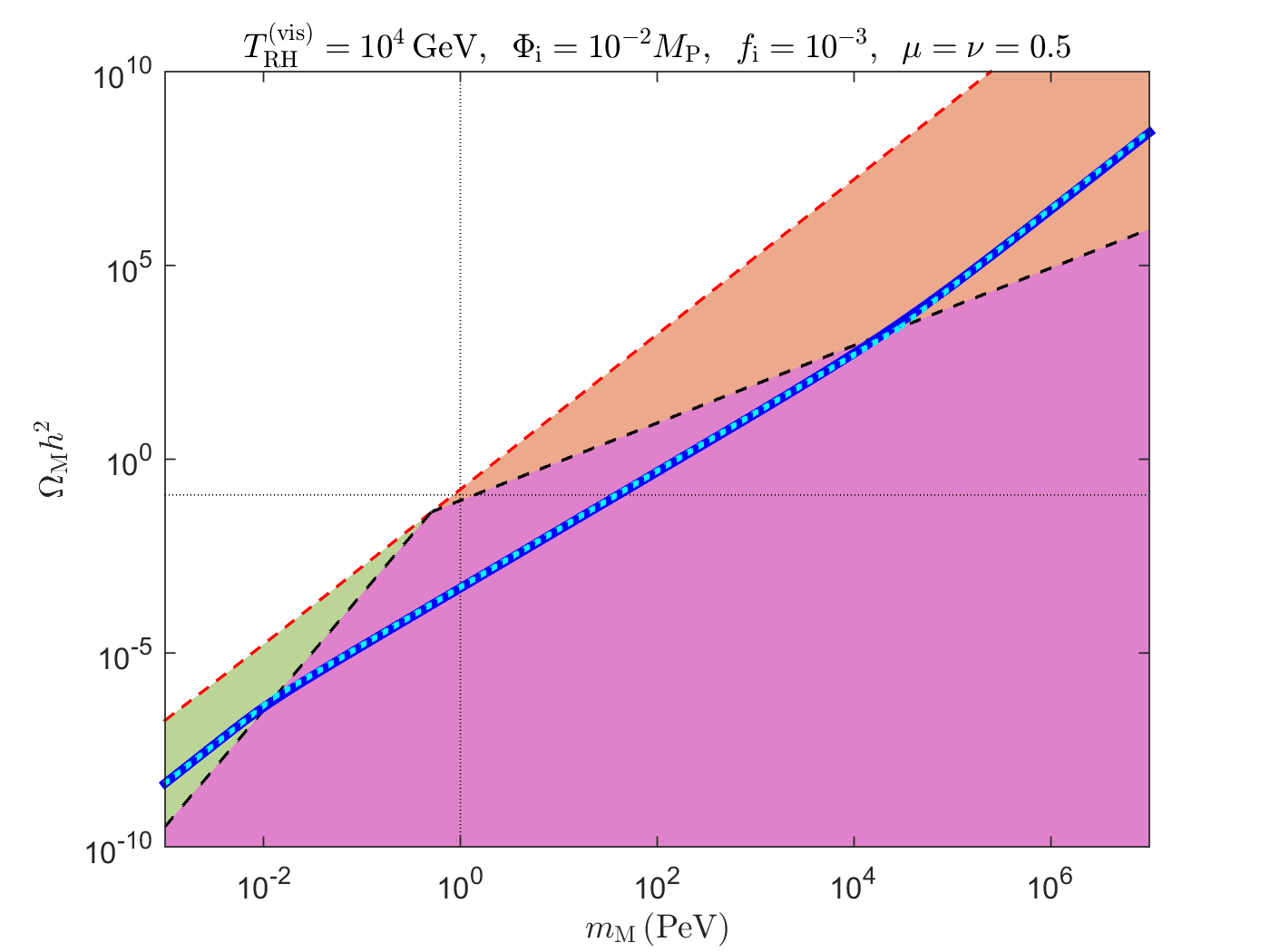}}
  \subfloat{\includegraphics[width=0.5\textwidth]{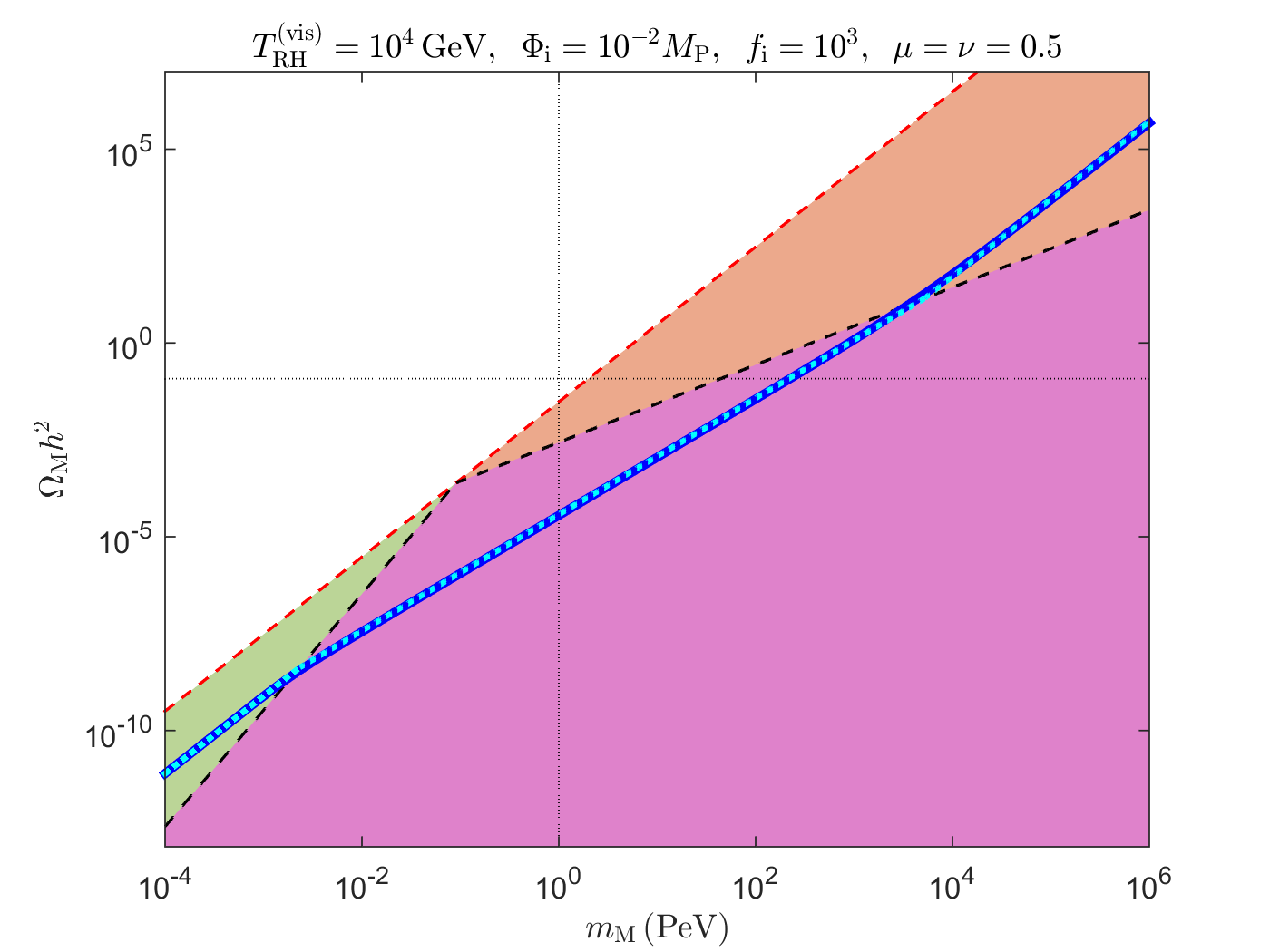}}
  \caption{Dependence of the present-day monopole relic abundance on the monopole mass, for an example set of parameter values, in the case of modulus-driven EMD. The parameters of the cosmological background are the same as in the two panels of Figure \ref{fig:rhovstmodulus-4}. Left: initial RD occurring in the hidden sector. Right: initial RD occurring in the visible sector. The solid curves (blue) are obtained from a numerical evolution of the background, while the dotted lines (light blue) lying on top of the numerical curves are the analytical expressions (\ref{omegabefore}), (\ref{omegaduring}), and (\ref{omegaafter}). The red dashed line in both panels marks the purely RD equivalent scenario for comparison. The two black dashed lines, separating the three shaded regions, indicate the relic abundance for that mass if monopole production occurs at the start (top) or end (bottom) of EMD. The upper-right shaded region (orange) corresponds to monopole production having occurred during the initial RD phase prior to EMD; the large central/lower-right region (magenta) corresponds to production during the EMD phase; and the small lower-left region (green) corresponds to production in the RD epoch after EMD has ended. Where the blue lines overlap with these three regions specifies the period in which monopole production occurred. For reference across the two panels, the dotted horizontal and vertical lines in both panels mark \(\Omega_{\rm M}h^2 = 0.12\) and \(m_{\rm M} = 1\;{\rm PeV}\) respectively. The entire set of curves and region boundaries in the right panel is shifted downward and to the left relative to the left panel, along the RD equivalent line due to the larger final offset between the visible and hidden radiation energy densities after reheating (see Figure \ref{fig:rhovstmodulus-4}).}
  \label{fig:omegamm}
\end{figure}

\newpage 
We now turn to our numerical results. 
In Figure \ref{fig:omegamm} we plot the present-day relic monopole abundance, \(\Omega_{\rm M}h^2\), as a function of monopole mass, \(m_{\rm M}\), where we have taken \(x_{\rm M} = 50\) to be fixed, 
as well as \(\lambda = 1\). In what follows we will set  \(x_{\rm M} = 50\) and \(\lambda = 1\)
throughout unless otherwise noted. The other parameter values match those of Figure \ref{fig:rhovstmodulus-4}. We show both numerical results, obtained from numerically solving the Boltzmann equations, and the three analytical approximations of Section \ref{n/s}, (\ref{omegabefore}), (\ref{omegaduring}), and (\ref{omegaafter}). The numerical curve, shown in dark blue, has three distinct segments corresponding to the three regimes of production time: in the top right, monopoles are produced in the RD period before EMD - the slope of the curve in this region is the same as that of a pure RD monopole production scenario; the central segment of the curve corresponds to production during EMD, with a slope given by \eqref{omegaduring}; and in the bottom left section, production after EMD recovers the RD slope. As can be seen by inspection, the analytic approximations, (\ref{omegabefore}), (\ref{omegaduring}), and (\ref{omegaafter}), have extremely good agreement with the numerical results -- the analytic results correspond to the light-blue dotted line ``lying inside" the numerical curve. 

\begin{figure}[!p]
   \centering
    \subfloat{\includegraphics[trim=0cm 0cm 0cm 0.3cm, clip=true, width=0.5\textwidth]{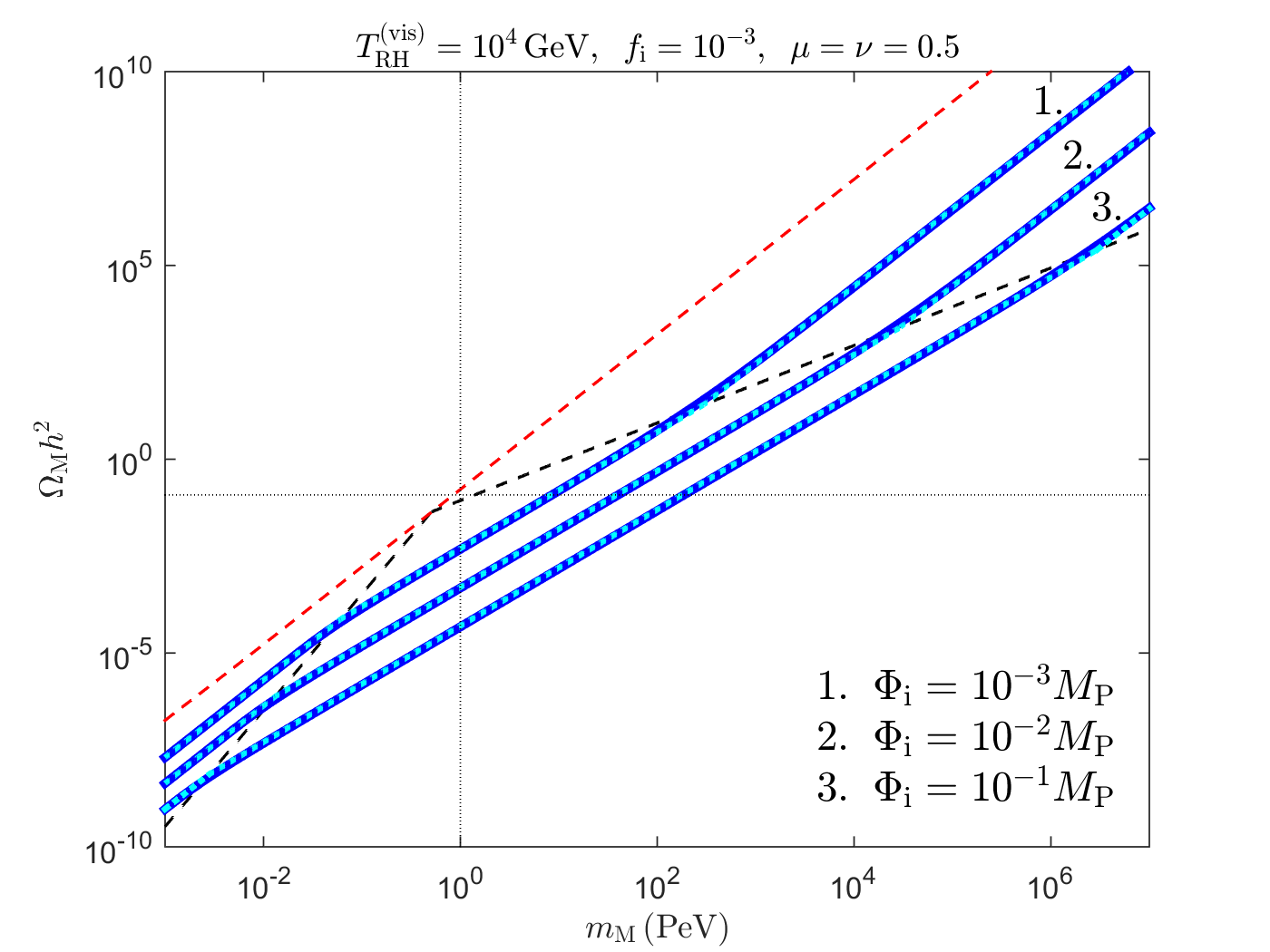}}
    \subfloat{\includegraphics[trim=0cm 0cm 0cm 0.3cm, clip=true, width=0.5\textwidth]{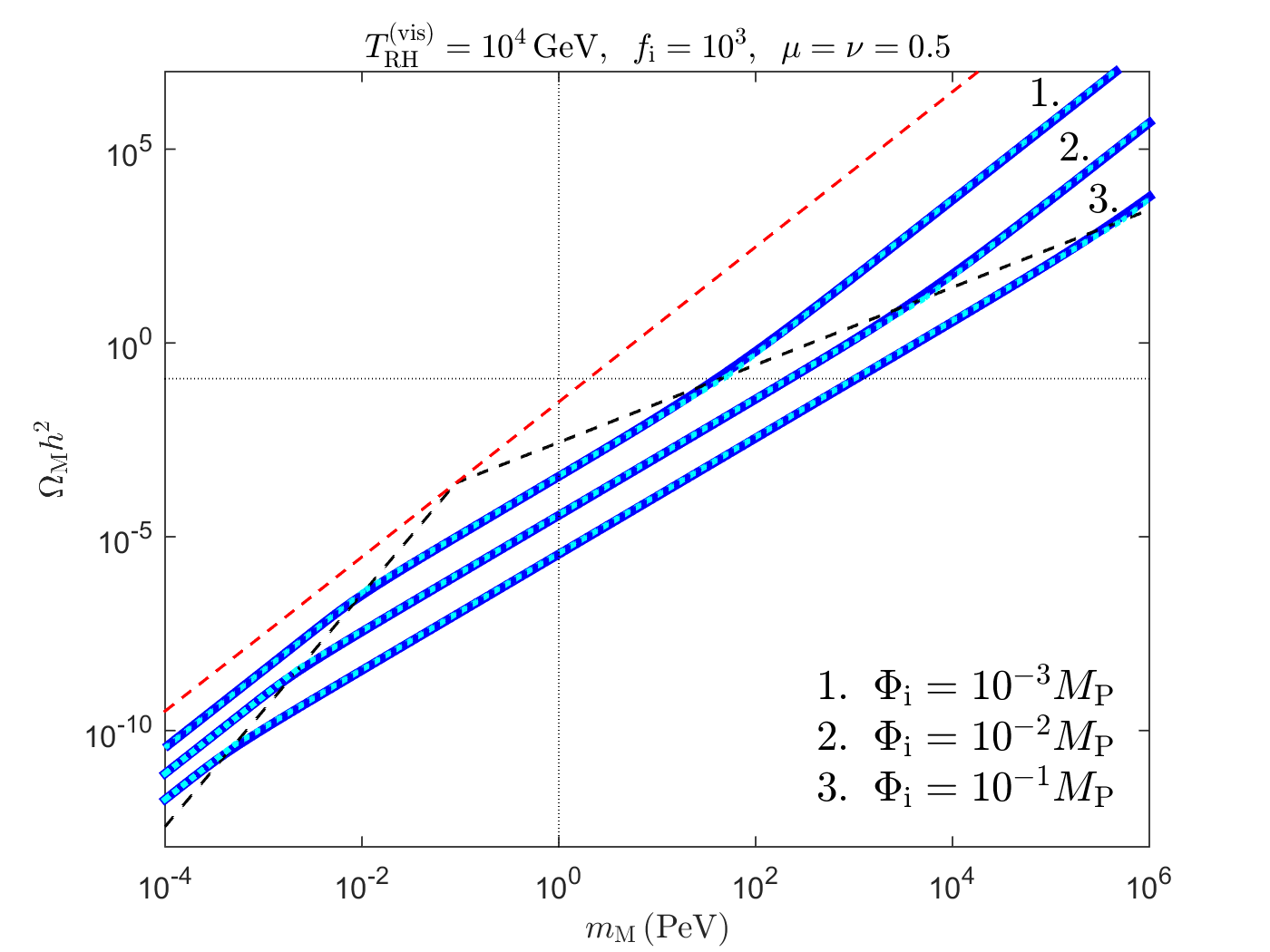}}\\\vskip -0.4cm
    \subfloat{\includegraphics[trim=0cm 0cm 0cm 0.3cm, clip=true, width=0.5\textwidth]{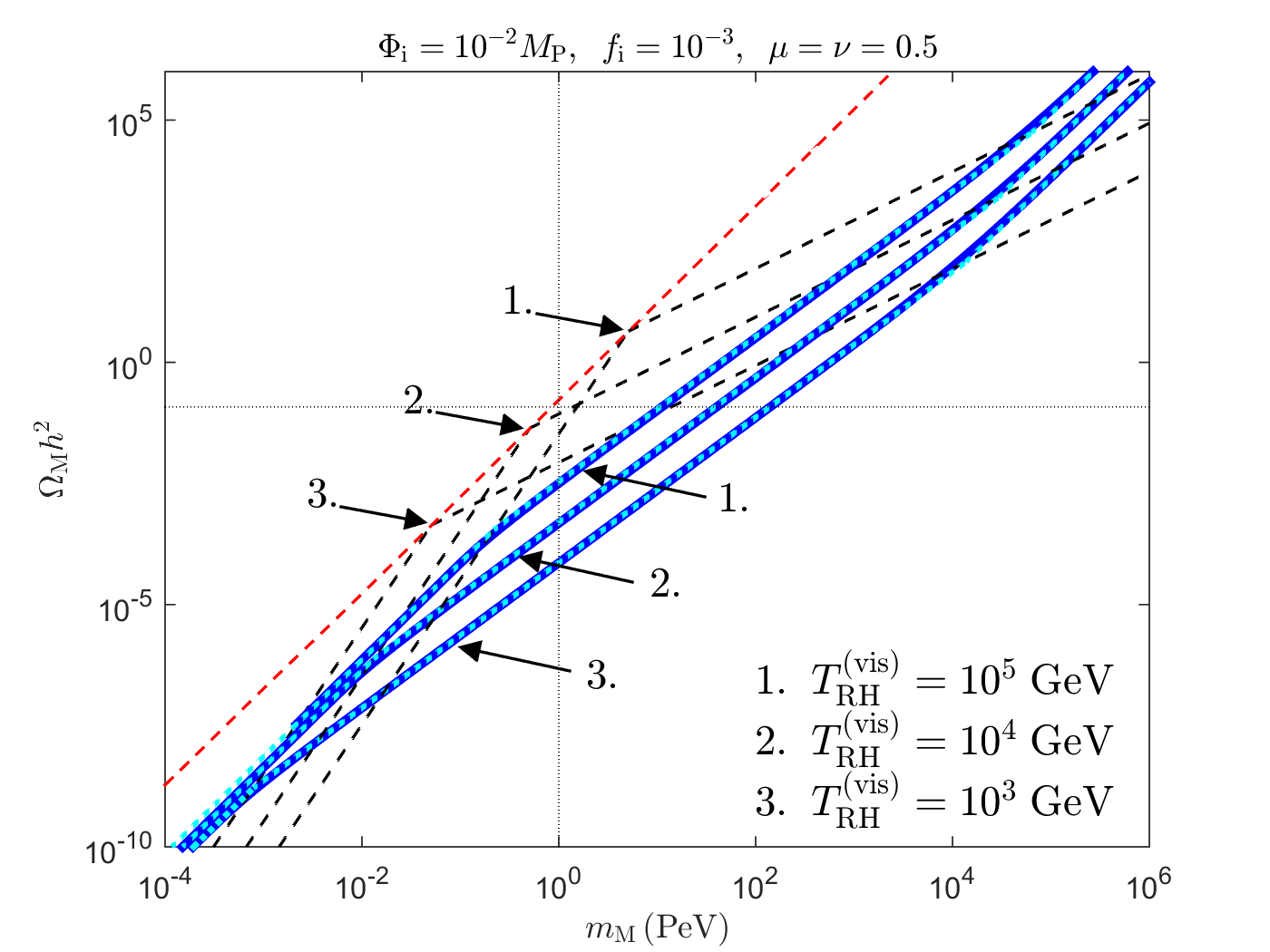}}
    \subfloat{\includegraphics[trim=0cm 0cm 0cm 0.3cm, clip=true, width=0.5\textwidth]{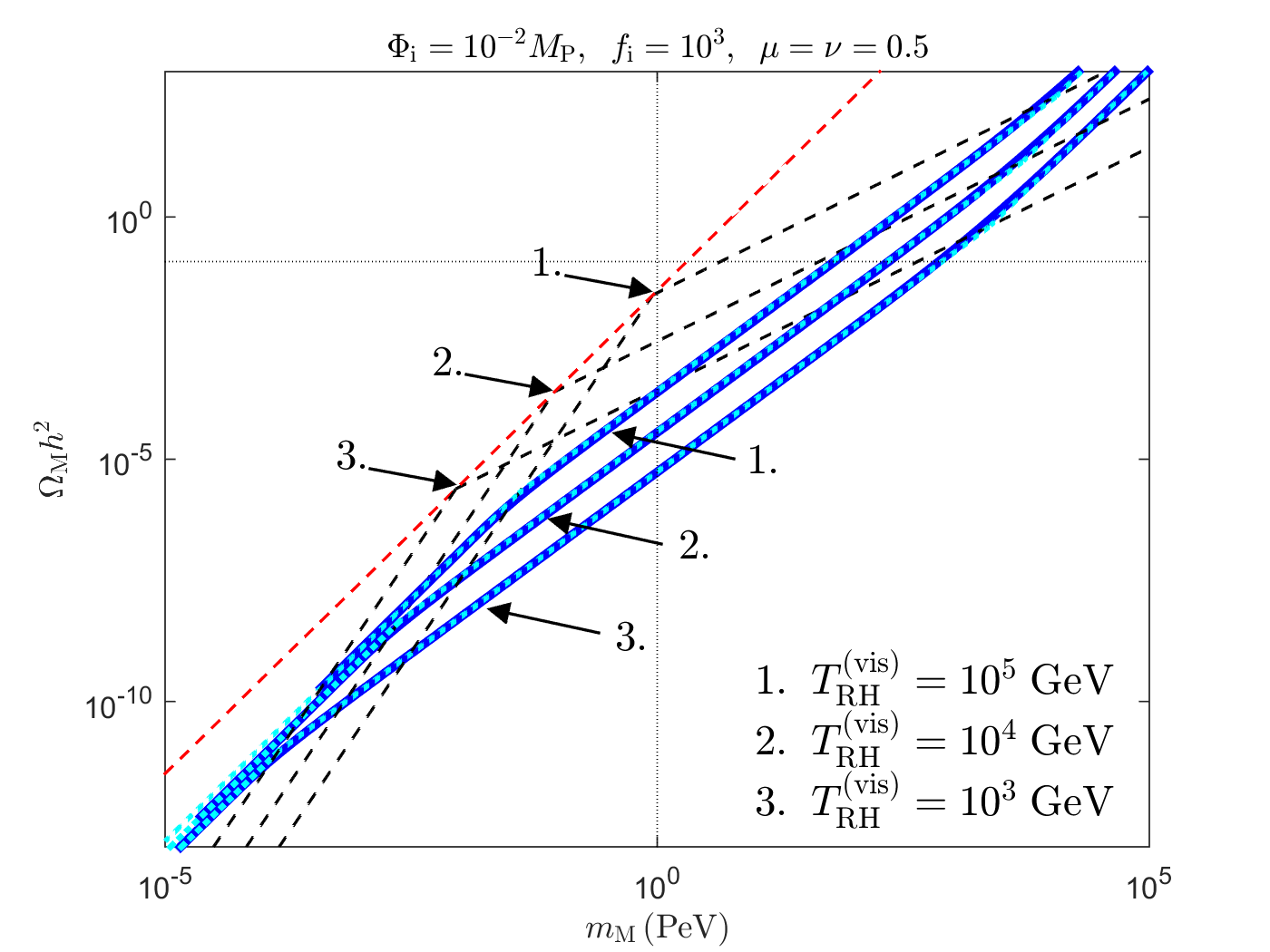}}\\\vskip -0.4cm
    \subfloat{\includegraphics[trim=0cm 0cm 0cm 0.3cm, clip=true, width=0.5\textwidth]{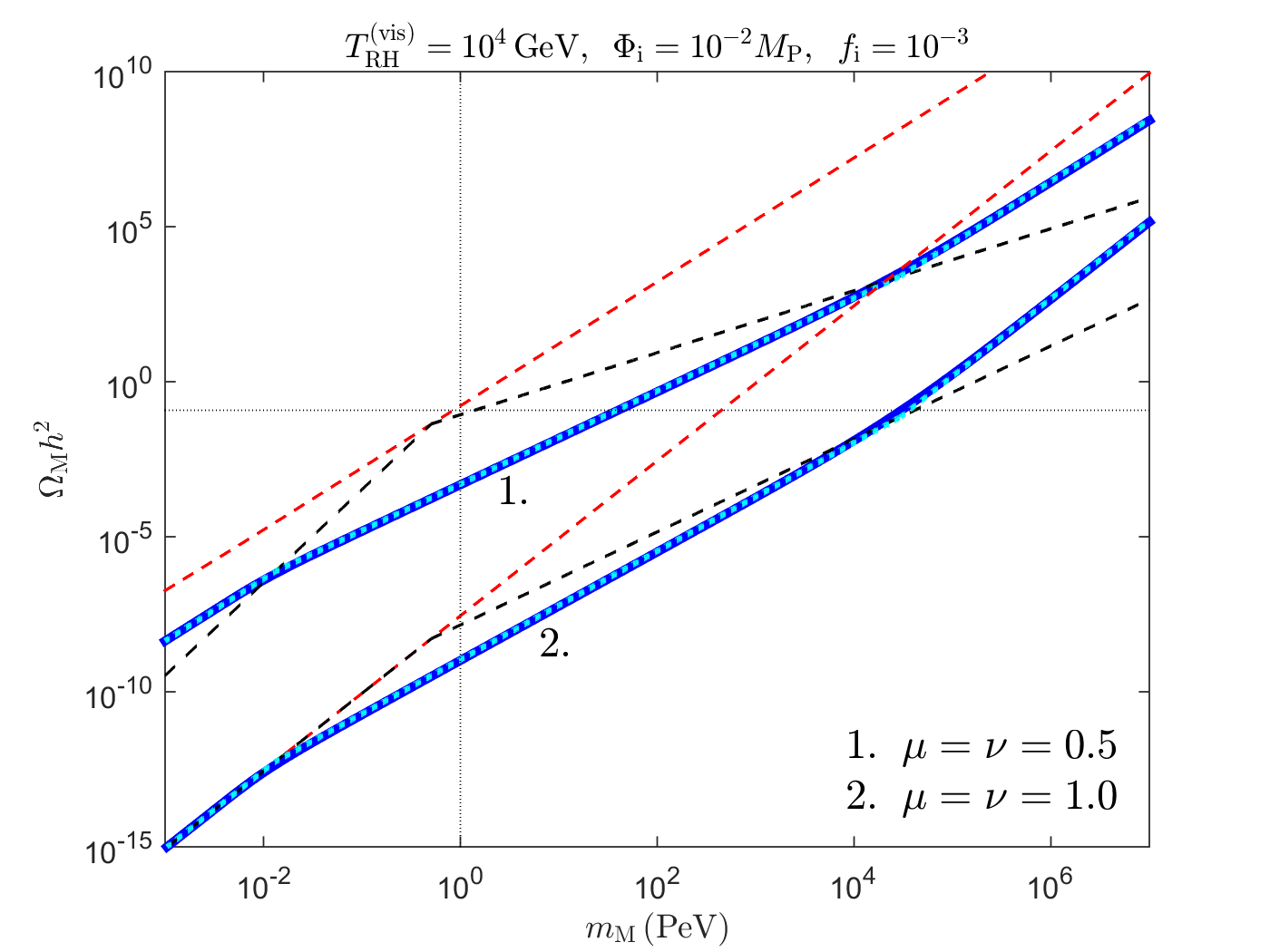}}
    \subfloat{\includegraphics[trim=0cm 0cm 0cm 0.3cm, clip=true, width=0.5\textwidth]{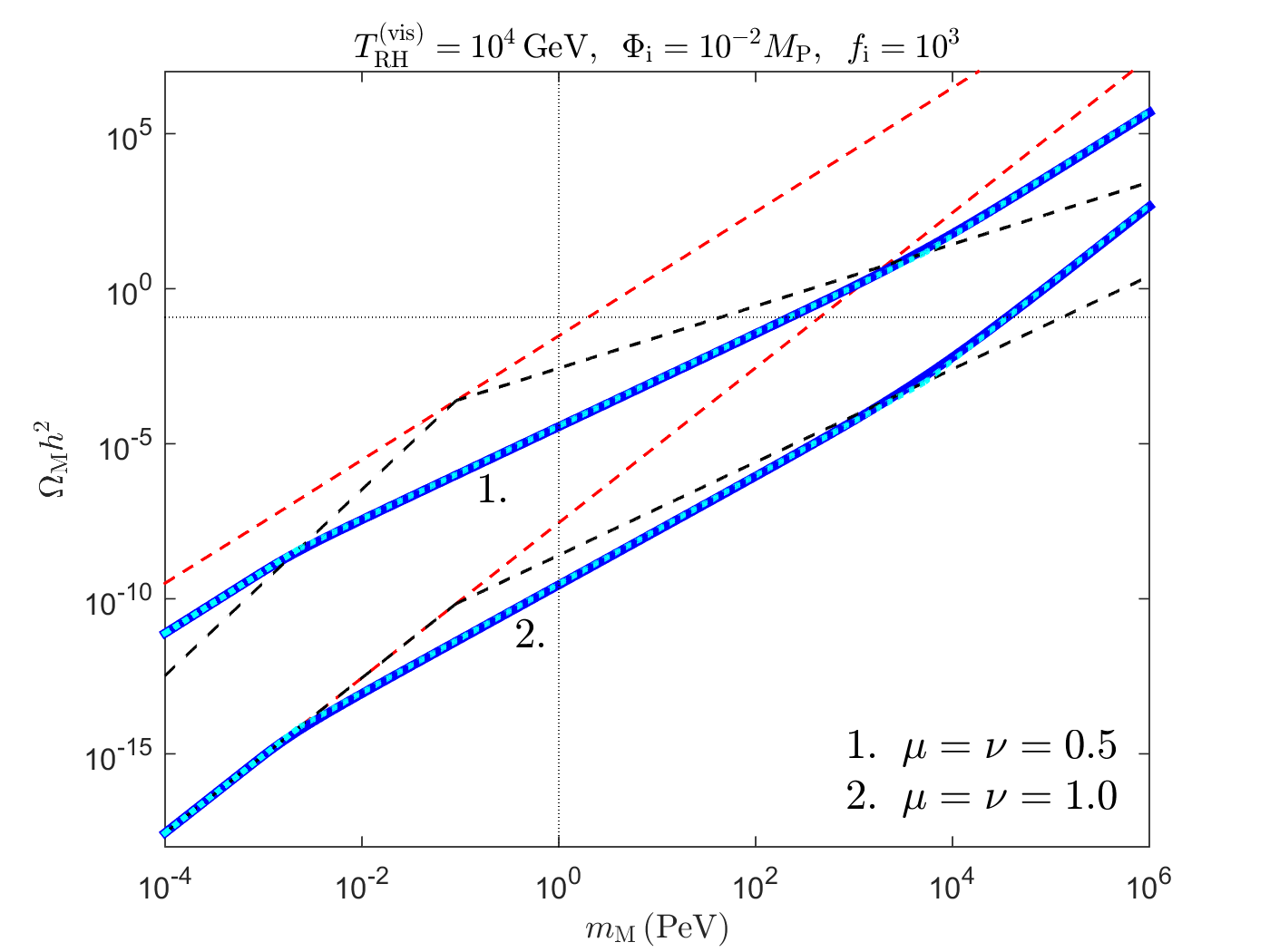}}
    \vspace{-0.3cm}
    \caption{Dependence of the present-day monopole relic abundance on the monopole mass, for a variety of parameter values, with fixed $x_{\rm M}=50$ and $\alpha=\lambda=1$. The solid curves (blue) are obtained from a numerical evolution of the background, while the dotted lines (light blue) on top of the numerical curves are the analytical expressions (\ref{omegabefore}) --
    (\ref{omegaafter}). Red and black dashed lines are as in Figure \ref{fig:omegamm}.
    For reference, the dotted horizontal and vertical lines in all panels mark \(\Omega_{\rm M}h^2 = 0.12\) and \(m_{\rm M} = 1\;{\rm PeV}\) respectively.
    The curves labeled by `2.' in the top panels, `2.' in the middle panels, and `1.' in the bottom panels correspond to the curves of Figure \ref{fig:omegamm}.
    Left panels: initial RD in the HS. Right panels: initial RD in the VS.}
    \label{fig:monopmod}
\end{figure}

Figure \ref{fig:omegamm} also shows colored regions depicting the three regimes of monopole production time. 
A given parameter set $\{m_{\rm M}$, $T^{\rm (vis)}_{\rm RH}$, $\Phi_{\rm i}$, 
$f_{\rm i}$, $\alpha$, $x_{\rm M}$, $\lambda$, $\mu$, $\nu\}$ corresponds to a single point on Figure \ref{fig:omegamm}, so that 
as $m_{\rm M}$ is varied, a single (blue) curve is traced out, passing through 
the colored regions that correspond to production after, during, or before the time 
of the phase transition. In this way only a subset of the colored regions are accessed. However, other points in the colored regions can be accessed by varying 
$m_{\rm M}$ together with one or more of these other parameters. This behavior can be seen in Figure \ref{fig:monopmod}, which we discuss in 
more detail below. 

Figure \ref{fig:omegamm} also shows as black dashed lines the two analytical expressions for production at the beginning \eqref{n/sMDstart} and end \eqref{n/sMDend} of EMD, separating these three regimes. 
One way to interpret the boundary curves is the following. 
These two lines give analytic predictions for monopole production if, for a given monopole mass, production occurs at the end of initial RD and start of EMD (upper), or end of EMD and start of second RD (lower). The intersection of either of these dashed lines and the solid blue (numerical) line gives the mass for which production did occur at cross-over, for the parameters assumed for the solid line. These intersection points therefore mark the transitions between the three behaviors of the numerical line discussed in the previous paragraphs. 

Lastly, we note that the entire numerical curve sits at higher monopole masses when compared to a pure RD production scenario (shown by the red dashed line) because of the offset of the hidden and visible sector energy densities. This is consistent with the behavior of \eqref{Omega-mass} and \eqref{EMD/RD}, specifically that the right-side of \eqref{EMD/RD} is always less than one when \(2 \nu \leq 1+\mu\). 

In Figure \ref{fig:monopmod} we show how the curves of Figure \ref{fig:omegamm} change for a variety of parameter values. As the beginning of EMD is placed earlier (by increasing the initial modulus amplitude \(\Phi_{\rm i}\)) while keeping the VS reheat temperature \(T_{\rm RH}^{\rm(vis)}\) fixed, the numerical curves (along with their analytical counterparts) shift farther away from the RD line toward larger monopole masses due to the increased amount of dilution from a progressively longer EMD period. If instead the end time of EMD is placed later (by decreasing \(T_{\rm RH}^{\rm(vis)}\)) while holding the start time fixed, the curves again shift toward higher monopole masses due to the longer EMD period, but now the corresponding dashed boundary lines shift downward due to their dependence on the reheat temperature. Finally, as the critical exponents, \(\mu\) and \(\nu\), are varied, the slopes of the curves change as expected. 

In all panels of Figure \ref{fig:monopmod}, all of the numerical curves retain the three-region slope behavior displayed in Figure \ref{fig:omegamm}, with the regions separated by the two dashed boundary lines regardless of the specific parameter values, as expected. We note that the change in slope between the three regimes of production time is most noticeable in the bottom blue curve of the bottom two panels, for which $\mu=\nu=1$. As in Figure \ref{fig:omegamm}, the left panels correspond to initial RD by HS radiation (with \(f_{\rm i} < 1\)), while the right panels correspond to initial VS domination (\(f_{\rm i} > 1\)). The full set of lines shown in each right panel is shifted downward and to the left as \(f_{\rm i}\) is increased above 1 relative to the corresponding left panels. Otherwise, the scale and orientation is the same between the left and right panels.

\clearpage

\section{EMD by a decoupled particle: numerical results}
\label{decp}
\begin{figure}[ht!]
    \centering
    \includegraphics[width=.5\textwidth]{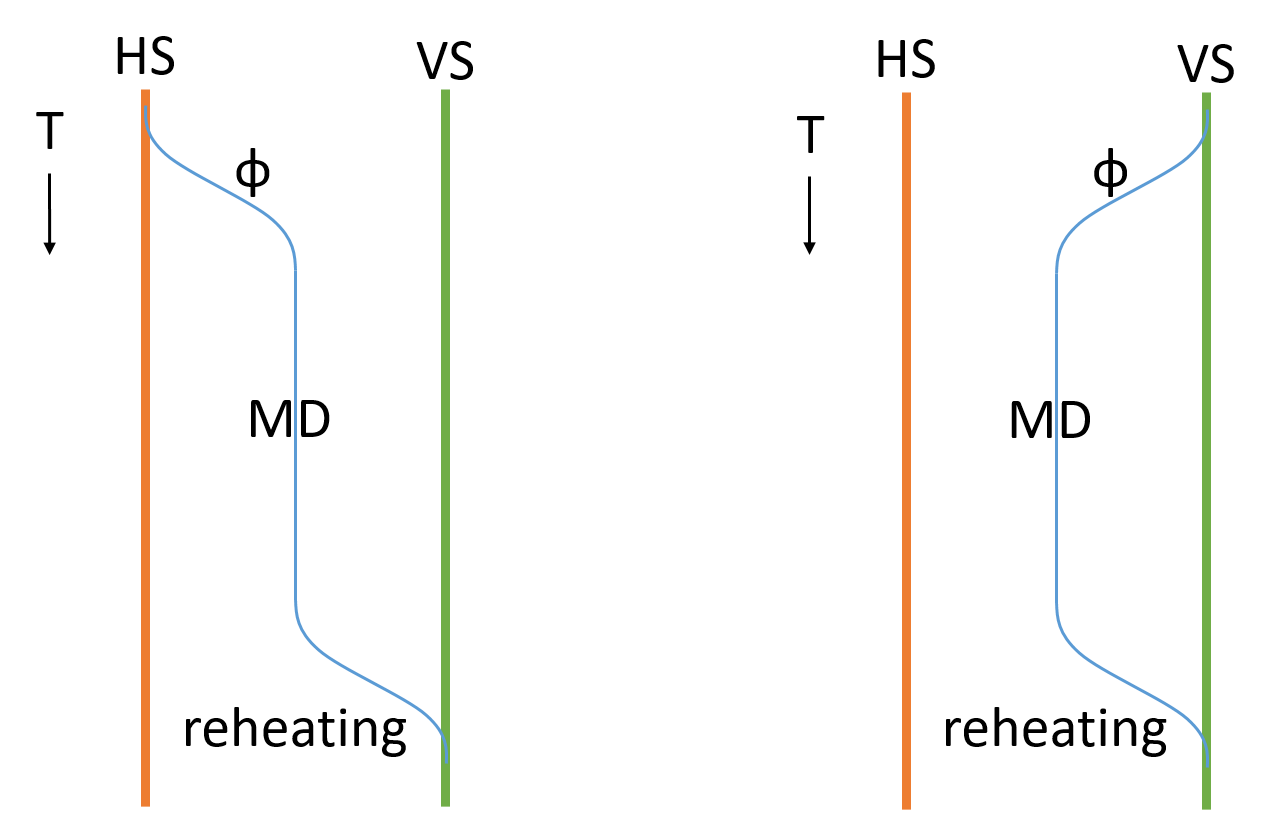}
    \caption{Diagram of particle \(\Phi\) decoupling from either sector while always reheating to the visible sector.}
    \label{fig:decoupling}
\end{figure}

Rather than being a modulus, the field \(\Phi\) that drives EMD can instead be a heavy particle which decouples from either the hidden or visible sector at a very early time and subsequently dominates the energy density of the Universe as a non-relativistic matter component before eventually decaying (see Figure \ref{fig:decoupling}). We will parameterize the interaction rate of \(\Phi\) with the sector from which it is decoupling (the ``host" sector) by the thermally averaged annihilation cross-section times relative velocity, \(\left<\sigma_\Phi v\right>\).\footnote{For simplicity, we assume
that \(\sigma_\Phi v\) is independent of velocity, so 
that \(\left<\sigma_\Phi v\right>\) is independent of temperature, as the details of the \(\Phi\) field and its interactions are not the focus of this work. However more general forms can and should be considered in a realistic model.} The Boltzmann equation for the number density of \(\Phi\) is then
\begin{equation}\label{BoltzmannPhiDec}
    \frac{dn_\Phi}{dt} + 3Hn_\Phi = \left<\sigma_\Phi v\right>(n_{\rm\Phi,eq}^2 - n_\Phi^2) - \Gamma_\Phi n_\Phi ~,
\end{equation}
where \(\Gamma_\Phi\) is the decay rate given in \eqref{gammaphi}, and the Hubble parameter \(H\) is again given by the sum of all energy density components. In our numerical calculations, we use the integral expression
\begin{equation} \label{neq}
    n_{\rm\Phi,eq} = \frac{g_\Phi}{(2\pi)^3}\int{\frac{d^3p}{e^{E(p)/T} \pm 1}}~,
\end{equation}
for the equilibrium number density, where \(+\) is for fermions, \(-\) is for bosons, \(E(p)^2 = m_\Phi^2 + |p|^2\), \(g_\Phi\) is the number of internal degrees of freedom for \(\Phi\), and the temperature \(T\) is of the host sector. 

If \(\Phi\) decouples from the HS, the remaining two Boltzmann equations for the radiation components are 
 \begin{align}\label{BoltzmannVS}
  &  \frac{d\rho_{\rm r}^{\rm(vis)}}{dt} + 4H\rho_{\rm r}^{\rm(vis)} = 
  \Gamma_\Phi \rho_\Phi~,
\\
&  \frac{d\rho_{\rm r}^{\rm(hid)}}{dt} + 4H\rho_{\rm r}^{\rm(hid)} = \left<\sigma_\Phi v\right>\left<E_{\Phi}\right>(n_\Phi^2 - n_{\rm\Phi,eq}^2) ~,\label{BoltzmannHSdec}
\end{align}
while if it decouples from the visible sector, we have 
\begin{align}\label{BoltzmannVSdec}
 &   \frac{d\rho_{\rm r}^{\rm(vis)}}{dt} + 4H\rho_{\rm r}^{\rm(vis)} = \Gamma_\Phi \rho_\Phi + \left<\sigma_\Phi v\right>\left<E_{\Phi}\right>(n_\Phi^2 - n_{\rm\Phi,eq}^2) ~,
\\
&    \frac{d\rho_{\rm r}^{\rm(hid)}}{dt} + 4H\rho_{\rm r}^{\rm(hid)} = 0 ~.\label{BoltzmannHS}
\end{align}
\begin{figure}[hb!]
    \centering
    \subfloat{\includegraphics[trim=0cm 0cm 0cm 0.3cm, clip=true, width=0.5\textwidth]{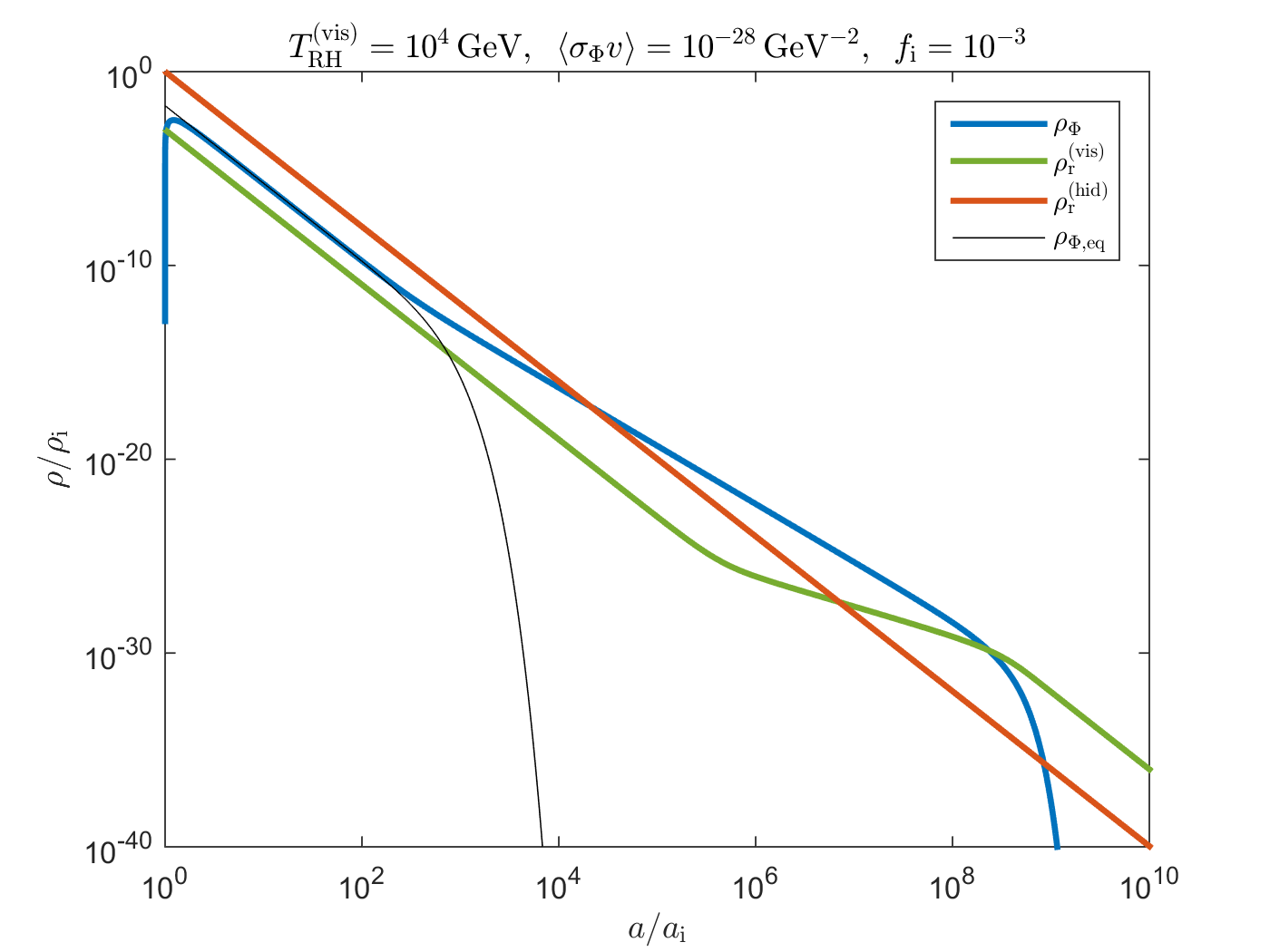}}
    \subfloat{\includegraphics[trim=0cm 0cm 0cm 0.3cm, clip=true, width=0.5\textwidth]{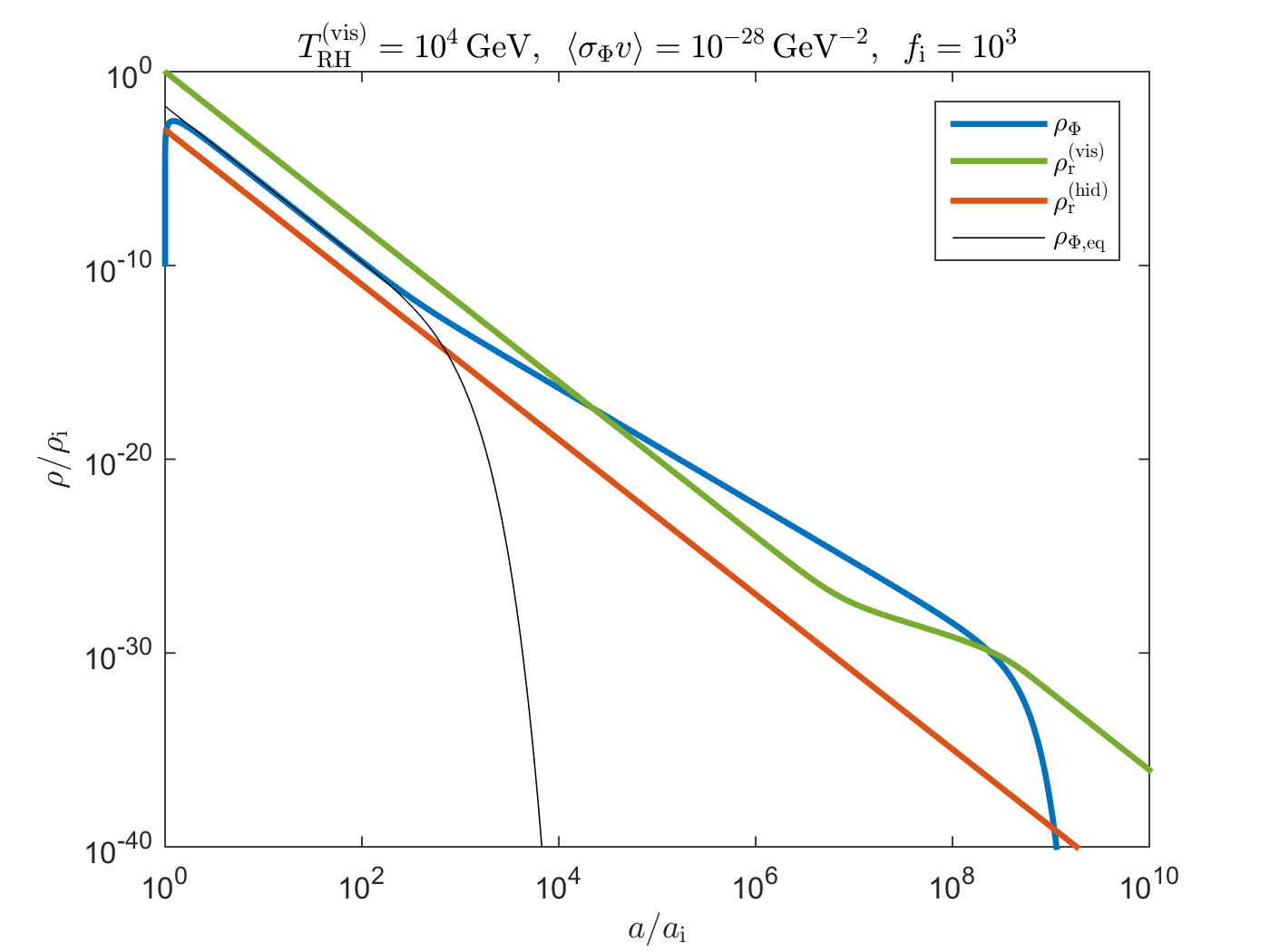}}\\\vskip -0.4cm
    \subfloat{\includegraphics[trim=0cm 0cm 0cm 0.3cm, clip=true, width=0.5\textwidth]{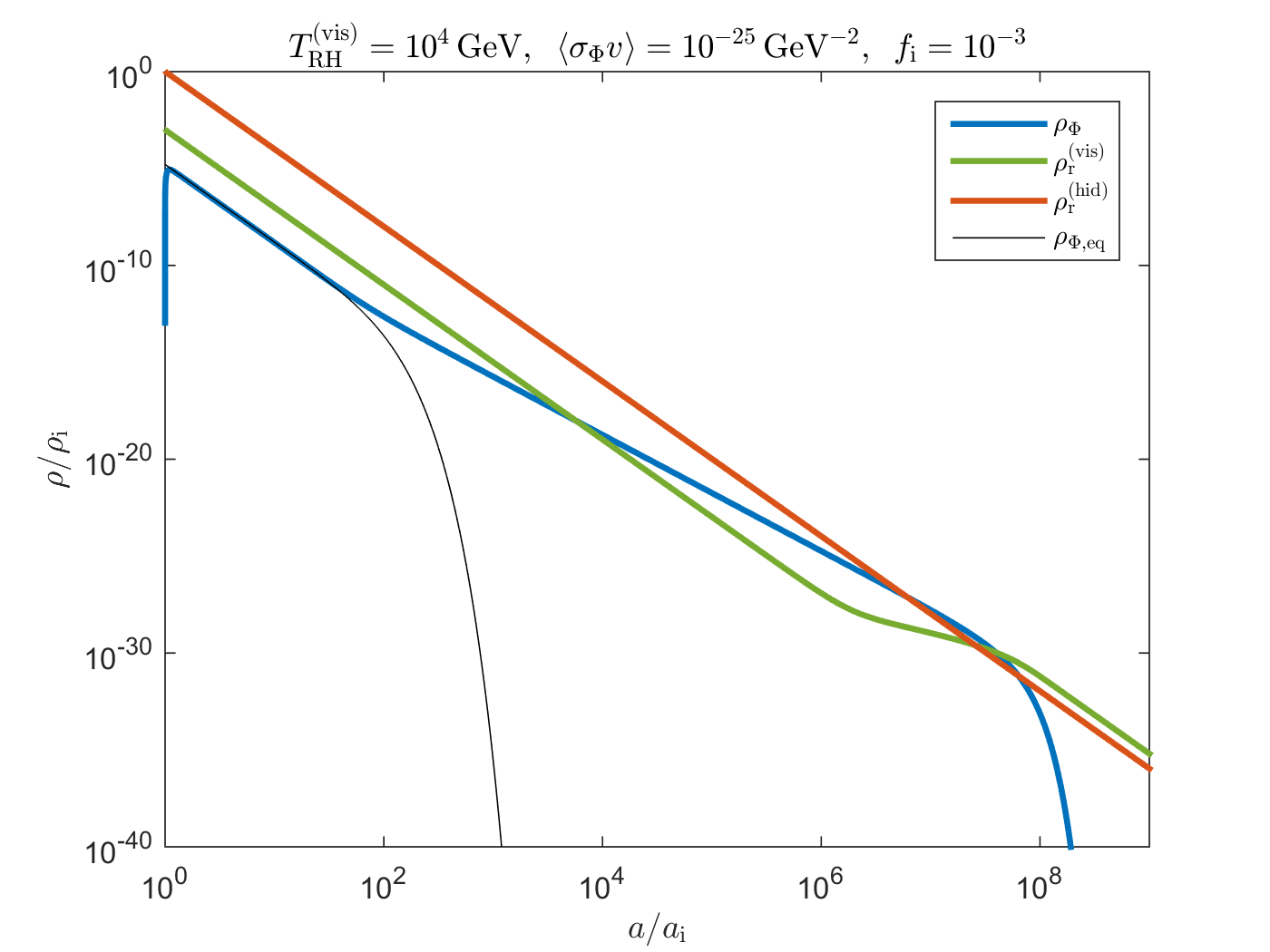}}
    \subfloat{\includegraphics[trim=0cm 0cm 0cm 0.3cm, clip=true, width=0.5\textwidth]{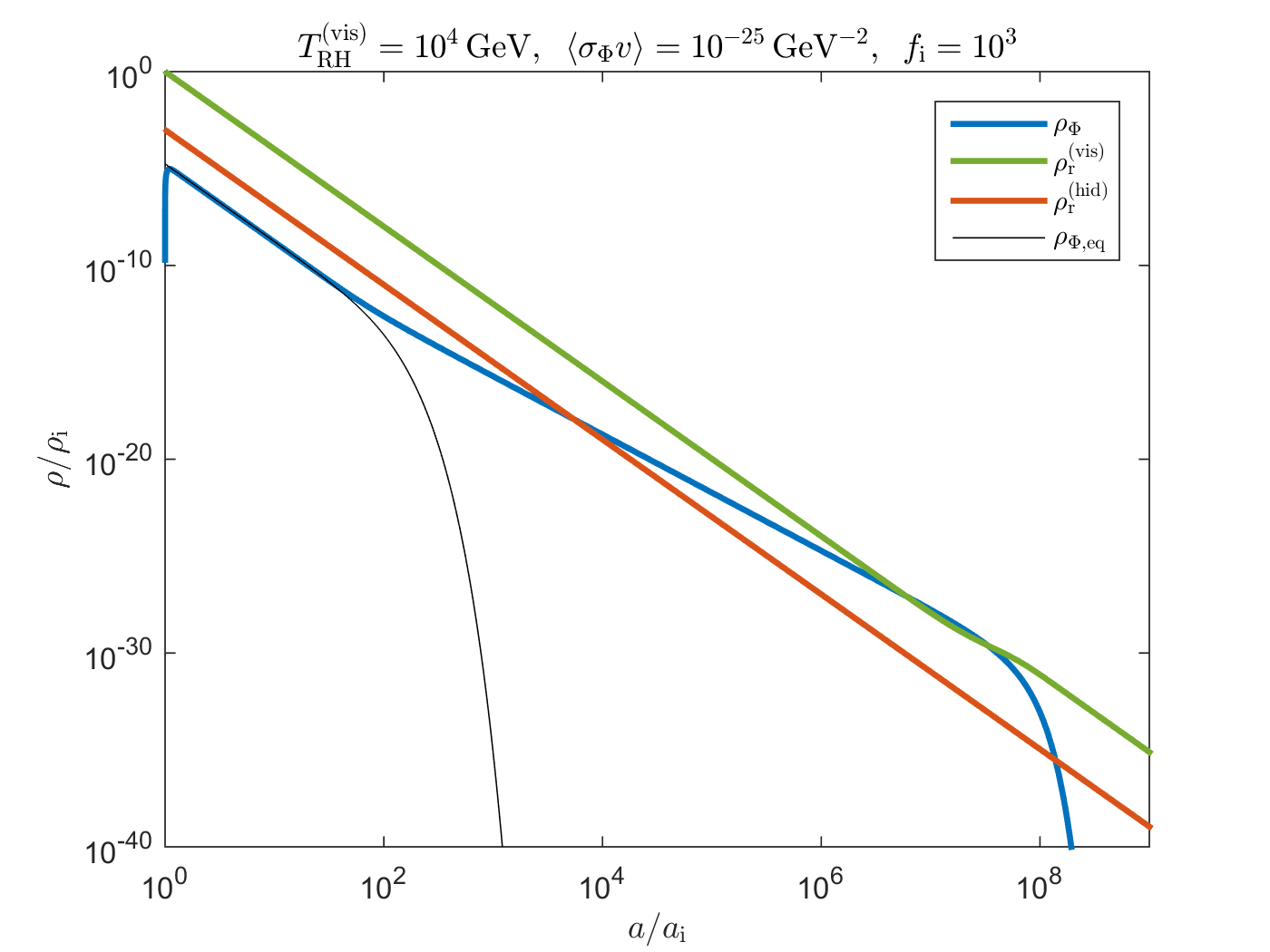}}
    \vspace{-0.4cm}
    \caption{Numerical evolution of the background energy density components with scale factor in the case of EMD by a decoupled particle \(\Phi\). EMD begins once \(\rho_\Phi\) dominates over both radiation components, and lasts until \(\Phi\) decays. Left panels: initial RD by the hidden sector. Right panels: initial RD by the visible sector. Top panels: \(\Phi\) decoupling from the dominant sector. Bottom panels: \(\Phi\) decoupling from the subdominant sector. The values of \(\left<\sigma_\Phi v\right>\) in each panel are chosen to correspond to relativistic freeze-out, thus yielding the longest possible EMD phase for the chosen background parameters. Larger values of \(\left<\sigma_\Phi v\right>\) will result in nonrelativistic freeze-out of \(\Phi\) while smaller values lead to freeze-in, both of which reduce the duration of EMD by lowering the frozen \(\Phi\) abundance and hence delaying the start time. Note that in the bottom two panels, relativistic freeze-out of \(\Phi\) essentially results in the limiting EMD case where the start and end are nearly coincident. The mass of \(\Phi\) in all panels is \(m_\Phi \approx 10^9\;{\rm GeV}\), due primarily to the value of \(T_{\rm RH}^{\rm(vis)} = 10^4\;{\rm GeV}\), as in Figure \ref{fig:rhovstmodulus-4}.}
    \label{fig:monoprhodec}
\end{figure}
The energy density of \(\Phi\) is given by \(\rho_\Phi =\left<E_{\Phi} f_\Phi \right> \), which 
we have approximated as \(\left<E_{\Phi}\right> n_\Phi\), with the average energy per particle given 
approximately as \(\left<E_{\Phi}\right> \approx \sqrt{m_\Phi^2 + 9T^2}\) \cite{Erickcek:2015jza, Giudice:2000ex}. The temperature \(T\) is that of the host sector. 
Note that we retain the decay of \(\Phi\) predominantly to the visible sector in order to preserve the standard history from BBN onward.\footnote{In the Boltzmann equations we do not include the possibility of \(\Phi\) decay to the HS, though one can easily include it by introducing branching fractions for both sectors. } 
We numerically solve the Boltzmann equations, in both decoupling cases, for the background energy densities, as shown in
Figure \ref{fig:monoprhodec}. As before, we use a smooth function for the temperature dependence of the relativistic degrees of freedom in the VS, \(g_*^{\rm(vis)}\), 
shown in Figure \ref{fig:gstar}.

To obtain the energy density evolution, we start in RD at some initial early time, with the HS and VS radiation related by the factor \(f_{\rm i}\), and with negligible \(\Phi\) energy density.\footnote{One can consider a non-negligible initial energy density for \(\Phi\), which will depend on the details of specific models, and we do not consider it any further here.} 
As the Universe cools, \(\Phi\) decouples from its host sector via freeze-out or freeze-in, leaving a frozen energy density that redshifts like matter once \(\Phi\) becomes non-relativistic. This matter energy density can then dominate over radiation, provided that the frozen energy density is high enough for domination to occur before the eventual decay of \(\Phi\). 
The decay completes near \(H \approx \Gamma_\Phi\), and we are subsequently left with the standard phase of domination by visible sector radiation.

The evolution of the equilibrium number density for \(\Phi\) transitions from relativistic to non-relativistic when the temperature of the host sector drops below \(m_\Phi\). Because of this transition, there is a maximum frozen number density for a given \(m_\Phi\), which is achieved through the decoupling of \(\Phi\) while it is relativistic and in chemical equilibrium with its host sector. This is relativistic freeze-out. If \(\Phi\) were to start with a number density larger than equilibrium, annihilations would drive it down to the equilibrium density, unless the annihilation rate was too small, which is not a scenario we will consider here because we assume RD at the initial time in order to justify an origin for the intervening EMD phase. Decoupling through relativistic freeze-out results in the earliest possible start time for the EMD phase caused by \(\Phi\) of a given mass, and requires the annihilation rate to be large enough such that \(\Phi\) reaches equilibrium while still relativistic, but not too large such that it remains in equilibrium after becoming non-relativistic. The largest value of \(\left<\sigma_\Phi v\right>\) that corresponds to relativistic freeze-out (which is the transition between relativistic and non-relativistic freeze-out) can be approximated by 

\begin{equation}\label{sigmavFO_R/NR}
    \left<\sigma_\Phi v\right> \approx \begin{dcases}
        \frac{\pi^3{g_*^{\rm(hid)}}^{1/2}(1+f_{\rm i})^{1/2}}{\sqrt{90}\zeta(3)g_\Phi M_{\rm P}m_\Phi} & {\rm HS\;decoupling}~,\\[5pt]
        \frac{\pi^3{g_{\rm*F}^{\rm(vis)}}^{1/2}\left(1+\frac{1}{f_{\rm i}}\right)^{1/2}}{\sqrt{90}\zeta(3)g_\Phi M_{\rm P}m_\Phi} & {\rm VS\;decoupling} ~,
    \end{dcases}
\end{equation}
assuming $x_{\rm F} \equiv m_\Phi/T_{\rm F}^{\rm(hid/vis)} \simeq O(1)$ for relativistic decoupling, and where 
$\zeta(s)$ is the Riemann zeta function of $s$.\footnote{We obtain these expressions by setting \(x_{\rm f} \simeq 1\) in the usual freeze-out condition using the relativistic expression for the equilibrium number density of \(\Phi\) (see Appendix \ref{Appendix:Freeze-out} for more on freeze-out decoupling).}

If instead the annihilation rate of \(\Phi\) is large enough to maintain equilibrium with its host sector below \(T \approx m_\Phi\), then decoupling will occur via non-relativistic freeze-out, resulting in a smaller frozen number density and thus a later start time for EMD. As the annihilation rate increases further, the frozen \(\Phi\) energy density decreases and the start of EMD approaches the time of reheating, resulting in a shorter duration for the EMD phase. This gives an upper limit, corresponding to \(H_{\rm MD} \gtrsim \Gamma_\Phi\), on the value of \(\left<\sigma_\Phi v\right>\), for a given mass and decay rate (or equivalently visible sector reheat temperature) for EMD to happen at all: 
\begin{equation}\label{sigmavFO_EMD}
    \left<\sigma_\Phi v\right> \lesssim \frac{m_\Phi}{3\Gamma_\Phi^{1/2}M_{\rm P}^2H_{\rm F}^{1/2}}\,,
\end{equation}
where \(H_{\rm F}\) is the expansion rate at freeze-out and given in Appendix \ref{Appendix:Freeze-out}, 
and we have used (\ref{monop:FOHMD}) for the expansion rate $H_{\rm MD}$ at the time of matter domination. 

Now going in the other direction, if the annihilation rate is smaller than that needed for relativistic freeze-out, \(\Phi\) will never reach local chemical and thermal equilibrium, which may possibly lead to a freeze-in process \cite{Hall:2009bx}. If freeze-in does occur, lowering \(\left<\sigma_\Phi v\right>\) further reduces the out-of-equilibrium number density, and thus the duration of EMD, down to a minimum value corresponding to the absence of EMD altogether. The value of \(\left<\sigma_\Phi v\right>\) corresponding to the transition between freeze-in and relativistic freeze-out (which defines the lower limit of the range of values leading to relativistic freeze-out) is approximately

\begin{equation}\label{sigmavFO/FI}
    \left<\sigma_\Phi v\right> \approx \begin{dcases}
        \frac{\pi^3{g_*^{\rm(hid)}}^{1/2}(1+f_{\rm i})^{1/2}}{\sqrt{90}\zeta(3)g_\Phi M_{\rm P}T_{\rm i}^{\rm(hid)}} & {\rm HS\;decoupling}~,\\[5pt]
        \frac{\pi^3{g_{\rm*i}^{\rm(vis)}}^{1/2}\left(1+\frac{1}{f_{\rm i}}\right)^{1/2}}{\sqrt{90}\zeta(3)g_\Phi M_{\rm P}T_{\rm i}^{\rm(vis)}} & {\rm VS\;decoupling}\,,
    \end{dcases}
\end{equation}
and the minimum value corresponding to \(H_{\rm MD} \gtrsim \Gamma_\Phi\) is (see Appendix \ref{Appendix:Freeze-in})
\begin{equation}\label{sigmavFI_EMD}
    \left<\sigma_\Phi v\right> \gtrsim
    \begin{dcases}
        \frac{3\pi^7{g_*^{\rm(hid)}}^{3/2}(1+f_{\rm i})^{3/2}\Gamma_\Phi^{1/2}}{90^{3/2}\zeta(3)^2g_\Phi^2M_{\rm P}m_\Phi H_{\rm i}^{1/2}} & {\rm HS\;decoupling}~,\\[5pt]
        \frac{3\pi^7{g_{\rm*i}^{\rm(vis)}}^{3/2}\left(1+\frac{1}{f_{\rm i}}\right)^{3/2}\Gamma_\Phi^{1/2}}{90^{3/2}\zeta(3)^2g_\Phi^2M_{\rm P}m_\Phi H_{\rm i}^{1/2}} & {\rm VS\;decoupling}\,.
    \end{dcases}
\end{equation}
We summarize these three different regimes of the annihilation rate. Starting with small annihilation rates, the decoupling of \(\Phi\) proceeds as follows. For \(\left<\sigma_\Phi v\right>\) less than the right-side of \eqref{sigmavFI_EMD}, \(\Phi\) decouples via freeze-in at such low energy densities that it will never dominate over radiation before decaying. For rates that satisfy \eqref{sigmavFI_EMD} but are less than \eqref{sigmavFO/FI}, the frozen-in energy density of \(\Phi\) is large enough to dominate, leading to longer EMD durations as \(\left<\sigma_\Phi v\right>\), and thus the frozen-in energy density, is increased. Between \eqref{sigmavFO/FI} and \eqref{sigmavFO_R/NR}, decoupling occurs via relativistic freeze-out, which yields the largest frozen \(\Phi\) energy density and the longest possible EMD duration, independent of \(\left<\sigma_\Phi v\right>\). We note that essentially the only difference in \eqref{sigmavFO/FI} and \eqref{sigmavFO_R/NR} is the presence of the initial host sector temperature or the \(\Phi\) mass in the denominator. Because the initial temperature can in general be quite large compared to \(m_\Phi\), the regime of \(\left<\sigma_\Phi v\right>\) corresponding to relativistic freeze-out can extend for many orders of magnitude. For \(\left<\sigma_\Phi v\right>\) larger than \eqref{sigmavFO_R/NR} but satisfying \eqref{sigmavFO_EMD}, \(\Phi\) decouples via nonrelativistic freeze-out, resulting in smaller frozen-out energy densities, and thus shorter EMD durations, as \(\left<\sigma_\Phi v\right>\) is increased. Finally, for rates larger than the right-side of \eqref{sigmavFO_EMD}, the frozen-out energy density is again too small to establish EMD before \(\Phi\) decays. 

Other than defining the range of annihilation rates that can yield an EMD phase\footnote{We include an additional constraint in Appendix \ref{Appendix:Additional-constraint} on the parameter values that must hold for an EMD phase to have nonzero duration.}, the significance of these regimes of \(\left<\sigma_\Phi v\right>\) is that a particular EMD phase, with a fixed start time and end time, can be established by two different values of \(\left<\sigma_\Phi v\right>\), one corresponding to freeze-out and the other to freeze-in.

The abundance of monopoles produced by the HS phase transition is determined by using
\eqref{omegabefore}--\eqref{omegaafter}, 
which are given 
in Section \ref{n/s}. These expressions were obtained in a model-independent context
and are valid in the cases presented in this section, provided that we use the appropriate expressions for quantities such as \(H_{\rm MD}\). 

\begin{figure}[ht!]
    \centering
    \subfloat{\includegraphics[width=0.5\textwidth]{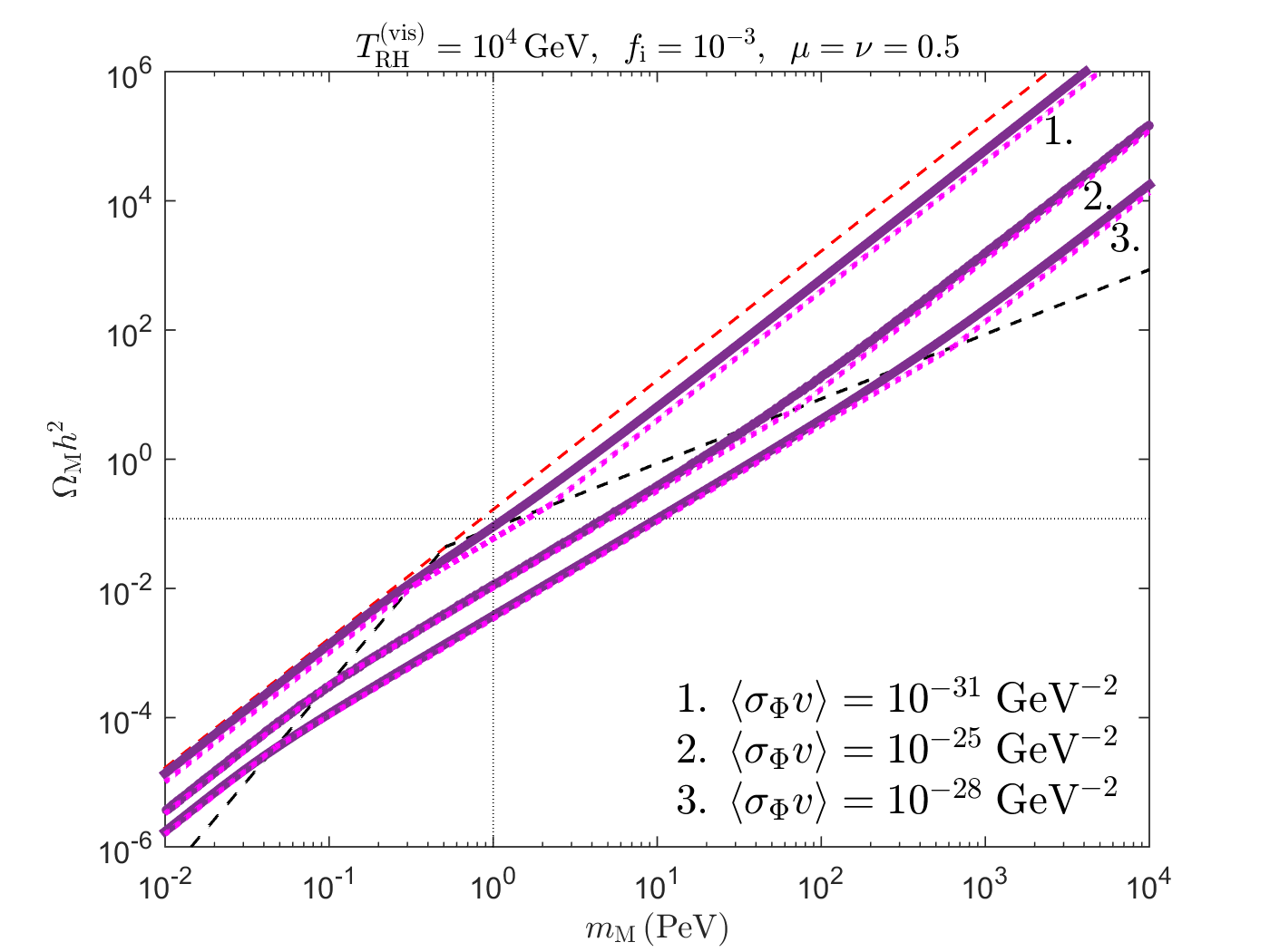}}
    \subfloat{\includegraphics[width=0.5\textwidth]{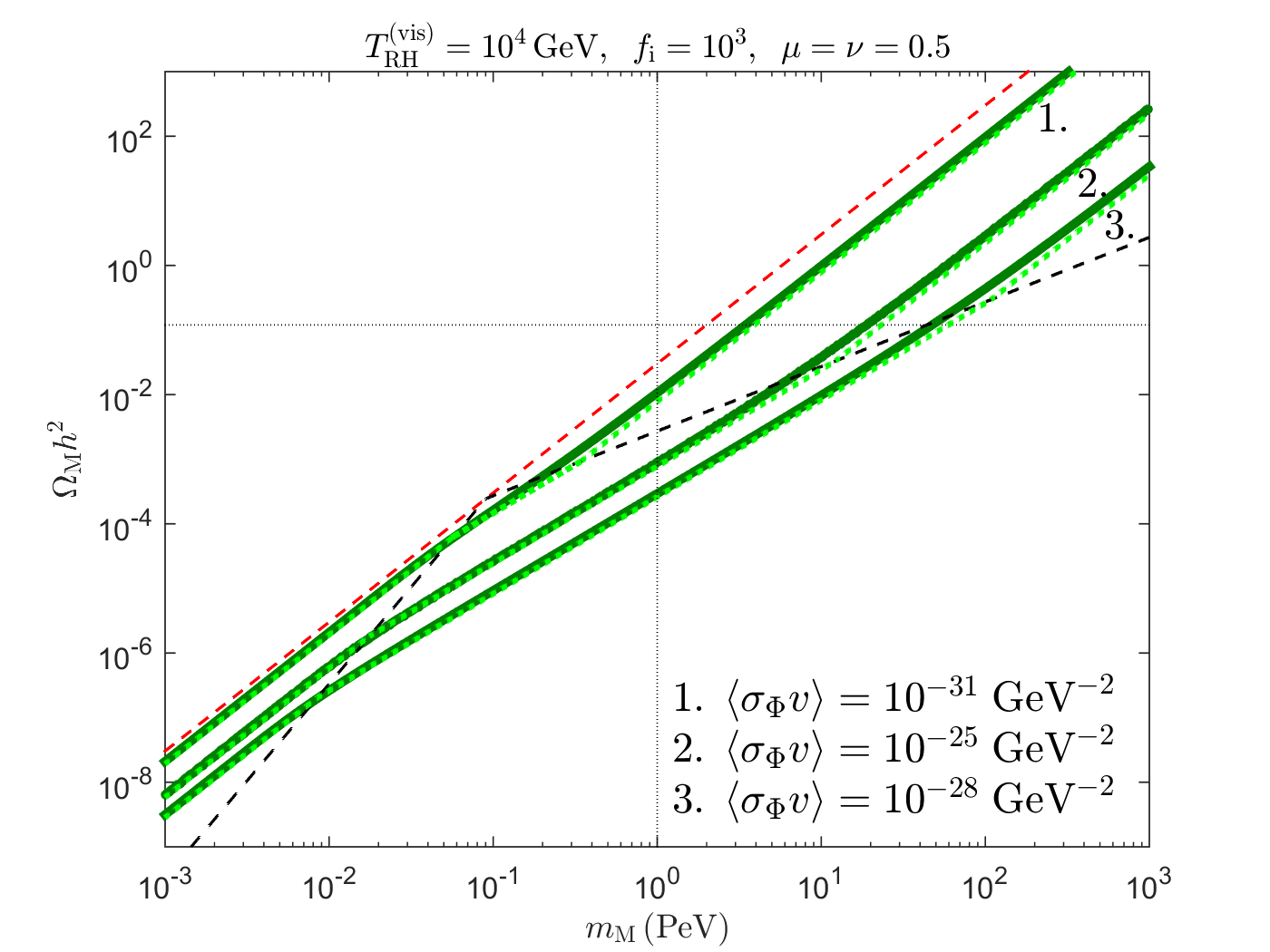}}\\\vskip -0.4cm
    \subfloat{\includegraphics[width=0.5\textwidth]{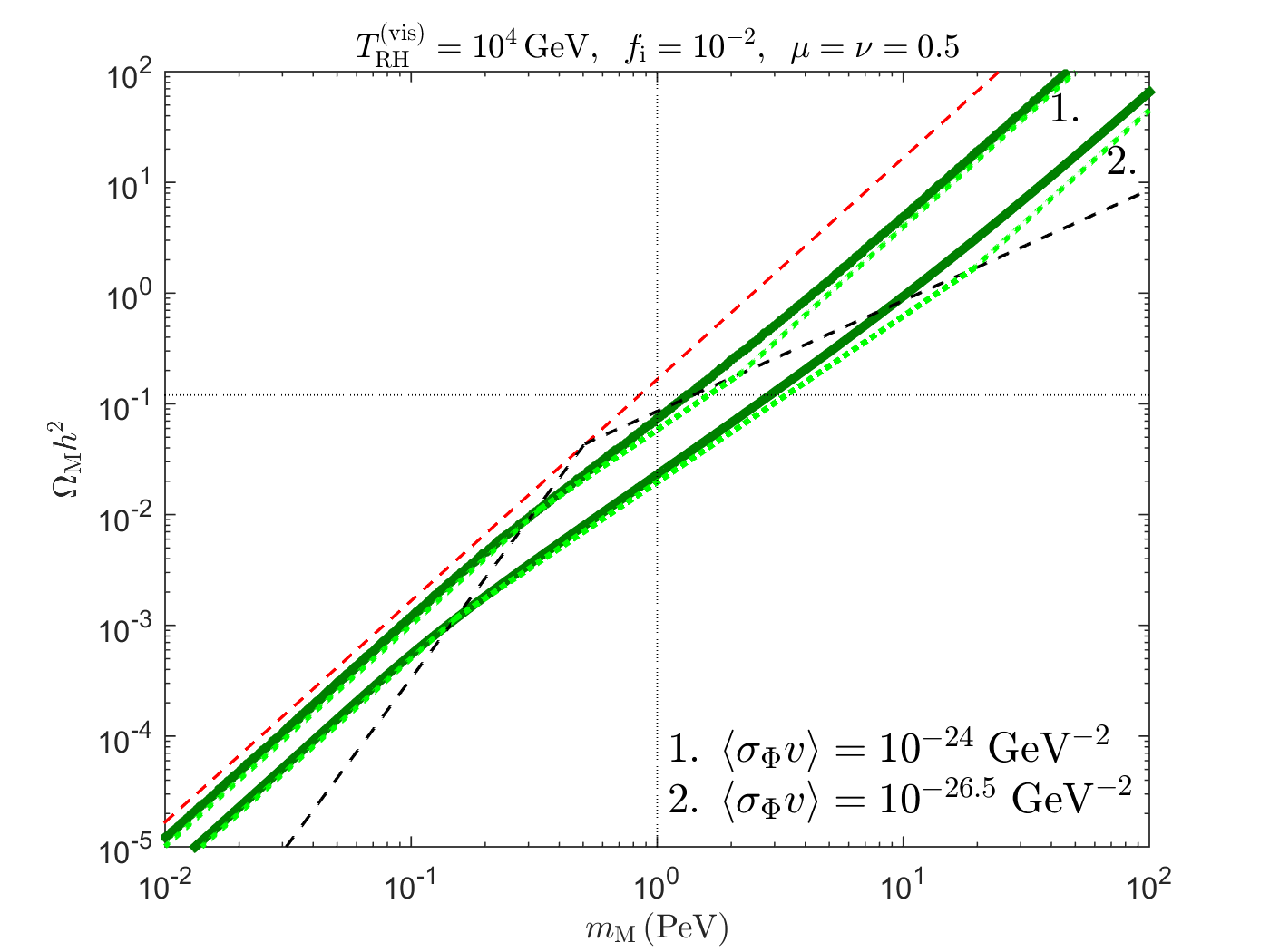}}
    \subfloat{\includegraphics[width=0.5\textwidth]{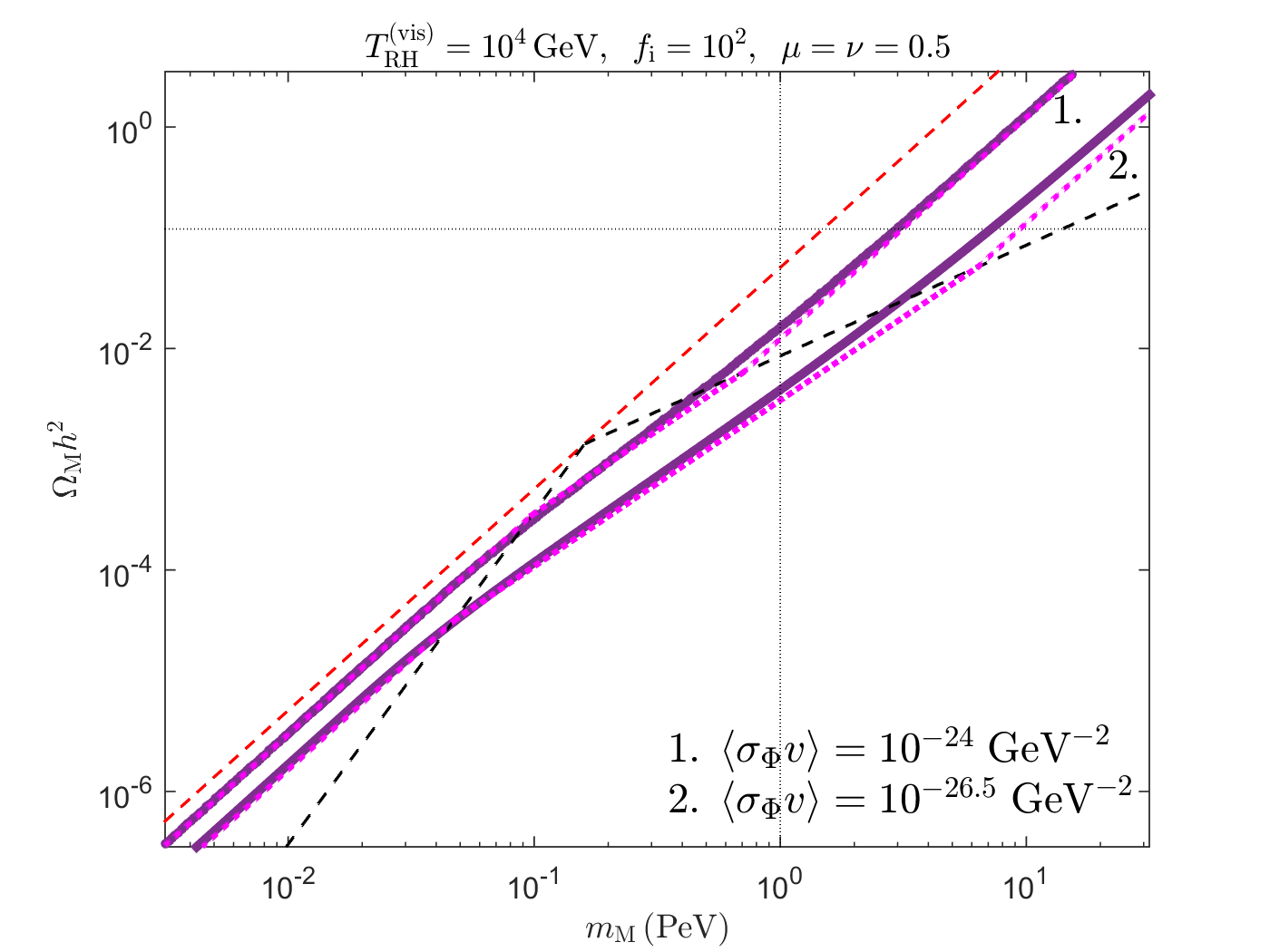}}\\
    \caption{Dependence of the present-day monopole relic abundance on the monopole mass in the case of EMD driven by a decoupled particle. As in Figures \ref{fig:omegamm} and \ref{fig:monopmod}, the solid curves (purple and green) are obtained from a numerical evolution of the background, while the dotted lines (light purple and light green) on top of the numerical curves are the analytical expressions (\ref{omegabefore}), (\ref{omegaduring}), and (\ref{omegaafter}). The purple color denotes \(\Phi\) decoupling from the HS, while the green color corresponds to decoupling from the VS. All other lines have the same meaning as in Figures \ref{fig:omegamm} and \ref{fig:monopmod}, which we repeat here. The red dashed line in all panels marks the purely RD equivalent scenario. The two black dashed lines in all panels indicate monopole production occurring at the start or end of EMD. The dotted horizontal and vertical lines in all panels mark \(\Omega_{\rm M}h^2 = 0.12\) and \(m_{\rm M} = 1\;{\rm PeV}\) respectively. Left panels: initial RD occurring in the hidden sector. Right panels: initial RD occurring in the visible sector. Top panels: \(\Phi\) decoupling from the dominant sector. Bottom panels: \(\Phi\) decoupling from the subdominant sector. In each panel, the curves which sit farthest to the right correspond to relativistic freeze-out of \(\Phi\) from its host sector and thus mark the largest monopole masses accessible for the chosen parameters. The dependence on \(T_{\rm RH}^{\rm(vis)}\) and the critical exponents \(\mu\) and \(\nu\) is the same as in Figure \ref{fig:monopmod}.}
    \label{fig:monopdec}
\end{figure}

The present-day relic monopole abundance is shown in Figure \ref{fig:monopdec} as a function of monopole mass for some example parameter values, and we have again taken \(x_{\rm M} \equiv m_{\rm M}/T_{\rm C}^{\rm(hid)} = 50\) 
and $\alpha=\lambda=1$. 
We in particular consider several values for $\langle \sigma_\Phi v \rangle$, and 
we have checked that these values are well-below the perturbativity limit 
for the $\Phi$ mass inferred from \eqref{gammaphi},~\eqref{HRH}, and the assumed reheat temperature.
As in the modulus case, there are three regions corresponding to monopole production before, during, and after EMD, and the curves have the same behavior as before. The main feature that sets the decoupled-particle case apart from the modulus case is that any particular curve can be obtained be either non-relativistic freeze-out or freeze-in, meaning the value of the annihilation rate of \(\Phi\) can be quite different while still reproducing the same curve. Otherwise, the same regions are generally accessible to a modulus or decoupled-particle scenario, where the maximum extent toward larger monopole masses is set by either the maximum initial modulus amplitude or by relativistic freeze-out in the two cases respectively. 

We finally note that the case of freeze-in depends on the initial host-sector temperature because freeze-in of \(\Phi\) occurs in RD, such that the time of peak \(\Phi\) production from the background occurs at the initial time (see \cite{Allahverdi:2019jsc} for details of freeze-in during RD before EMD). In our numerical calculations, we chose the initial time arbitrarily, with an initial energy density configuration consisting of dominant radiation and negligible \(\Phi\). For a given initial time, there is a unique annihilation rate that results in a particular freeze-in \(\Phi\) energy density, provided that we remain within the freeze-in regime of the annihilation rate. The important thing to note is that the accessible region in \(\Omega_{\rm M}h^2\) vs \(m_{\rm M}\) is generally independent of the initial time because it is determined by the start and end of EMD, which can be obtained by multiple values of the initial time and annihilation rate. 

\clearpage

\newpage
\section{Parameter values giving observed dark matter relic abundance}
\label{sec:monopole-mass-for-relic-abundance}
In this section, we will consider the values of our various parameters that result in the observed present-day DM abundance of \(\Omega_{\rm M}h^2 = 0.12\). As we've seen in the two previous sections, our analytical and numerical results agree very well, and we will therefore present an analytical analysis of the main parameters of our scenario, rather than a full numerical parameter scan. 

We will primarily use \eqref{omegabefore}--\eqref{omegaafter} as well as  \eqref{eq:fRH-ef} which 
gives $f_{\rm RH} \propto e_{\rm f}$, requiring that the observed DM relic abundance is achieved. For clarity in the analysis below, we will not specify the identity of the field \(\Phi\), taking the beginning and end of EMD as the more fundamental parameters. We will use the VS reheat temperature \(T_{\rm RH}^{\rm(vis)}\) to set the end of EMD, and the factor \(e_{\rm f} = a_{\rm RH}/a_{\rm MD}\) to fix the duration of EMD.
% (see Eq.~\eqref{e_f})
Recall that \(e_{\rm f}\) can be expressed as (see Appendix \ref{appendix:e_f}): 

\begin{equation}\label{e_f_H}
    e_{\rm f} = \left(\frac{H_{\rm MD}}{\Gamma_\Phi}\right)^{2/3}
    \,.
\end{equation}
The remaining parameters are the initial ratio of the VS to HS radiation energy density \(f_{\rm i}\), the monopole mass \(m_{\rm M}\), as well as the various parameters associated with the details of the phase transition, \(x_{\rm M}\), \(\lambda\), \(\mu\), and \(\nu\). Four of these eight parameters can vary by many orders of magnitude in the cosmological histories we have been considering: \(m_{\rm M}\), \(T_{\rm RH}^{\rm(vis)}\), \(e_{\rm f}\), and \(f_{\rm i}\), 
so here we will focus on those as they lead to a more direct effect 
on the resulting cosmology. The others have much narrower ranges, and for these we will 
consider a discrete set of possibilities. Also, we will not vary parameters such as $\alpha$, 
$m_\Phi$, $\Phi_{\rm i}$ (in the case of the modulus), or $\langle \sigma_\Phi v \rangle$ (in the case of the decoupled particle), 
as including variations in   
these parameters is degenerate, in the sense that they lead to the same cosmology, as discussed in Section \ref{mod}.

\begin{figure}[ht!]
    \centering
    \subfloat{\includegraphics[trim=0cm 0cm 0cm 0.3cm, clip=true, width=0.5\textwidth]{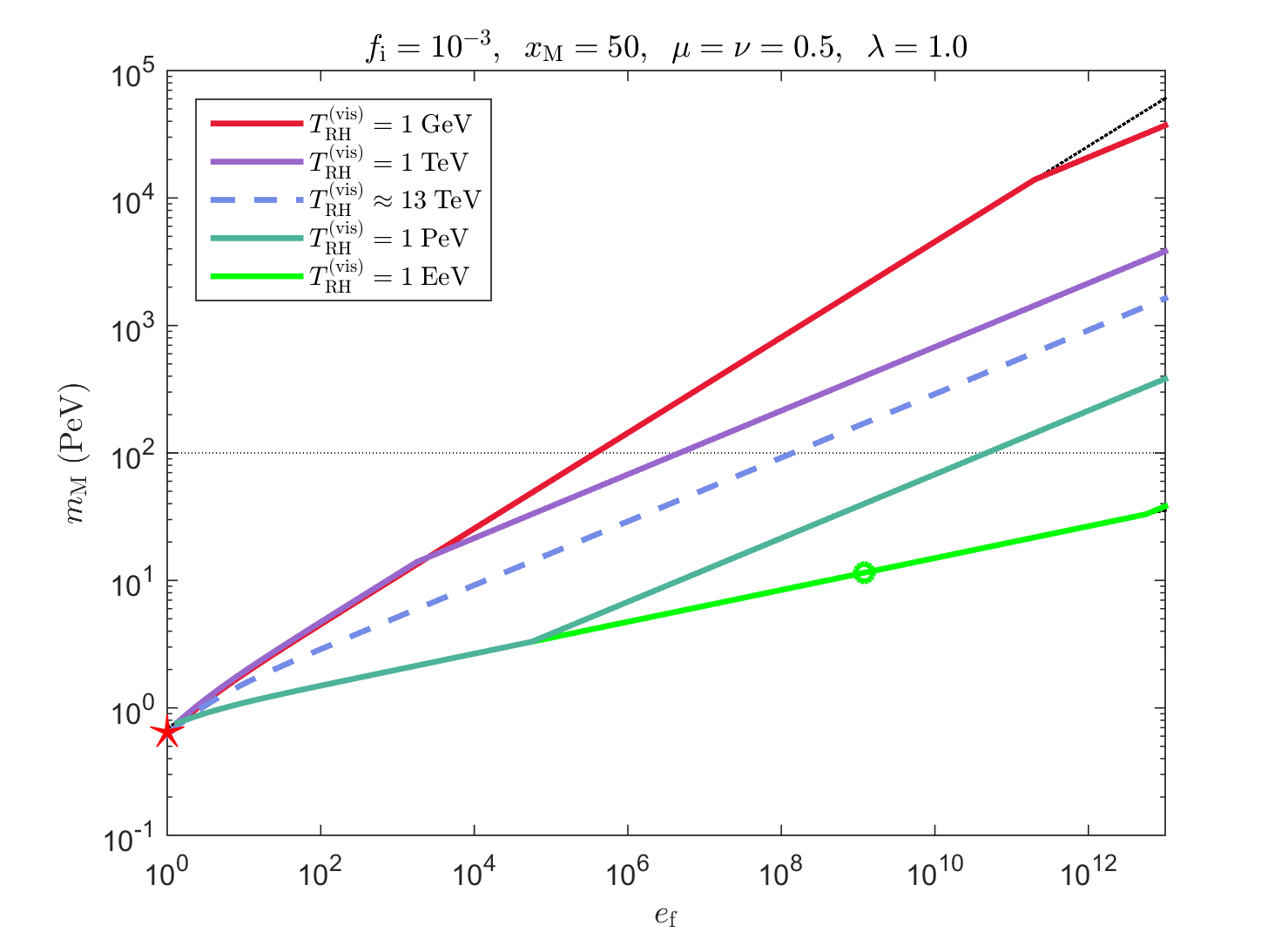}}
    \subfloat{\includegraphics[trim=0cm 0cm 0cm 0.3cm, clip=true, width=0.5\textwidth]{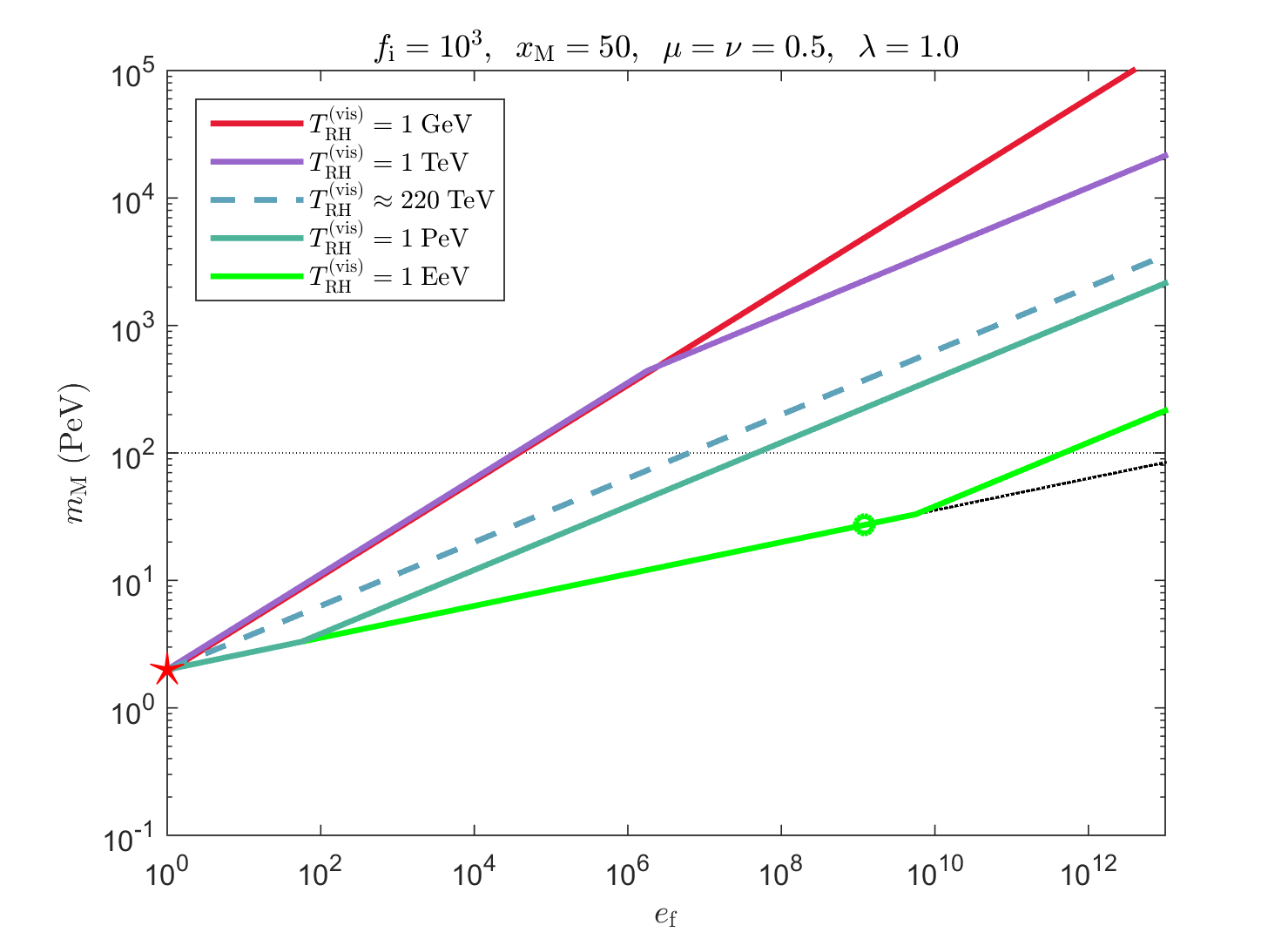}}\\
    \vspace{-0.4cm}
    \subfloat{\includegraphics[trim=0cm 0cm 0cm 0.3cm, clip=true, width=0.5\textwidth]{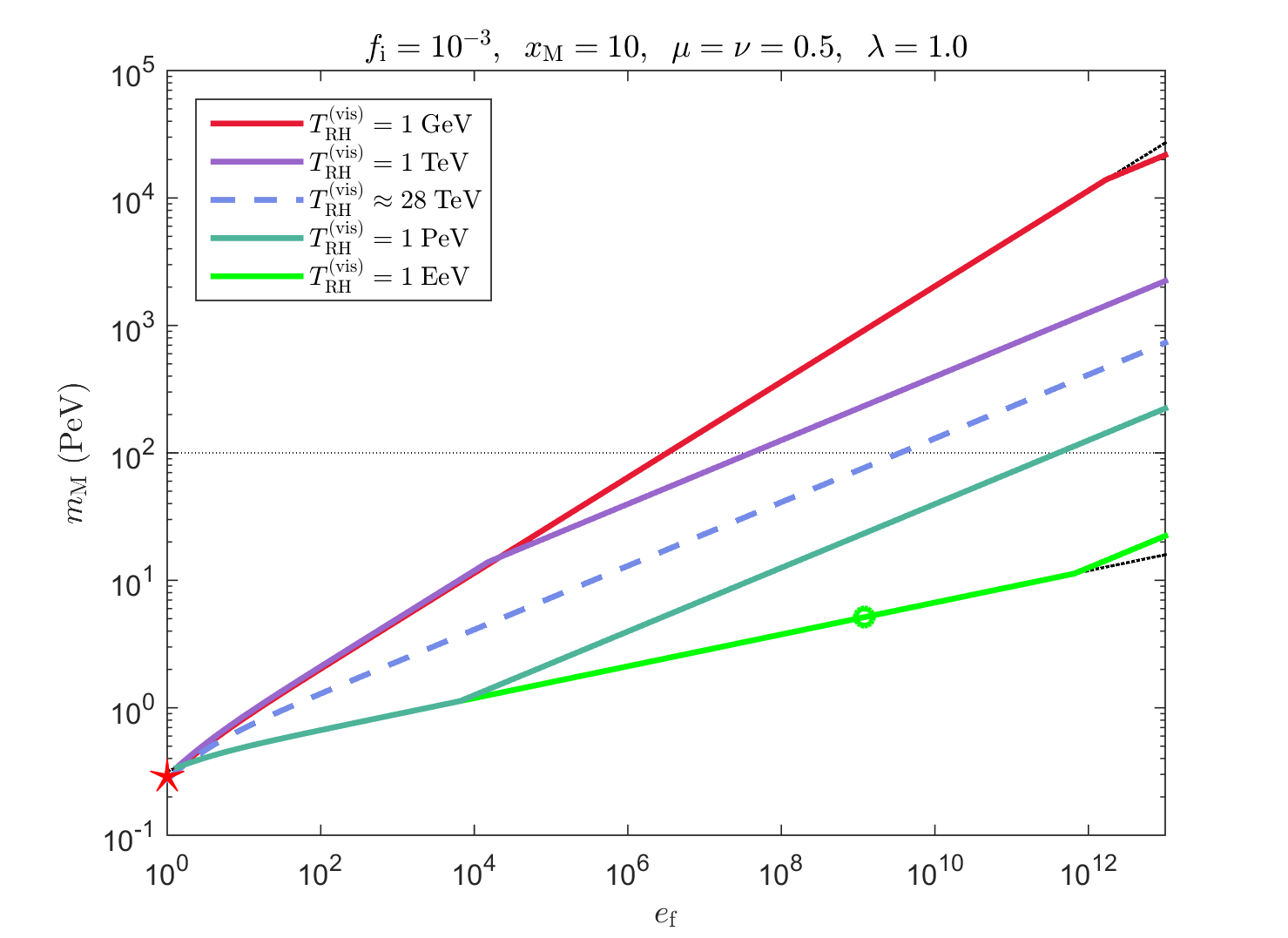}}
    \subfloat{\includegraphics[trim=0cm 0cm 0cm 0.3cm, clip=true, width=0.5\textwidth]{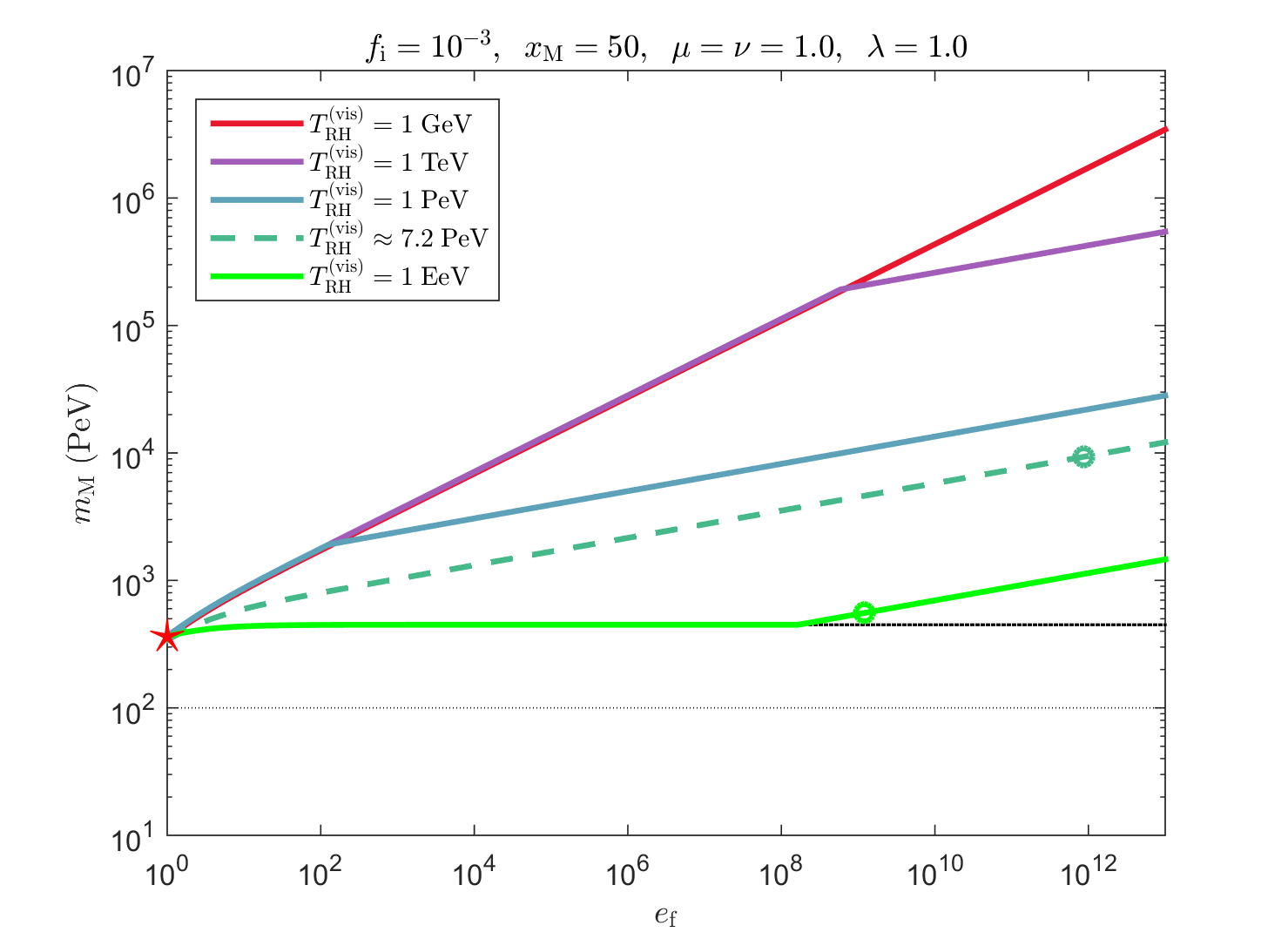}} 
    \vspace{-0.4cm}
    \caption{Contours of \(T_{\rm RH}^{\rm(vis)}\) in the \(m_{\rm M} - e_{\rm f}\) plane holding the monopole relic abundance fixed at the observed value for DM of \(\Omega_{\rm M}h^2 = 0.12\), obtained using \eqref{omegabefore}--\eqref{omegaafter}, and setting $\alpha=1$.
    Each panel corresponds to parameter variation relative to the top left panel. Larger values of \(T_{\rm RH}^{\rm(vis)}\) indicate that EMD ends at an earlier time, while larger values of \(e_{\rm f}\) correspond to longer EMD durations. Each solid contour has two segments with different slopes (a few of which occur beyond the range shown in the figure). Contours above the dashed contour overlap in their steeper segments, which follow the upper dotted black boundary line corresponding to monopole production before EMD, while those below overlap in their shallower segments, and follow the lower dotted black boundary line corresponding to production after EMD. Segments that are parallel to the dashed contour indicate monopole production during EMD. 
    The region above a given contour results in overproduction of DM, while the region below results in underproduction. The slight differences in the overlap of the upper contours are due to changes in \(g_{\rm*RH}^{\rm(vis)}\). We include a horizontal line at \(m_{\rm M} = 100\;{\rm PeV}\) for reference across panels, as well as a red `star' which marks the pure RD scenario at \(e_{\rm f} = 1\). The green circles located at \(e_{\rm f} \approx 1.2\times 10^9\) along the 
    EeV contour, and \(e_{\rm f} \approx 8.7\times 10^{11}\) along the 
    7.2 PeV contour, correspond to the bound on \(H_{\rm MD}\) from \eqref{Planck_H_MD}. 
    Please see the text for more details.}
    \label{fig:parameters}
\end{figure}

Figure \ref{fig:parameters} shows contours of \(T_{\rm RH}^{\rm(vis)}\) in the \(m_{\rm M} - e_{\rm f}\) plane, with the monopole abundance held fixed at \(\Omega_{\rm M}h^2 = 0.12\). The region above each contour results in overproduction of DM, while the region below results in underproduction. What can immediately be seen from the figure is that most lines shown have positive slopes in this plane, meaning that a longer EMD duration (i.e.~a larger value of \(e_{\rm f}\)) requires a larger monopole mass in order to achieve the same monopole abundance. This is consistent with the behavior in Figures \ref{fig:monopmod} and \ref{fig:monopdec}, where the curves corresponding to longer EMD periods cross the \(\Omega_{\rm M}h^2 = 0.12\) line at larger monopole masses. Furthermore, for fixed monopole mass, a longer EMD duration results in too much dilution and thus underproduction of DM, while a shorter duration doesn't dilute the monopole abundance enough, leading to overproduction.

In each panel of the figure, the region accessible to the 
\(T_{\rm RH}^{\rm (vis)}\) contours is bounded by two 
black dotted lines: an upper line with slope given by 
\(m_{\rm M}^{1+\frac{3\nu}{1+\mu}} \propto e_{\rm f}^{\frac{3}{4}}\) 
corresponding to monopole production before EMD; and a lower line with slope given by 
\(m_{\rm M}^{1+\frac{3\nu}{1+\mu}} \propto e_{\rm f}^{\frac{3}{4}-\frac{3\nu}{2(1+\mu)}}\) 
corresponding to production after EMD. Note that only segments of these lines are visible in the figure, as they extend underneath the main contours. The boundary lines meet at the left edge of each figure panel, where \(e_{\rm f} \simeq 1\), which corresponds to the absence of an EMD phase, 
denoted by a `red star' in the figure. The monopole mass at this point agrees with the mass at which the RD line crosses \(\Omega_{\rm M}h^2 = 0.12\) in Figures \ref{fig:monopmod} and \ref{fig:monopdec}, for corresponding parameter values.\footnote{As \(e_{\rm f}\) approaches 1, which corresponds to shorter and shorter EMD periods until EMD is no longer well defined, the power-law behavior of the contours in Figure \ref{fig:parameters} breaks down. This can be seen in the slight curvature of the contours near \(e_{\rm f} = 1\), and one must be more careful when using approximate expressions for \(e_{\rm f}\) in this region. However, because this deviation is quite small, and only occurs for poorly-defined EMD periods, approximations based on large \(e_{\rm f}\) are sufficient when considering our EMD scenarios.} 

Each panel additionally shows a special, dashed blue-green contour which separates two regimes of \(T_{\rm RH}^{\rm(vis)}\), and passes through the RD point mentioned above without changing slope. Relative to Figures \ref{fig:monopmod} and \ref{fig:monopdec}, this contour corresponds to the special value of \(T_{\rm RH}^{\rm(vis)}\) which places the intersection of the two black dashed lines (representing the start and end of EMD) at \(\Omega_{\rm M}h^2 = 0.12\) (this is most easily seen in the middle panels of Figure \ref{fig:monopmod}, where the intersection point of the two black dashed lines shifts along the RD line as \(T_{\rm RH}^{\rm(vis)}\) is changed). As the duration of EMD is increased along this dashed contour, the contour rises away  from \(e_{\rm f} = 1\) with a slope given by \(m_{\rm M}^{1+\frac{3\nu}{2(1+\mu)}} \propto e_{\rm f}^{\frac{3}{4}-\frac{9\nu}{8(1+\mu)}}\), where we have assumed the large-\(e_{\rm f}\) behavior. Thus the entire dashed blue contour corresponds to monopole production occurring in EMD, consistent with Figures \ref{fig:monopmod} and \ref{fig:monopdec}. 

Each contour located above the dashed contour (with lower values of \(T_{\rm RH}^{\rm(vis)}\)) has two segments with different slopes: beginning on the left side at \(e_{\rm f} \simeq 1\), the contours rise along the upper boundary line, corresponding to monopole production before EMD, until they reach a point which corresponds to production at the start of EMD - beyond this point, the contours deviate from the upper boundary with a slope parallel to the dashed contour - this segment corresponds to monopole production during EMD. 

The contours located below the dashed contour (with higher values of \(T_{\rm RH}^{\rm(vis)}\)) have a similar two-segment behavior: beginning again at \(e_{\rm f} \simeq 1\), the contours rise at a shallow slope along the lower boundary line (monopole production after EMD), until they reach a point corresponding to production at the end of EMD - from here on the contours leave the lower boundary and continue with the same slope as the dashed contour - production in this region occurs during EMD. The region above the dashed contour can therefore only access monopole production before and during EMD, while the region below only accesses production during and after EMD. Additionally, we note that in the lower right panel, with \(\mu = \nu = 1\), the slope of the ``after EMD" segment is essentially independent of \(e_{\rm f}\), consistent with the lower panels of Figure \ref{fig:monopmod} where the segments of the numerical curves corresponding to monopole production after EMD coincide with the pure RD scenario, thus erasing any dependence on the prior EMD history. 

The boundaries of the accessible region in the \(m_{\rm M} - e_{\rm f}\) plane, which correspond to monopole production before and after EMD, are given by \eqref{omegabefore} and \eqref{omegaafter}, and are independent of \(T_{\rm RH}^{\rm(vis)}\). This can be trivially understood for production after EMD, while in the case of production before, the monopole abundance experiences dilution from the full EMD phase, regardless of it's specific timing. However, as \(e_{\rm f}\) increases, a given contour turns away from the boundary at a point that corresponds to the start (upper contours) or the end (lower contours) of EMD, which does depend on \(T_{\rm RH}^{\rm(vis)}\) (see \eqref{n/sMDstart} and \eqref{n/sMDend}). The location of these ``turn-off" points in the \(m_{\rm M} - e_{\rm f}\) plane can be obtained in the following way. 

For monopole production at the start of EMD, \eqref{HTRD}, \eqref{HRH}, and \eqref{e_f_H} lead to the relation 

\begin{equation}\label{turnoff_start}
    m_{\rm M}^{\rm(start)} \approx \left(\frac{g_{\rm*RH}^{\rm(vis)}}{g_{\rm*MD}^{\rm(hid)}(1+f_{\rm i})}\right)^{1/4} x_{\rm M} T_{\rm RH}^{\rm(vis)} e_{\rm f}^{3/4} \,,
\end{equation}
where we have additionally made use of \(f_{\rm RH} \gtrsim 1\). This expression can then be used along with \eqref{omegabefore} to locate the monopole mass and EMD duration which result in the observed DM abundance for monopole production at the start of EMD. 

For monopole production at the end of EMD, \eqref{HRH} can be expressed in terms of VS quantities and set equal to itself in terms of HS quantities to obtain 

\begin{equation}\label{turnoff_end}
    m_{\rm M}^{\rm(end)} \approx \left(\frac{g_{\rm*RH}^{\rm(vis)}}{g_{\rm*RH}^{\rm(hid)}(1+f_{\rm i})}\right)^{1/4} x_{\rm M} T_{\rm RH}^{\rm(vis)} e_{\rm f}^{-1/4} \,.
\end{equation}
This expression, together with \eqref{omegaafter}, then yields the monopole mass and EMD duration which result in the observed DM abundance for monopole production at the end of EMD.

The value of \(T_{\rm RH}^{\rm(vis)}\) for the dashed contour shown in Figure \ref{fig:parameters}, which separates the two sets of contours, can similarly be obtained by first eliminating \(e_{\rm f}\) in \eqref{turnoff_start} and \eqref{turnoff_end}. This corresponds to the RD point at \(e_{\rm f} \simeq 1\), marked by the red star, where the before and after boundaries meet (as do production at the start and end of EMD). Then, using either \eqref{omegabefore} or \eqref{omegaafter} gives the monopole mass required for \(\Omega_{\rm M}h^2 = 0.12\), which then yields the special value of \(T_{\rm RH}^{\rm(vis)}\) by direct substitution. 

Contours of \(\Omega_{\rm M}h^2\) in the \(m_{\rm M} - e_{\rm f}\) plane for 
values of \(\Omega_{\rm M}h^2\) not equal to 0.12 can be obtained by shifting the curves of Figure \ref{fig:parameters} according to Eq.~\eqref{Omega-mass}. Note that because the monopole abundance produced during the EMD and RD periods displays different power-law dependence on the monopole mass, the EMD and RD segments of the curves will shift by different amounts, resulting in movement of the turn-off points along the before and after boundaries.

As is evident from \eqref{e_f_H}, a 
long duration for the EMD phase requires a large separation 
between $H_{\rm MD}$ and $\Gamma_\Phi$. This is rather easy to achieve, even for high reheat temperatures.
However in inflationary models, the Hubble parameter at the start of EMD, $H_{\rm MD}$, is bounded 
from above by the value of the Hubble parameter $H_{\rm I}$ at the end of inflation. This would correspond to an 
interesting scenario in which after inflation the early Universe directly enters the EMD phase,  
with some reheating in the hidden sector so that the initial temperature in that sector is above the critical 
temperature. However, from the non-detection of tensor modes, PLANCK data gives an upper limit to $H_{\rm I}$
\cite{Akrami:2018odb}
\begin{equation}\label{Planck_H_MD}
    H_{\rm I} < 2.5 \times 10^{-5} M_{\rm P}~.
\end{equation}
Using \eqref{e_f_H} for a  given $T_{\rm RH}^{\rm(vis)}$, this limit translates into an upper bound on $e_{\rm f}$. For the highest $T_{\rm RH}^{\rm(vis)}$ considered in Figure \ref{fig:parameters}, which is EeV, the maximum value for $e_{\rm f}$ allowed by this bound is $e_{\rm f, max}\approx 10^9$. This maximum for the $T_{\rm RH}^{\rm(vis)}=1$ EeV contour is denoted in Figure \ref{fig:parameters} by a `green dot'. 
Along this contour larger values for $e_{\rm f}$ are excluded 
by \eqref{Planck_H_MD}. The only other $T_{\rm RH}^{\rm(vis)}$ contour affected is $T_{\rm RH}^{\rm(vis)}=$ 7.2 PeV, for which 
$e_{\rm f, max}\approx 10^{12}$. For all other values of $T_{\rm RH}^{\rm(vis)}$ considered, the maximum value of $e_{\rm f}$ is 
off the right edge of the plots.

Lastly, we comment on some interesting effects when the critical exponents, \(\mu\) and \(\nu\), satisfy \(\mu = \nu > 1\). Though we have specifically considered \(\mu = \nu = \{1/2,\,1\}\) in our figures, the expressions presented throughout the text are applicable to more general values of the critical exponents.\footnote{One has to be careful about possible modifications to the \(m_{\rm M} - T_{\rm C}\) relation in such cases as well. For the purposes of this discussion, we will assume the direct proportionality of a classical phase transition, though this can be generalized without too much effort.} In particular, we recall that as \(\mu\) and \(\nu\) approach 1, the case of monopole production after EMD (case III) approaches a purely RD scenario, so that when \(\mu = \nu = 1\), the dependence on the prior EMD history is completely removed. This suggests that for \(\mu = \nu > 1\), or more generally \(2\nu > 1 + \mu\), the monopole mass required for the observed dark matter abundance can actually be smaller than the RD case, at least for monopole production after EMD or shortly before its end. We have checked that this is indeed the case, however, the RD curve itself gets shifted to higher monopole masses when \(\mu = \nu > 1\) such that case III actually results in heavier masses as compared to \(\mu = \nu < 1\) (keeping the relic abundance fixed). This can be seen from expressions such as \eqref{nMsVSc} and \eqref{omegaafter}, where increasing the critical exponents above 1 results in an increase in the required monopole mass for both EMD and RD scenarios, but the increase is larger in the RD case. We also note that, from \eqref{corrlength}, the correlation length gets larger as the critical exponents are increased, resulting in less correlation volumes per Hubble volume, which in turn results in a smaller monopole number density at production. 

In this work we have broadened the scale for hidden sector monopoles masses to O(1--$10^5$) PeV. One may 
wonder how robust the lower limit of 1 PeV actually is. The effect of lowering the monopole mass relative to a RD scenario when \(\mu = \nu > 1\) is greater for a longer EMD duration, as the lower boundary line in Figure \ref{fig:parameters} acquires a negative slope. Additionally, the visible-sector reheat temperature needs to be larger than that for \(\mu = \nu < 1\) in order for contours of the observed dark matter relic abundance to access the lower boundary line -- note the different positioning of the PeV contour in the upper-left and lower-right panels of Figure \ref{fig:parameters}. 
Because of these two effects, 
an extended EMD period occurring very early will have the greatest effect in producing enough lower-mass monopoles to reproduce the observed DM abundance. Perhaps if the phase transition occurs toward the end of (or after) a period of EMD caused by inflationary reheating at very high temperatures, the monopole mass may be able to be brought below the PeV scale and still result in the full DM relic abundance. Furthermore, having the HS temperature be extremely suppressed below the VS actually helps lower the needed monopole mass significantly, as long as the VS reheat temperature is large enough to bring up the abundance. This suppression effect also applies to a purely RD scenario. 

In passing, we finally note that like \(\mu = \nu = 1\), setting \(\mu = \nu = 2\) is another special case in which the monopole abundance produced during EMD is now independent of the Hubble rate at the time of production, and only depends on the critical temperature. This can easily be seen in \eqref{nMsVSreh}, where the factor of \(H_{\rm C}^2\) in the denominator due to redshift cancels the dependence on the critical exponents. If the altered phase of expansion is instead caused by a form of energy density other than matter, this effect would occur for a different value of the critical exponents.

Overall, with the exception of the effect of the critical exponents 
discussed above, as we vary the parameters of our scenarios, the accessible regions which reproduce the observed DM relic abundance do not change drastically. As we saw in Figure \ref{fig:monopmod}, the largest shifts occur when the critical exponents are changed. Our main finding 
that for hidden sector monopoles to be dark matter candidates, their masses must be larger than O(PeV) scale appears generic, with longer EMD periods leading to larger monopole masses when \(2 \nu \leq 1+\mu\).

\section{Discussion}\label{disc}

In this work we have considered a scenario for dark matter production via a second order phase transition in the early Universe, where the dark matter (DM) candidate is a hidden-sector magnetic monopole. Such a topological dark matter scenario has been studied before, with the entire relic DM abundance being produced in the standard radiation-dominated (RD) era before BBN \cite{Zurek:1985qw,Zurek:1993ek, Zurek:1996sj,Murayama:2009nj}. 
We have expanded the parameter space region of viability to allow the different sectors to have different temperatures, and by generalizing the cosmological history to include a period of early 
matter domination (EMD). By allowing the phase transition to occur at any time before, during, or after EMD, we have shown that histories involving EMD generally require heavier monopole masses in order to produce the entire DM relic abundance. Along with this general result, we have considered two specific examples of how a period of EMD may be generated: by a modulus, or by a heavy decoupled particle. These examples illustrate how one can embed our scenario in a specific model, and how the underlying model parameters influence the monopole abundance. Our main results are summarized in  Figures \ref{fig:monopmod}, \ref{fig:monopdec}, and \ref{fig:parameters}. 
We generally find that hidden sector monopoles in the mass range O(1--$10^5$) PeV can be dark matter candidates.

We now summarize our main caveats, address some ways our scenario can be changed for future work, and what we 
expect that will do. 

Throughout this work we have assumed the number density for PeV scale monopoles is small enough to ignore the effects of monopole-anti-monopole annihilation, as shown in \cite{Murayama:2009nj} following \cite{Preskill:1979zi}. But because the scattering cross-section between fermions and monopoles is a strongly coupled problem, it is possible that the final monopole abundance is depleted 
more than the diffusion approximation studied in \cite{Preskill:1979zi}, due to 
the interaction with the hidden sector plasma (if present). The interaction of the monopole with the plasma may be 
more critical to understand if the monopole is a dyon, a possibility not considered here. Of course, if the number density decreases further due to annihilation, a higher monopole mass will be needed to get the same DM abundance.

Another key assumption pervading this work is that the second order phase transition is classical, although 
we have strayed from that strict assumption by allowing the critical exponents to have generic values.  But a consequence of assuming the monopole to be a classical topological object  
is that 
the monopole mass and the temperature of the phase 
transition are at similar mass scales, \(m_{\rm M} \sim T_{\rm C}^{\rm(hid)}\). Our conclusions will change substantially in  theories for which this relation no longer holds.  A prominent counter-example is provided by the $N=2$ Seiberg-Witten theory \cite{Seiberg:1994rs,Seiberg:1994aj} near the massless monopole or massless dyon points of the moduli space, 
in which the effective theory 
below the symmetry breaking scale contains nearly massless composites -- `mesons' and `baryons' of a 
magnetic $U(1)$. Additionally, here the effect of annihilations at energies near the scale of the transition
are expected to be important.

Another fundamental assumption in our work is the set-up of our sectors, where we have assumed the sector which hosts the phase transition to interact very weakly, if at all with the visible sector of standard model particles. This can in general be different, and can result in changes to the monopole abundance after their production. For example, kinetic mixing between the visible and hidden sectors can lead to a long-range force which can then deplete the monopole abundance via annihilation. We expect this to have a similar effect to the scattering of monopoles with a HS plasma followed by annihilating, but in this case the monopole abundance can depend more strongly on visible-sector as well as hidden-sector properties. See for example 
\cite{Sanchez:2011mf}.

Along these lines, we have also assumed that the energy density component driving the EMD period decays almost entirely to visible sector radiation. With additional interactions between the sectors, the EMD driving field may decay to hidden sector radiation as well. This can easily be incorporated into our analysis by generalizing the decay rate \(\Gamma_\Phi\) to include branching fractions to both visible and hidden radiation. One must then be careful to not produce too much hidden (or ``dark") radiation by restricting the branching fractions with current limits on dark radiation 
\cite{Ackerman:mha}.

Aside from the set-up of our sectors, another important generalization of our work is to allow for early domination by a component with a generic equation of state, rather than focusing on EMD alone. The redshift relation for the dominating energy density is then \(\rho \propto a^{-3(1+w)}\), with the parameter \(w\) determining the behavior, which modifies subsequent calculations. 

A specific alternative to EMD is a period of kination, where the kinetic energy of a scalar field dominates the energy density of the universe for a time. In such a period, the dominant form of energy density redshifts faster than radiation, with \(w = 1\) and \(\rho \propto a^{-6}\), which can have interesting consequences for the monopole abundance if the phase transition occurs during or before such a period. In fact, the phase transition occurring after a period of kination can also affect the resultant monopole abundance, for example by flipping the radiation energy densities of the two sectors. 
Kination would typically not last very long because it dilutes as \(a^6\), but if other components are suppressed, it can last longer - perhaps the same EMD driving field can have an early period of kination which later transitions to EMD before decaying. One should track the behavior of radiation in the two sectors during such a history to see how it affects the temperatures and thus the final monopole abundance. 

Lastly, in our decoupled particle example, the mechanism of \(\Phi\) decoupling need not be velocity independent. This can lead to temperature dependence in the interaction rate of \(\Phi\) with its host sector and can alter the details of the decoupling. Such effects, however, shouldn't change our main results, just the specifics of the particle decoupling models (what values of $\Phi$ mass and decoupling parameter lead to an EMD phase of a given start and end). 

We hope this work stimulates further research into topological dark matter scenarios.

\section{Acknowledgements}
The authors thank Lukasz Cincio, Jacek Dziarmaga, Erich Poppitz, Marek Rams, John Terning and Wojciech Zurek for useful discussions and bringing references to our 
attention. This research  was  supported  in  part  by  the  U.  S.  Department  of  Energy,  Office  of  High Energy Physics, under Contract No. DE-AC52-06NA25396, the LDRD program at Los Alamos National Laboratory, the Office of Workforce Development for Teachers  and  Scientists,  and the Office  of  Science  Graduate  Student  Research  (SCGSR)  program. 
The SCGSR program is administered by the Oak Ridge Institute for Science and Education (ORISE) for the DOE. ORISE is managed by ORAU under contract number \({\rm DE}\)-\({\rm SC}0014664\).

%\newpage
\appendix

\section{Table of notation}
\label{app:table}
In this Appendix we provide a table of notation. Subscripts generally label the time at which a quantity is evaluated, while superscripts generally label the sector to which a quantity belongs, unless stated otherwise.

\begin{center}
\begin{tabular}{ |l|l| } 
 \hline
 \(T_{\rm i}^{\rm(hid)}\) & Initial temperature of the hidden sector \\ 
 \(T_{\rm i}^{\rm(vis)}\) & Initial temperature of the visible sector \\ 
 \(H_{\rm i}\) & Initial Hubble expansion rate \\ 
 \hline
 \(T_{\rm MD}^{\rm(hid)}\) & Temperature of the hidden sector at the start of EMD \\
 \(T_{\rm MD}^{\rm(vis)}\) & Temperature of the visible sector at the start of EMD \\
 \(H_{\rm MD}\) & Hubble rate at the start of EMD \\
 \hline
 \(T_{\rm RH}^{\rm(hid)}\) & Temperature of the hidden sector at reheating \\
 \(T_{\rm RH}^{\rm(vis)}\) & Temperature of the visible sector at reheating \\
 \(H_{\rm RH}\) & Hubble rate at reheating, approximately equal to \(\Gamma_\Phi\) \\
 \hline
 \(T_{\rm C}^{\rm(hid)}\) & Temperature of the hidden sector at the critical time \\
 \(T_{\rm C}^{\rm(vis)}\) & Temperature of the visible sector at the critical time \\
 \(H_{\rm C}\) & Hubble rate at the critical time \\
 \hline
 \multicolumn{2}{|c|}{Decoupled Particle Case} \\
 \hline
 \(T_{\rm F}^{\rm(hid)}\) & Temperature of the hidden sector when \(\Phi\) decouples \\
 \(T_{\rm F}^{\rm(vis)}\) & Temperature of the visible sector when \(\Phi\) decouples \\
 \(H_{\rm F}\) & Hubble rate when \(\Phi\) decouples \\
 \hline
\end{tabular}
\end{center}

\section{The factors \texorpdfstring{\(e_{\rm f}\)}{} and \texorpdfstring{\(f_{\rm RH}\)}{}}\label{appendix:e_f}

The factor \(e_{\rm f}\), defined as 
\begin{equation}\label{e_f}
    e_{\rm f} \equiv \frac{a_{\rm RH}}{a_{\rm MD}}\,,
\end{equation}
is determined by the duration of the EMD phase, and we can approximate it in the following way. 

At the end of EMD, as \(\Phi\) completes its decay and reheats the visible sector, the ratio of the radiation energy densities of the two sectors becomes fixed as 
\begin{equation}
    f_{\rm RH} \equiv \frac{\rho_{\rm r,RH}^{\rm(vis)}}{\rho_{\rm r,RH}^{\rm(hid)}}\,,
\end{equation}
where the additional subscript `\({\rm RH}\)' on the energy densities indicates their value at reheating. 
At the onset of EMD, the energy densities of \(\Phi\) and radiation are close to equal and we have \(\rho_{\Phi,\rm MD} \approx \rho_{\rm r,MD}\approx 3H_{\rm MD}^2M_{\rm P}^2\), while at the end of EMD we have \(\rho_{\Phi,\rm RH} \approx \rho_{\rm r,RH}\approx 3\Gamma_\Phi^2M_{\rm P}^2\). In the case of initial HS domination, \(\rho_{\rm r,MD}\) is dominated by \(\rho_{\rm r,MD}^{\rm(hid)}\), while for initial visible sector domination it is dominated by \(\rho_{\rm r,MD}^{\rm(vis)}\). The energy density at reheating in both cases is dominated by the visible sector because of our decay requirement. Therefore, the ratio of the visible sector and HS radiation energy densities at reheating is 
\begin{equation}\label{f_R}
    f_{\rm RH} \approx (1+f_{\rm i})
        \frac{\Gamma_\Phi^2}{H_{\rm MD}^2}\left(\frac{H_{\rm MD}}{\Gamma_\Phi}\right)^{8/3}  \,
\end{equation}
where we have redshifted hidden sector quantities back to the start of EMD, and where $f_{\rm i}$ is defined as the ratio of the visible sector to hidden sector radiation energy densities 
at some time $t_{\rm i}$ prior to the onset of the EMD phase, 
\begin{equation}
    f_{\rm i} \equiv \frac{\rho^{\rm (vis)}_{\rm r,i}}{\rho^{\rm (hid)}_{\rm r,i}}~.
\end{equation}
During a MD era we have \(a \propto H^{-2/3}\), which combined with $H_{\rm RH} \simeq \Gamma_\Phi$ gives 
\begin{eqnarray}
     f_{\rm RH} &\simeq& (1+f_i) \frac{a_{\rm RH}}{a_{\rm MD}} \\
     & =  & (1+ f_i) e_{\rm f} \label{eq:fRH-ef}
\end{eqnarray}

To facilitate our comparison between scenarios which include a phase of EMD and those which remain purely RD, we make use of the double ratio 
\begin{eqnarray}\label{f_EMD/RD}
    \frac{f_{\rm RH}^{\rm(EMD)}}{f^{\rm(RD)}_{\rm RH}} &=& \begin{dcases} 
        f_{\rm RH} & \quad f_{\rm i} \ll 1\\
        \frac{f_{\rm RH}}{f_{\rm i}} & \quad f_{\rm i} \gg 1\,,
    \end{dcases} \\
    & \approx & \frac{a_{\rm RH}}{a_{\rm MD}} = e_{\rm f} \quad f_{\rm i} \ll 1 \quad {\rm or} \quad f_{\rm i} \gg 1\ 
    \,,\label{f_EMD/RD-2}
\end{eqnarray}
where in the second line we have made use of \eqref{eq:fRH-ef} and
where we have included superscripts on the two \(f_{\rm RH}\)'s on the left-side for clarity (whenever \(f\) appears without a superscript label, it refers to the EMD case). 
We note that since in any given RD-equivalent scenario $f_{\rm RH}^{\rm (RD)}$ is just a number, to simplify our notation we will often 
drop the subscript and 
just write this term as $f^{\rm (RD)}$. The energy density ratio in a purely RD scenario corresponding to an EMD scenario with initial domination by visible sector radiation is given by \(f^{\rm(RD)}_{\rm RH} = f_{\rm i}\), while in the case of an EMD scenario with initial domination by HS radiation, it is \(f^{\rm(RD)}_{\rm RH} = 1\).

We have additionally numerically verified the value of \(e_{\rm f}\) as the ratio of the scale factors at the end and beginning of the EMD period, as well as the double ratio of radiation energy densities.

\section{Decoupling of \texorpdfstring{\(\Phi\)}{} from either sector via freeze-out}
\label{Appendix:Freeze-out}
In order to analytically estimate the relic abundance of topological DM from \eqref{omegabefore}-\eqref{omegaafter}, we need to obtain an expression for the Hubble rate at the onset of EMD, \(H_{\rm MD}\). We do so by redshifting the frozen number density of \(\Phi\) at the time of freeze-out, given by \(n_{\rm\Phi,F}\), to the start of EMD: 
\begin{equation}
    n_{\rm\Phi,MD} = n_{\rm\Phi,F}\left(\frac{H_{\rm MD}}{H_{\rm F}}\right)^{3/2} = \frac{H_{\rm MD}^{3/2}}{\left<\sigma_\Phi v\right>H_{\rm F}^{1/2}}\,.
\end{equation}
Noting that we have \(m_\Phi n_{\rm\Phi,MD} \approx 3H_{\rm MD}^2M_{\rm P}^2\) at the onset of EMD, we are left with 
\begin{equation}\label{monop:FOHMD}
    H_{\rm MD} \approx \frac{m_\Phi^2}{9\left<\sigma_\Phi v\right>^2M_{\rm P}^4H_{\rm F}}\,.
\end{equation}
What remains is to specify \(H_{\rm F}\), which we do below for a number of cases. 

\noindent
\subsection{Non-relativistic freeze-out from hidden sector} Using the usual freeze-out condition of \(n_{\rm\Phi,eq}\left<\sigma_\Phi v\right> = H_{\rm F}\), with the non-relativistic form of the equilibrium number density for a boson $\Phi$, we have 
\begin{equation}
\label{NR-freeze-out-HS}
    g_\Phi\left(\frac{m_\Phi^2}{2\pi x_{\rm F}}\right)^{3/2}\,{\rm e}^{-x_{\rm F}}\left<\sigma_\Phi v\right> \approx \sqrt{\frac{\pi^2}{90}g_*^{\rm(hid)}(1+f_{\rm i})}\frac{m_\Phi^2}{M_{\rm P}x_{\rm F}^2}\,,
\end{equation}
where we have used \(H_{\rm F} \approx \sqrt{\frac{\pi^2}{90}g_*^{\rm(hid)}(1+f_{\rm i})}\frac{m_\Phi^2}{M_{\rm P}x_{\rm F}^2}\) with \(x_{\rm F} \equiv m_\Phi/T_{\rm F}^{\rm(hid)}\). Rearranging yields an expression that can be solved for \(x_{\rm F}\): 
\begin{equation}
    x_{\rm F} \approx \ln\left(\frac{3\sqrt{5}g_\Phi\left<\sigma_\Phi v\right>m_\Phi M_{\rm P}x_{\rm F}^{1/2}}{2\pi^{5/2}{g_*^{\rm(hid)}}^{1/2}(1+f_{\rm i})^{1/2}}\right)\,.
\end{equation}
If $\Phi$ is instead a fermion, the left-side of (\ref{NR-freeze-out-HS}) is multiplied by a factor of $3/4$, with a corresponding change in the expression for $x_{\rm F}$.
The solution to this can then be used in the expression for \(H_{\rm F}\) above to complete its specification in terms of the parameters of our scenario. 

\noindent
\subsection{Non-relativistic freeze-out from visible sector} Here we define \(x_{\rm F} \equiv m_\Phi/T_{\rm F}^{\rm(vis)}\), resulting in 
\begin{equation}
\label{NR-freeze-out-VS}
    g_\Phi\left(\frac{m_\Phi^2}{2\pi x_{\rm F}}\right)^{3/2}\,{\rm e}^{-x_{\rm F}}\left<\sigma_\Phi v\right> \approx \sqrt{\frac{\pi^2}{90}g_{\rm*F}^{\rm(vis)}\left(1+\frac{1}{f_{\rm i}}\right)}\frac{m_\Phi^2}{M_{\rm P}x_{\rm F}^2}\,.
\end{equation}
and 
\begin{equation}
    x_{\rm F} \approx \ln\left(\frac{3\sqrt{5}g_\Phi\left<\sigma_\Phi v\right>m_\Phi M_{\rm P}x_{\rm F}^{1/2}}{2\pi^{5/2}{g_{\rm*F}^{\rm(vis)}}^{1/2}\left(1+\frac{1}{f_{\rm i}}\right)^{1/2}}\right)\,.
\end{equation}
Otherwise, this case is the same as above. 

\noindent
\subsection{Relativistic freeze-out from hidden sector} In this case, we use the relativistic expression for the equilibrium number density, giving  
\begin{equation}
    \frac{\zeta(3)g_\Phi m_\Phi^3}{\pi^2x_{\rm F}^3}\left<\sigma_\Phi v\right> \approx \sqrt{\frac{\pi^2}{90}g_*^{\rm(hid)}(1+f_{\rm i})}\frac{m_\Phi^2}{M_{\rm P}x_{\rm F}^2}\,,
\end{equation}
and 
\begin{equation}\label{monop:xfRFOHS}
    x_{\rm F} \approx \frac{\sqrt{90}\zeta(3)g_\Phi\left<\sigma_\Phi v\right>M_{\rm P}m_\Phi}{\pi^3{g_*^{\rm(hid)}}^{1/2}(1+f_{\rm i})^{1/2}}\,.
\end{equation}

\noindent
\subsection{Relativistic freeze-out from visible sector} In this case, we have  
\begin{equation}
    \frac{\zeta(3)g_\Phi m_\Phi^3}{\pi^2x_{\rm F}^3}\left<\sigma_\Phi v\right> \approx \sqrt{\frac{\pi^2}{90}g_{\rm*F}^{\rm(vis)}\left(1+\frac{1}{f_{\rm i}}\right)}\frac{m_\Phi^2}{M_{\rm P}x_{\rm F}^2}\,,
\end{equation}
and 
\begin{equation}\label{monop:xfRFOVS}
    x_{\rm F} \approx \frac{\sqrt{90}\zeta(3)g_\Phi\left<\sigma_\Phi v\right>M_{\rm P}m_\Phi}{\pi^3{g_{\rm*F}^{\rm(vis)}}^{1/2}\left(1+\frac{1}{f_{\rm i}}\right)^{1/2}}\,.
\end{equation}

\section{Decoupling of \texorpdfstring{\(\Phi\)}{} from either sector via freeze-in}\label{Appendix:Freeze-in}
Because \(\Phi\) is the source of the EMD period, at some point it decouples in the prior RD phase.
If the annihilation rate to produce $\Phi$ is too tiny, $\Phi$ may never reach local, chemical and thermal equilibrium with the ambient radiation. However, the produced number density of $\Phi$ particles may be large enough to eventually dominate the energy density. This is known as {\em freeze-in} \cite{Hall:2009bx}. 
In this case, 
freeze-in in a RD period is dominated by the relativistic component and the abundance is set at the initial time. We begin with 
\begin{equation}
    \frac{d(a^3n_\Phi)}{dt} = a^3\left<\sigma_\Phi v\right>(n_{\rm\Phi,eq}^2 - n_\Phi^2) - a^3\Gamma_\Phi n_\Phi\,,
\end{equation}
We are interested in the early evolution of the \(\Phi\) number density in a freeze-in scenario well-before it decays, as well as well-before it reaches equilibrium. Thus we may drop the decay term relative to the decoupling term above, as well as the actual number density relative to the thermal equilibrium value. With these approximations we have 
\begin{equation}
    \frac{d(a^3n_\Phi)}{dH} = -\frac{a^3 \left<\sigma_\Phi v\right>n_{\rm\Phi,eq}^2}{2H^{2}}\,, 
\end{equation}
which for $a=a_{\rm i} (t/t_{\rm i})^{1/2}$ and $H=1/2t$, appropriate for RD, one has 
\begin{equation}
 \frac{d(a^3n_\Phi)}{dH}   = -\frac{a_{\rm i}^3H_{\rm i}^{3/2}\left<\sigma_\Phi v\right>n_{\rm\Phi,eq}^2}{2H^{7/2}}\,.
\end{equation}
To continue, we must express the temperature dependence of the equilibrium number density in terms of \(H\), which is most easily done by specializing to the two decoupling cases. 

\noindent
\subsection{Freeze-in from hidden sector} If \(\Phi\) is produced from the HS, we have 
\begin{equation}
    a_{\rm F}^3n_{\rm\Phi,F} \approx -\frac{90^{3/2}\zeta(3)^2g_\Phi^2a_{\rm i}^3H_{\rm i}^{3/2}\left<\sigma_\Phi v\right>M_{\rm P}^3}{2\pi^7{g_*^{\rm(hid)}}^{3/2}(1+f_{\rm i})^{3/2}}\int_{H_{\rm i}}^{H_{\rm F}}\frac{dH}{H^{1/2}}\,,
\end{equation}
which results in a produced freeze-in number density of 
\begin{equation}
    n_{\rm\Phi,F} \approx \frac{90^{3/2}\zeta(3)^2g_\Phi^2H_{\rm F}^{3/2}\left<\sigma_\Phi v\right>M_{\rm P}^3H_{\rm i}^{1/2}}{\pi^7{g_*^{\rm(hid)}}^{3/2}(1+f_{\rm i})^{3/2}}\,.
\end{equation}
Assuming this can be large enough to dominate the energy density at, by definition, the beginning of EMD, and using \(m_\Phi n_{\rm\Phi,MD} \approx 3H_{\rm MD}^2M_{\rm P}^2\), setting $H_{\rm F}=H_{\rm MD}$ gives 
\begin{equation}
    H_{\rm MD} \approx \frac{90^3\zeta(3)^4g_\Phi^4M_{\rm P}^2\left<\sigma_\Phi v\right>^2m_\Phi^2H_{\rm i}}{9\pi^{14}{g_*^{\rm(hid)}}^3(1+f_{\rm i})^3}\,.
\end{equation}

\noindent
\subsection{Freeze-in from visible sector} If \(\Phi\) is produced from the visible sector, we similarly have 
\begin{equation}
    n_{\rm\Phi,F} \approx \frac{90^{3/2}\zeta(3)^2g_\Phi^2H_{\rm F}^{3/2}\left<\sigma_\Phi v\right>M_{\rm P}^3H_{\rm i}^{1/2}}{\pi^7{g_{\rm*i}^{\rm(vis)}}^{3/2}\left(1+\frac{1}{f_{\rm i}}\right)^{3/2}}\,,
\end{equation}
and  
\begin{equation}
    H_{\rm MD} \approx \frac{90^3\zeta(3)^4g_\Phi^4M_{\rm P}^2\left<\sigma_\Phi v\right>^2m_\Phi^2H_{\rm i}}{9\pi^{14}{g_{\rm*i}^{\rm(vis)}}^3\left(1+\frac{1}{f_{\rm i}}\right)^3}\,.
\end{equation}

In sum, the equations in this Appendix give the number density $n_{\rm F}$ of $\Phi$ particles in a freeze-in scenario, assuming it is produced in the early Universe from either the hidden or visible sectors, evaluated well-before it decays. And by definition of the freeze-in scenario,  
the number density $n_{\rm F}$ is assumed to be well-below its equilibrium number density. 

\section{Additional consistency constraint for the decoupled \texorpdfstring{\(\Phi\)}{} scenario}
\label{Appendix:Additional-constraint}
We obtain another constrain that must be satisfied in order for the EMD phase caused by the decoupled \(\Phi\) to have nonzero duration. If \(\Phi\) decouples from the subdominant sector, the value of \(f_{\rm i}\) must be such that the decoupled number density is large enough to lead to EMD. Using \eqref{sigmavFO_R/NR} for an annihilation rate that achieves relativistic freeze-out (which corresponds to the maximum frozen number density and thus longest possible duration for EMD), we require \(H_{\rm MD} \gtrsim \Gamma_\Phi\). Using \eqref{monop:FOHMD} for \(H_{\rm MD}\) and \eqref{monop:xfRFOHS} and \eqref{monop:xfRFOVS} for \(x_{\rm F}\) in their respective cases, we have 
\begin{equation}
    f_{\rm i} \lesssim \left(\frac{30\sqrt{10}\zeta(3)^2g_\Phi^2m_\Phi^2}{\pi^7{g_*^{\rm(hid)}}^{3/2}M_{\rm P}\Gamma_\Phi}\right)^{2/3}\,,
\end{equation}
in the case of decoupling from the HS while the VS is dominant, and 
\begin{equation}
    f_{\rm i} \gtrsim \left(\frac{\pi^7{g_{\rm*F}^{\rm(vis)}}^{3/2}M_{\rm P}\Gamma_\Phi}{30\sqrt{10}\zeta(3)^2g_\Phi^2m_\Phi^2}\right)^{2/3}\,,
\end{equation}
in the case of decoupling from the VS while the HS is dominant. 

\bibliographystyle{JHEP}
\bibliography{bibliography}

\end{document}